\documentclass[
12pt, % The default document font size, options: 10pt, 11pt, 12pt
oneside, % Two side (alternating margins) for binding by default, uncomment to switch to one side
english, % ngerman for German
singlespacing, % Single line spacing, alternatives: onehalfspacing or doublespacing
headsepline, % Uncomment to get a line under the header
]{MastersDoctoralThesis} % The class file specifying the document structure
\usepackage[utf8]{inputenc} % Required for inputting international characters
\usepackage[T1]{fontenc} % Output font encoding for international characters
\usepackage{verbatim} % For commenting out large sections if needed
\usepackage{mathpazo} % Use the Palatino font by default - OVERRIDE FOR DEFAULT
\usepackage{lmodern}
\usepackage{sansmath}
\usepackage[utf8]{inputenc}
\usepackage{amsmath}
\usepackage{bm}
\usepackage{amssymb}
\usepackage{float}
\usepackage{textcomp}
\usepackage{multirow}
\usepackage{booktabs} %For tables
\usepackage{ragged2e}
\usepackage[obeyspaces,hyphens,spaces]{url}

\usepackage{hyperref}
\usepackage{cleveref}
\crefformat{section}{\S#2#1#3} % see manual of cleveref, section 8.2.1
\crefformat{subsection}{\S#2#1#3}
\crefformat{subsubsection}{\S#2#1#3}
\usepackage[backend=biber,citestyle=authoryear-comp,url=false,natbib=true,sorting=ynt]{biblatex}
\newcommand*{\figref}[2][]{%
  \hyperref[{fig:#2}]{%
    Figure~\ref*{fig:#2}%
    \ifx\\#1\\%
    \else
      \,#1%
    \fi
  }%
}
\DeclareRobustCommand{\uvec}[1]{{%
  \ifcat\relax\noexpand#1%
    \bm{\hat{#1}}%
  \else
    \ifcsname uvec#1\endcsname
      \csname uvec#1\endcsname
    \else
      \bm{\hat{\mathbf{#1}}}%
     \fi
   \fi
}}
\NewDocumentCommand{\evalat}{sO{\big}mm}{%
  \IfBooleanTF{#1}
   {\mleft. #3 \mright|_{#4}}
   {#3#2|_{#4}}%
}
\usepackage{hyphenat}
\usepackage{xparse}
\usepackage{enumitem}
\DeclareMathOperator{\sech}{sech}
\graphicspath{
    {Figures/Ch0/}
    {Figures/Ch1/}
    {Figures/Ch2/}
    {Figures/Ch3/}
    }

\usepackage{subcaption}
\usepackage[labelfont=bf]{caption}
\usepackage[autostyle=true]{csquotes} % Required to generate language-dependent quotes in the bibliography

\thesistitle{Star formation in disk galaxies and \\
its relation with spiral structure \\ in numerical simulations} % Your thesis title, this is used in the title and abstract, print it elsewhere with \ttitle
\supervisor{Dr. Juan Carlos \textsc{Muñoz-Cuartas}} % Your supervisor's name, this is used in the title page, print it elsewhere with \supname
\examiner{} % Your examiner's name, this is not currently used anywhere in the template, print it elsewhere with \examname
\degree{Astronomer} % Your degree name, this is used in the title page and abstract, print it elsewhere with \degreename
\author{B.Sc. David Felipe\\ \textsc{Barros Ramirez}} % Your name, this is used in the title page and abstract, print it elsewhere with \authorname
\addresses{} % Your address, this is not currently used anywhere in the template, print it elsewhere with \addressname

\subject{Astronomy} % Your subject area, this is not currently used anywhere in the template, print it elsewhere with \subjectname
 % Keywords for your thesis, this is not currently used anywhere in the template, print it elsewhere with \keywordnames
\university{\href{http://www.udea.edu.co}{Universidad de Antioquia}} % Your university's name and URL, this is used in the title page and abstract, print it elsewhere with \univname
\department{\href{http://www.udea.edu.co/wps/portal/udea/web/inicio/unidades-academicas/ciencias-exactas-naturales/acerca-facultad/institutos/instituto-fisica}{Instituto de Física}} % Your department's name and URL, this is used in the title page and abstract, print it elsewhere with \deptname
\group{\href{http://www.udea.edu.co/wps/portal/udea/web/inicio/investigacion/grupos-investigacion/ciencias-naturales-exactas/facom/}{Grupo de Física y Astrofísica Computacional (FACom)}} % Your research group's name and URL, this is used in the title page, print it elsewhere with \groupname
\faculty{\href{http://www.udea.edu.co/wps/portal/udea/web/inicio/unidades-academicas/ciencias-exactas-naturales}{Facultad de Ciencias Exactas y Naturales}} % Your faculty's name and URL, this is used in the title page and abstract, print it elsewhere with \facname

\AtBeginDocument{
\hypersetup{pdftitle=\ttitle} % Set the PDF's title to your title
\hypersetup{pdfauthor=\authorname} % Set the PDF's author to your name
\hypersetup{pdfkeywords=\keywordnames} % Set the PDF's keywords to your keywords
\hypersetup{pdfborder={0 0 1}}
}

\begin{document}

\frontmatter % Use roman page numbering style (i, ii, iii, iv...) for the pre-content pages

\pagestyle{plain} % Default to the plain heading style until the thesis style is called for the body content
\mainmatter
%----------------------------------------------------------------------------------------
%	TITLE PAGE  (Commented = ON)
%----------------------------------------------------------------------------------------

\begin{titlepage}
\begin{center}

\begin{figure}[ht]
    \centering
    \includegraphics[width=0.35\textwidth]{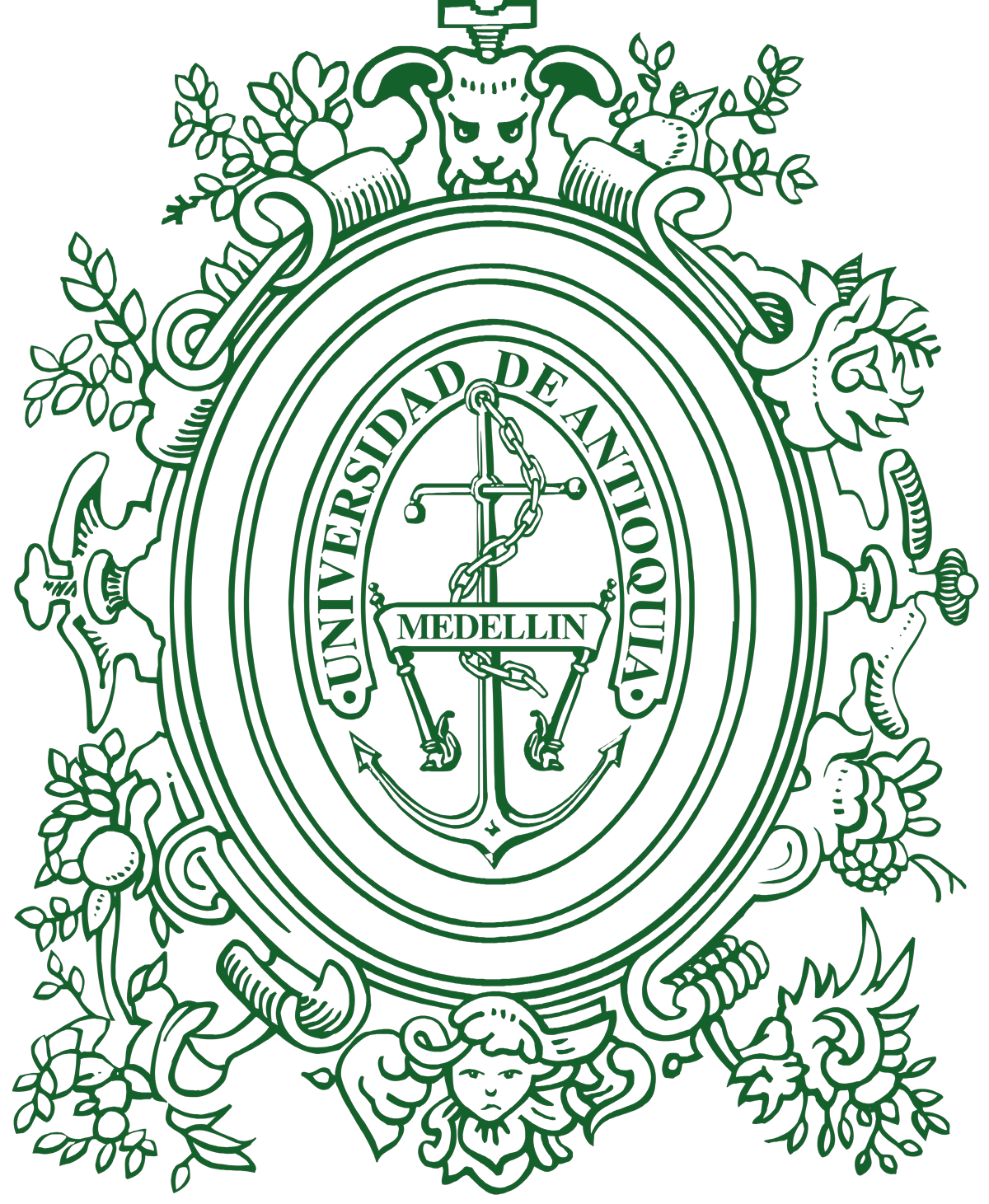} % University/department logo - uncomment to place
\end{figure}
%\vspace*{.06\textheight}
{\scshape\LARGE \univname\par}\vspace{1.5cm} % University name
\textsc{\Large Bachelor Thesis}\\[0.5cm] % Thesis type

\HRule \\[0.4cm] % Horizontal line
{\huge \bfseries \ttitle\par}\vspace{0.4cm} % Thesis title
\HRule \\[1.5cm] % Horizontal line
 
\begin{minipage}[t]{0.4\textwidth}
\begin{flushleft} \large
\emph{Author:}\\
\href{https://scienti.minciencias.gov.co/cvlac/visualizador/generarCurriculoCv.do?cod_rh=0001665707}{\authorname} % Author name - remove the \href bracket to remove the link
\end{flushleft}
\end{minipage}
\begin{minipage}[t]{0.4\textwidth}
\begin{flushright} \large
\emph{Supervisor:} \\
\href{http://scienti.colciencias.gov.co:8081/cvlac/visualizador/generarCurriculoCv.do?cod_rh=0000517127}{\supname} % Supervisor name - remove the \href bracket to remove the link  
\end{flushright}
\end{minipage}\\[2cm]
 
\vfill
\large \textit{A thesis submitted in fulfillment of the requirements\\ for the degree of \degreename}\\[0.3cm] % University requirement text
%\textit{in the}\\[0.4cm]
\groupname\\\facname\\\deptname\\[2cm] % Research group name and department name
 
\vfill

%{\large \today}\\[4cm] % Date
 
\vfill
\end{center}
\end{titlepage}

\begin{acknowledgements}
\addchaptertocentry{\acknowledgementname} % Add the acknowledgements to the table of contents
The completion of this work would not have been possible without the support of many people that helped me during the process. I want to thank specially to my parents who always supported me and my dream of becoming an astronomer. Their love and support during all my life are the foundation of this achievement. I would also like to thank my brother and sister for their constant support. 
\\[12pt]
I want to thank all my friends in the astronomy program at the university, with whom I went through all the ups and downs that academia brings. They definitely made this journey more enjoyable and fun. 
To my girlfriend and best friend Teresa, for giving me energy and motivation to finish this work, and for being the best company during all the time I spent developing this work and writing this manuscript. 
\\[12pt]
I also want to thank my advisor Dr. Juan Carlos Muñoz Cuartas for his support during all the stages of this work. I have learned a lot under his direction and, most importantly, I have been able to trust my potential and capabilities. 
\\[12pt]
This research was funded by COLCIENCIAS. Project FP44842- 285-2016 – Code: 111571250082 and by the program "Joven Investigador Universidad de Antioquia". \par 
\end{acknowledgements}

%----------------------------------------------------------------------------------------
%	LIST OF CONTENTS/FIGURES/TABLES PAGES
%----------------------------------------------------------------------------------------
\hypersetup{linkcolor=mdtRed}
\tableofcontents % Prints the main table of contents

\listoffigures % Prints the list of figures

\listoftables % Prints the list of tables

\dedicatory{A Bismark y Martha.}

%----------------------------------------------------------------------------------------
%	THESIS CONTENT - CHAPTERS
%----------------------------------------------------------------------------------------

%\mainmatter % Begin numeric (1,2,3...) page numbering

\pagestyle{thesis} % Return the page headers back to the "thesis" style

% Include the chapters of the thesis as separate files from the Chapters folder
% Uncomment the lines as you write the chapters

% Chapter Template

\chapter{Introduction} % Main chapter title
\label{Chapter1} % Change X to a consecutive number; for referencing this chapter elsewhere, use \ref{ChapterX}

\begin{figure}[H]
\begin{subfigure}{0.49\textwidth}
  %\captionsetup[subfigure]{aboveskip=-1pt,belowskip=-1pt}
  \centering
  \includegraphics[width=0.6\linewidth]{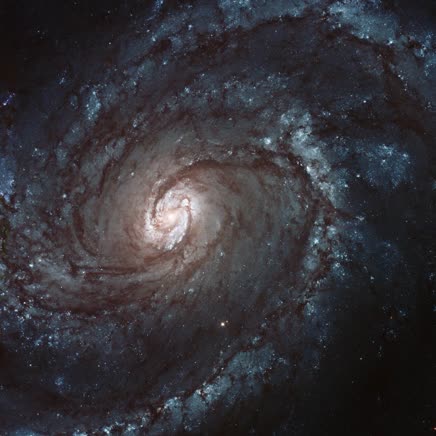}
  %\caption{}
  \label{fig:1.1a}
\end{subfigure}%
\begin{subfigure}{.5\textwidth}
  \centering
  \includegraphics[width=0.6\linewidth]{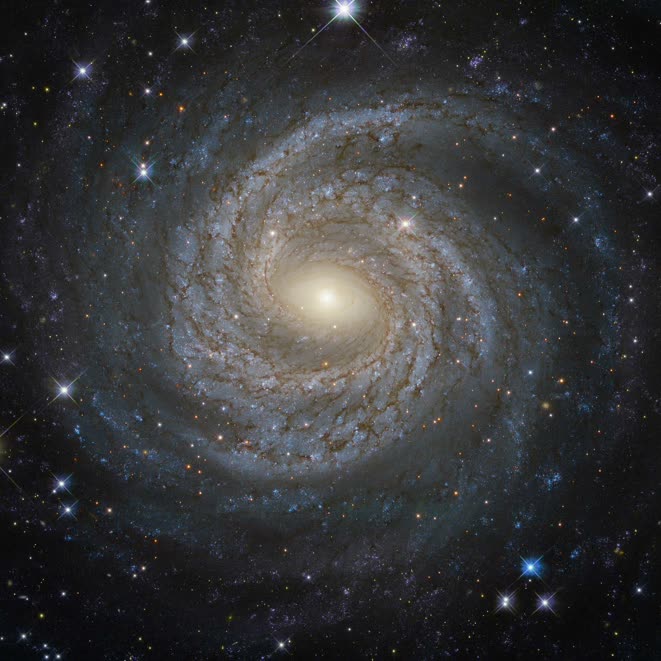}
  %\caption{}
  \label{fig:1.1b}
\end{subfigure}
%\vspace{-1\baselineskip}
\caption[M100 and NGC 6814 spiral galaxies.]{Left, M100 galaxy. Right, NGC 6814 galaxy. Two examples of "Grand Design" spirals. Credits: ESA/Hubble.}
\label{fig:1.1}
\end{figure}

%----------------------------------------------------------------------------------------
%	SECTION 1
%----------------------------------------------------------------------------------------
\noindent
The implementation of computational simulations in the solution of the N-body problem is a milestone in modern astrophysics. The use of computers has allowed us to study complex systems for which analytical solutions are impossible to obtain. One of the applications of these type of simulations is the modelling of galaxies, systems which are of particular interest due to their internal dynamics, evolution and structure. 
\\[12pt]
\noindent
In the large scale structure of the universe, galaxies constitute the building blocks and thus they are subject of careful study. Galaxies play a fundamental role in the study of cosmological theories such as the existence of dark matter, dark energy and early-universe hypothesis like inflation; in this sense they are not only interesting astronomical objects per se but they also serve as enormous laboratories to further understand the Universe at its greatest scale.  
\\[12pt]
\noindent
There exists a rich variety of galaxies, most of them classified in Hubble's sequence. We can roughly separate them in elliptical and disk galaxies. In this work, the subject of study is the latter, more specifically, spiral galaxies, like the ones showed in \figref{1.1} This type of galaxy is, morphologically, a three component system: galactic halo, bulge and disk. Disks are present everywhere in our Universe, accretion disks around compact objects such as black holes or neutron star binaries, planetary systems, planetary rings, amongst others. They originate in systems driven by gravity where angular momentum is conserved. All of these systems present a net rotation, and as the gas radiates energy while rotating, a flat disk is the configuration in which energy is minimized. 

\begin{figure}[ht]
    \centering
    \includegraphics[width=1\textwidth]{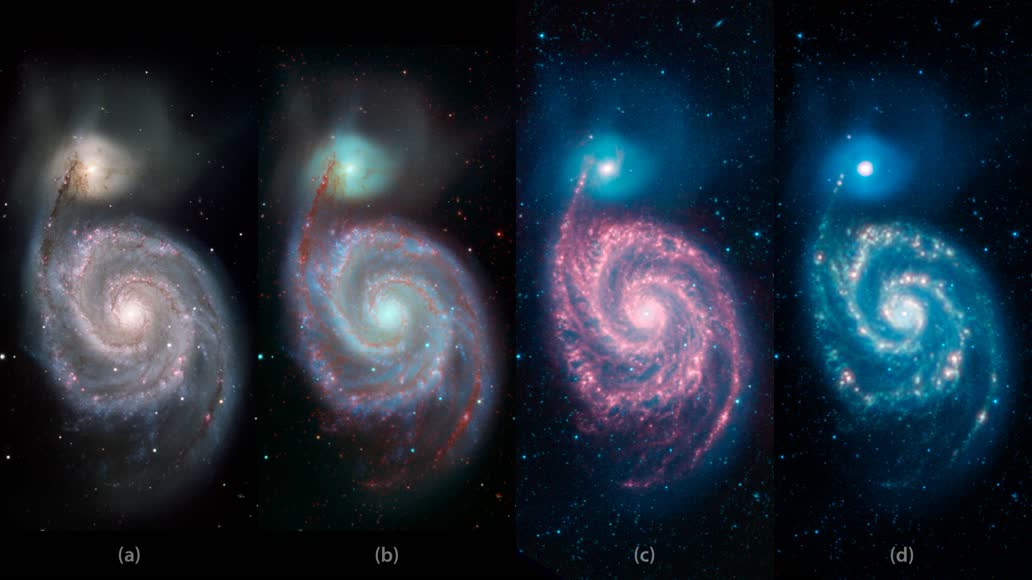}
    \caption[M51 galaxy viewed at different wavelengths.]{The Whirpool galaxy (M51) and its companion NGC 5194/5195 viewed at different wavelengths . (a) Combination of two wavelengths at 0.4 microns (blue) and 0.7 microns (green) captured at  Kitt Peak National Observatory. (b) Combination of visible light and infrared captured by NASA's Spitzer Space Telescope. (c) Combination of infrared light: 3.6 microns (shown in blue), 4.5 microns (green) and 8 microns (red) and finally (d) includes light at a wavelength of 24 microns (in red). Credits: KPNO, Nasa/Spitzer}
    \label{fig:1.2}
\end{figure}

\noindent
When simulating these type of objects, we have to account for multiple type of physics beyond gravitational interactions between particles. Modern computational codes implement hydrodynamics to model the gas' dynamics, some include ideal magneto-hydrodynamics to incorporate the effects of magnetic fields in the dynamics of the gas. Additional physics such as radiative heating, gas cooling or energy feedback are some of the complexities that have been recently added to the mix (\citealt{springel2005cosmological,hopkins2015new}). In this work we are particularly interested in the simulations of isolated disk galaxies. Said simulations, given certain conditions, give rise to spiral-arm structures and, in general, region of overdensities in the gas disk.  
\\[12pt]
\noindent
Despite that nowadays a fairly good understanding of the physics inside disk galaxies exists, the mechanisms under which the stability of the spiral structure is sustained remains as an open problem. Different hypothesis have been proposed to explain the origin of the spiral structures, one of the most important is the one in which spiral arms propagate inside the disk as density waves (\citealt{lin1987spiral}). The problem of the origin of the spiral arms is not the main goal of the present work, instead, we want to characterize the overdensity regions using different properties like the gas density, the gas' internal energy and the star formation rate.
\\[12pt]
\noindent
The process of star formation involves complex physics and it constitutes a frontier problem on which new advances are constantly being made. The way in which the Star Formation Rate (SFR) is measured is a difficult task, however, some tracers such as H$\alpha$ emission and gas surface density have been used to estimate its value from observations (\citealt{kennicutt1989star,kennicutt1998global}). When it comes to simulations, they try to reproduce some of the observed properties of the gas, such as surface density, star formation history, amongst other (\citealt{springel2003history,ceverino2009role,khalatyan2008agn,gnedin2010kennicutt,hopkins2011self,hopkins2012stellar,sales2009origin}). As of today, there is not enough work that seeks to reproduce the properties (structure, SFR, mass and gas distribution) of observed systems. Some work in this direction, however, has been made by \citealt{krabbe2011effects}, \citealt{quiroga2017comparing}, to name a few.
\\[12pt]
\noindent
Galaxies can vary its features and morphology depending on the wavelength they are observed at. Different physical phenomena emits different types of light, depending on how energetic the processes involved are. If we observe a galaxy in near infrared it will show more visible stars or giant post-main sequence stars; on the other hand if we observe in Ultraviolet we will get information on recent star formation. The spiral structure, however, can be seen through the "eyes" of different features such as dust, free electrons, molecular gas, neutral gas or H$\alpha$ regions. How simulated galaxies "seem" in those properties?, current star formation models account for the observational results?, does the spiral structure changes significantly? In \figref{1.2} we see the spiral galaxy M51 viewed at different wavelengths, do the spiral arms in the simulated galaxy change its shape, depending on what property we are "looking" through?, can we somehow model the shape of the arms in a consistent manner?, does this model apply for all the properties?, those, amongst other questions, we will try to answer in this work. 

\noindent
Being clear about the problem's context and its theoretical framework, we present an extensive analysis on a series of numerical simulations using the GIZMO multi-physics code, in which we analyse the possible correlation between the characteristics of the star formation process and the spiral structure. In Chapter 1 some theoretical background is given, including some theory on morphology and structure of disk galaxies, their mass distributions, the dynamics of the spiral structure and some of the related work on the origin of spiral structure; in Chapter 2 we describe the  simulations, the code  implemented  for  the  analysis and the complete process to extract the spiral structure; in Chapter 3 we show the results of this process for some properties of the gas (density, SFR, free electrons per number proton, internal energy and Neutral Hydrogen fraction) and compare its features and properties. Finally we give some conclusions. 
% Chapter 1

\chapter{Theoretical Framework} % Main chapter title

\label{Chapter2} % For referencing the chapter elsewhere, use \ref{Chapter1} 

%----------------------------------------------------------------------------------------

% Define some commands to keep the formatting separated from the content 
\newcommand{\keyword}[1]{\textbf{#1}}
\newcommand{\tabhead}[1]{\textbf{#1}}
\newcommand{\code}[1]{\texttt{#1}}
\newcommand{\file}[1]{\texttt{\bfseries#1}}
\newcommand{\option}[1]{\texttt{\itshape#1}}

%----------------------------------------------------------------------------------------
In this Chapter some basic concepts and theory regarding disk galaxies will be presented. We will also give a general overview on spiral wave theory that shall help us grasp a general picture of the problem. The concepts provided in this chapter will be of relevance in the posterior analysis of the problem. 

%%%%%%%%%%%%%%% MORPHOLOGY AND STRUCTURE OF DISK GALAXIES %%%%%%%%%%%%%%%
\section{Morphology and Structure of Disk Galaxies}
Since the subject of study of this work is spiral galaxies we will only discuss their properties, ignoring the theory of elliptical systems. When the Hubble sequence appeared it was understood as a diagram that pictured the evolution of galaxies from elliptical (early-type), passing through spirals (early and late-type) and finishing in irregulars, as seen in Figure 2.1. This nomenclature (early/late) is of widespread use in literature, although it does not account for the real process of evolution which is believed to actually be backwards from what Hubble initially proposed. We will start by giving a brief overview on the morphology of spiral galaxies and then talk about the differences between them that make the Sa/Sb/Sc classification. \par

%%%%%%%%%%%%%%% SPIRAL GALAXIES: A THREE-COMPONENT SYSTEM %%%%%%%%%%%%%%%
\subsection{Spiral Galaxies: a three-component system}
\label{section2.1.1}
The first morphological models of spirals were made by studying our Galaxy, the Milky Way. The first approach based on star counts was made by William Herschel  in the 1780s, his model was purely qualitative and did not gave any numerical scale for distances. Shortly after, Jacobus Kapteyn used a quantitative approach based on star counts to confirm Hershel's model and concluded that our Galaxy had $\sim$20kpc in diameter with a disk-like shape. Harlow Shapley used variable stars located in globular clusters and estimated the diameter to be $\sim$100kpc. \par
\noindent
As it was later shown, both estimations were wrong because they did not account for interstellar extinction in their calculations. Although their models were numerically inaccurate, they all agreed that the Milky way possesses a disk of stars, and the Sun makes part of it.
\\[12pt]
\noindent
%%%%%%%%%%%%%%% DISK %%%%%%%%%%%%%%%
\textbf{(a) Disk}. The cold disk is perhaps the most interesting part of the whole galaxy, the majority of the gas is located there, and for that reason, almost the totality of the star formation processes take place in this region. Disks are modelled as axially-symmetric systems (bars and spiral arms can be treated as perturbations) and observations of galactic disk at various inclinations show that its radial brightness profile is well-fitted by a function of the form
\begin{equation}
\label{eq:brightness_disk_radial}
    I(R) = I_0e^{-R/h}
\end{equation}
\noindent
where $h$ is the radial scale-length of the disk, the point at which the central brightness $L_0$ is reduced by a factor of $1/e$. As for the vertical structure of the disk, photometry performed on edge-on disk galaxies shows that their brightness can be fitted with a hyperbolic secant function

\begin{equation}
\label{eq:brightness_disk_vertical}
    I(z) = I_0 \sech^2(z/z_0),
\end{equation}

\noindent
where $z_0$ is the vertical scale. To make a description of the brightness density, understood as the brightness per unit volume at any point of the disk, we can model the whole disk as the product of equations (\ref{eq:brightness_disk_radial}) and (\ref{eq:brightness_disk_vertical}): 

\begin{equation}
\label{brightness_combination}
    I(R,z) = I_0 e^{-R/h} \sech^2(z/z_0).
\end{equation}

The fact that this is possible implies that the vertical scale height does not depend on the radius, i.e, the vertical shape does not change, independent of $R$. 
Astronomers have observed a wide range of sizes when it comes to radial and vertical scales of disk galaxies. Usual values of $h$ lie in the range of $1 \lesssim h \lesssim 14$ kpc, with most galaxies around $h \approx $ 6.5 kpc; for the vertical scale, typical values are around $0.3 \lesssim z_0 \lesssim 4$ kpc, with a mean value around 1.5 kpc (\citealt{yoachim2006structural,bizyaev2014catalog}).
\\[12pt]

\noindent
%%%%%%%%%%%%%%% BULGE %%%%%%%%%%%%%%%
\noindent
\textbf{(b) Bulge}. The bulge is a spheroidal distribution of stars and gas located at the center of the galaxy. It is dominated by an old stellar population, although a wide variety of stars from different ages have been observed, at least in the case of our Galaxy. It is observed with more or less prominence between spiral galaxies and its brightness is one of the criteria to label spiral galaxies. Bulges' brightness profile is usually described like an elliptical galaxy, as their morphology are basically the same. Such brightness profile is a $r^{1/4}$ law called the de Vaucouleurs Profile:
\begin{equation}
\label{eq:de_vaucouleurs}
    \log \left(\frac{I(R)}{I_{\mathrm{e}}}\right)=-3.3307\left[\left(\frac{R}{R_{\mathrm{e}}}\right)^{1 / 4}-1\right]
\end{equation}
where $R_e$ is defined as the effective radius, i.e, the radius that contains half of the total brightness $I_e$. In the case of barred spiral galaxies, this bulge has an elongated or ellipsoidal shape. On the origin of bulges in spiral galaxies there are two main theories: secular instabilities in the rotating disk or through minor mergers.
\\[12pt]
%%%%%%%%%%%%%%% HALO %%%%%%%%%%%%%%%
\noindent
\textbf{(c) Halo}. The galactic halo is usually divided in two components: the stellar and dark halo. The stellar halo is the less luminous of the three components, it is composed of globular clusters and field stars that are not part of any particular group. Stars of the Halo are characterized by their high velocities and they usually reach great distances above and below the galactic plane. The density distribution for the globular clusters and field stars in the halo, in the case of our Galaxy, follows a power law of the form $\rho \propto r^{-3}$ at large radii. \par
\noindent
On the other hand, the dark halo is more of a theoretical consequence of observation; it is also the least understood component since the nature of its only constituent, dark matter, is still unknown. The first evidence on the existence of dark matter halos around spiral galaxies came from analysing circular-speed curves, which appear to be flat at large galactocentric radii, which suggests that the mass in the outer regions is spherically distributed. Indeed, a flat rotation curve implies that $\rho \propto r^{-2}$, which clearly can not be attributed to stars or gas at such large distances. 

%%%%%%%%%%%%%%% HUBBLE CLASSIFICATION %%%%%%%%%%%%%%%
\subsection{Hubble classification}
\label{section_hubble_classification}
\begin{figure}[ht]
    \centering
    \includegraphics[width=0.6\textwidth]{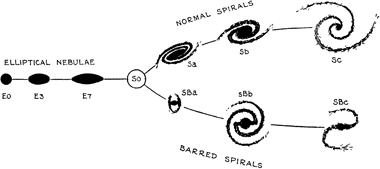}
    \caption[Hubble sequence]{The Hubble sequence as it appears in The Realm of the Nebulae (\citealt{hubble1936realm})}
    %\caption[Hubble sequence]{The Hubble sequence depicting elliptical, lenticular, spiral and barred spirals. From %Ville Koistinen, licensed under \href{https://creativecommons.org/licenses/by-sa/3.0/}{CC BY-SA 3.0}}
    \label{fig:hubble_sequence}
\end{figure}

With the main aspects of the morphology covered we can discuss the different types of spiral galaxies in the Hubble sequence. Hubble classification, as was originally conceived by him, depicts an evolutionary scheme from left to right, as seen in \figref{hubble_sequence}. The far left are elliptical galaxies E0-E7 where the number is a measure of ellipticity. The S0 galaxies are called lenticular and are an intermediate state between ellipticals and spirals, they contain a large bulge and a disk-like structure in its surroundings that lacks a spiral structure. Spiral galaxies are subdivided into two groups, "normal" spirals (S's) in the upper branch and barred spirals (SB's) in the lower branch. \par
\noindent
%%%%%%%%%%%%%%% EARLY AND LATE-TYPE SPIRALS %%%%%%%%%%%%%%%
\textbf{- Early and late-type spirals} Spirals are usually divided into early and late. Sa-Sab are called early and Sb-Sbc-Sc late. The classification of the spiral galaxies is made based on two observational properties: the bulge/disk brightness ratio and the pitch angle. The bulge/disk brightness is the percentage of the total brightness coming from the bulge and the pitch angle is a measure of how open or wound the spiral arms are. As we advance the sequence from Sa to Sc the bulge/disk ratio decreases from $L_{bulge}/L_{disk} \approx 0.3$ for Sa's to $L_{bulge}/L_{disk} \approx 0.05$ for Sc's. Contrarily, the pitch angle increases from $\alpha \sim 6^{\circ}$ to $\alpha \sim 18^{\circ}$ meaning that the arms are more open for Sc galaxies than for Sa. Another characteristic of Sa-SBa's galaxies is that their spiral structure has a smoother brightness distribution, while the late-type spirals show a more clumpsy distribution with resolved H II regions and star clumps.

%%%%%%%%%%%%%%% MASS DISTRIBUTIONS AND THEORY OF POTENTIALS %%%%%%%%%%%%%%% 
\section{Mass distributions and theory of gravitational potentials}
The influence of gravity in the universe is ubiquitous, in all physical scales. Galaxies are, of course, no exception, and their shape and structure is molded by self-gravity, i.e, the gravitational interactions between all its elements. Of course, interactions between galaxies, both through minor and major mergers, are more the rule than the exception, and these interactions can have great influence in the way a galaxy looks, but we are not interested in this type of scenario. \par
\noindent
To compute the gravitational potential of a collection of mass points (stars) it will suffice to add the potential generated by each one of them. Galaxies, however, contain a huge amount of stars, roughly $10^{10}-10^{12}$, which makes such calculation extremely inefficient and time consuming. We will discuss now some details of how potential theory deals with this problem and some applications to systems of our interest. \par   

%%%%%%%%%%%%%%% GENERAL ASPECTS OF POTENTIAL THEORY %%%%%%%%%%%%%%% 
\subsection{General aspects of potential theory}
The idea behind this theory is to get rid of the discrete sum to compute the gravitational interaction between particles. The way to do this is considering that the mass distribution of the galaxy can be described by a continuous function $\rho(\bm{r})$. By doing this, we can approximate the discrete summation over all the particles to compute the force by a volume integral, like this: \par
\begin{equation}
\label{eq:discrete_to_cont}
    \bm{{F}}_{j}=-G m_{j} \sum_{i \neq j}^{N-1} \frac{m_{i} (\bm{r}_{i}-\bm{r}_{j})}{|\bm{r}_{i}-\bm{r}_{j}|^{3}} \rightarrow -G m \int \frac{\bm{r}^{\prime}-\bm{r}}{\left|\bm{r}^{\prime}-\bm{r}\right|^{3}} \rho\left(\bm{r}^{\prime}\right) d^{3} \bm{r^{\prime}},
\end{equation}
\noindent
where $G$ is the gravitational constant, $\bm{{F}}_{j}$ is the force that the j-th particle feels due to the rest of the $N-1$ particles in the system. In the volume integral, $\rho(\bm{r}^{\prime})$ is a smooth density distribution of the mass of the galaxy and the integration is performed over all space.  The gravitational force per unit mass is known as as the \textbf{gravitational field}:
$$ \bm{g}(\bm{r}) = G \int \frac{\bm{r}^{\prime}-\bm{r}}{\left|\bm{r}^{\prime}-\bm{r}\right|^{3}} \rho\left(\bm{r}^{\prime}\right) d^{3} \bm{r^{\prime}}. $$
\noindent
We can define the \textbf{gravitational potential} as

$$ \Phi(\bm{r}) \equiv-G \int \mathrm{d}^{3} \bm{r}^{\prime} \frac{\rho\left(\bm{r}^{\prime}\right)}{\left|\bm{r}^{\prime}-\bm{r}\right|} $$
\noindent
and hence rewrite $\bm{g}(\bm{r})$ using the fact that $\nabla \left(\frac{1}{\left|\bm{r}^{\prime}-\bm{r}\right|}\right)=\frac{\bm{r}^{\prime}-\bm{r}}{\left|\bm{r}^{\prime}-\bm{r}\right|^{3}}$ as

$$\bm{g}(\bm{r}) = -\nabla \Phi.$$
\noindent
The potential is an scalar field, for that reason it is easier to manipulate than vector fields such as $\bm{F}$ or $\bm{g}(\bm{r})$, and it contains the same information. This scalar field, satisfies Poisson's equation:
\begin{equation}
\label{eq:poisson_equation}
\nabla^{2} \Phi=4 \pi G \rho
\end{equation}
which can be solved for $\Phi$ given a density distribution $\rho$ and appropriate boundary conditions for the potential, i.e, that $\lim_{|\bm{r}|\to\infty} \Phi = 0$.

%%%%%%%%%%%%%%% POTENTIAL OF AXISYMMETRIC DISKS %%%%%%%%%%%%%%%
\subsection{Potential of axisymmetric disks}  

%%%%%%%%%%%%%%% KUZMIN DISK %%%%%%%%%%%%%%%
\textbf{(a) Kuzmin Disk}. \noindent \citet{kuzmin1956model} introduced a potential of the form
\begin{equation}
\label{eq:kuzmin_potential}
    \Phi_{K}(R, z)=-\frac{G M}{\sqrt{R^{2}+(a+|z|)^{2}}} \quad(a \geq 0),
\end{equation}
such potential describes an infinitesimally thin disk with infinite extension. Notice that if we set $a=0$ the potential becomes spherically symmetric.\par
\begin{figure}[ht]
    \centering
    \includegraphics[width=0.6\textwidth]{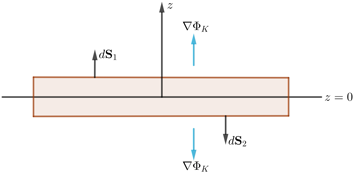}
    \caption[Gauss' theorem on Kuzmin Disk]{Application of Gauss' theorem to get surface density of the Kuzmin disk.}
    \label{fig:gauss_kuzmin}
\end{figure}
For a thin disk, the density is non-zero only at $z=0$, applying Gauss' theorem with a gaussian surface like \figref{gauss_kuzmin} we get: \par
$$\int_{S} \nabla \Phi_{K} \cdot d \bm{S}=4 \pi G M$$
$$\int_{S_{1}} \nabla \Phi_{K} \cdot d \bm{S_{1}}+\int_{S_{2}} \nabla \Phi_{K} \cdot d \bm{S_{2}}=4 \pi G M.$$
Since for both surfaces $d\bm{S} = dS \uvec{k} = 2 \pi R dR \uvec{k}$, the only non-zero component of the gradient is the one perpendicular to the disk; the total flux is the sum of the fluxes through surfaces 1 and 2, so
$$\int_{0}^{R} R\left|\nabla_{z} \Phi_{K}\right| d R=G M$$
$$\int_{0}^{R} R\left|\nabla_{z} \Phi_{K}\right| d R=G \int_{0}^{R} 2 \pi R \Sigma(R) d R,$$
solving the gradient inside the integral in the left hand side and equating the integrands, we can solve for the surface density:
\begin{equation}
\label{eq:surface_kuzmin}
    \Sigma(R)=\frac{M}{2 \pi} \cdot \frac{a}{({R^{2}+a^{2}})^{3/2}}.
\end{equation}

\begin{figure}[ht]
    \centering
    \includegraphics[width=1\textwidth]{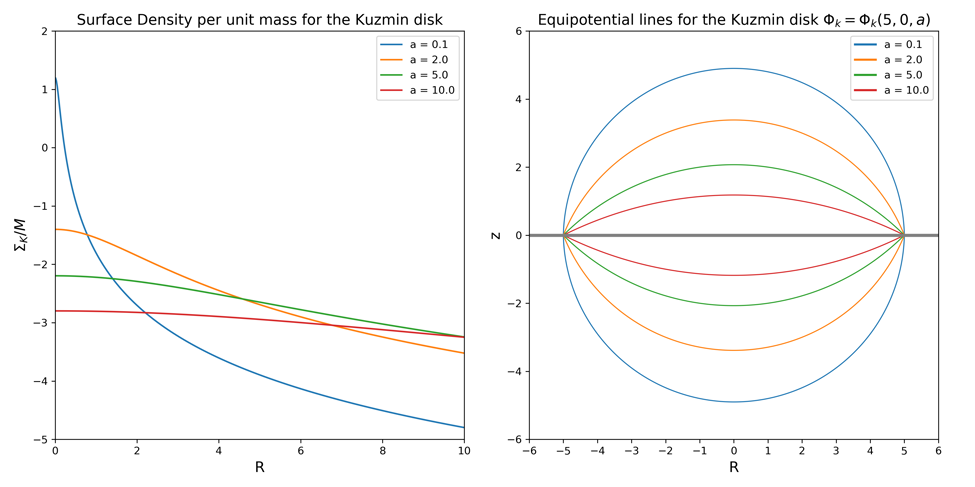}
    \caption[Surface density and equipotential lines for the Kuzmin disk]{(a) Surface density per unit mass for \autoref{eq:surface_kuzmin}. (b) Equipotential lines at $\Phi_{K}(R=5,z=0)$ for different values of a.}
    \label{fig:surfa+pot_kuzmin}
\end{figure}

\noindent
In \figref{surfa+pot_kuzmin}a we can see the surface density per unit mass $\Sigma_{k}(R)/M$ for different values of the parameter $a$, in \figref{surfa+pot_kuzmin}b the equipotential lines for $\Phi_{K}(R=5,z=0)$ are plotted for the same values of $a$ and with the same color scheme. Notice that for small $a$ the equipotential lines are closer to spherical symmetry, while for larger values of $a$ the potential gets more flattened.   
\noindent
A generalization for the potential \autoref{eq:kuzmin_potential} is the \citet{miyamoto1975three} potential:
\begin{equation}
\label{eq:potential_miyamoto}
    \Phi_{\mathrm{M}}(R, z)=-\frac{G M}{\sqrt{R^{2}+(a+\sqrt{z^{2}+b^{2}})^{2}}},
\end{equation}
\noindent
if $a=0$ it reduces to a spherical potential (\citealt{plummer1911problem}) and when $b=0$ it reduces to the Kuzmin potential. Depending on the values of $a$ and $b$ it can represent anything in between a thin disk and a spherical system.
\\[12pt]
%%%%%%%%%%%%%%% LOGARITHMIC POTENTIALS %%%%%%%%%%%%%%%
\noindent
\textbf{(b) Logarithmic Potentials} This type of potentials have non-keplerian circular speed curves at large R, trying to account for the observational evidence of flat rotation curves in galaxies. The circular velocity in a disk is defined as
\begin{equation}
\label{eq:circular_velocity}
    v_{c}^{2} = R \evalat[\bigg]{\frac{d \Phi}{d R}}{z=0},
\end{equation}
if $v_{c} = v_{0} = \text{constant}$ then $\Phi \propto v_{0}^{2} \ln{R} + C$, based on this argument a potential of the following form is sometimes used: 
\begin{equation}
\label{eq:potential_log}
    \Phi_{\mathrm{L}}=\frac{1}{2} v_{0}^{2} \ln \left(R_{\mathrm{c}}^{2}+R^{2}+\frac{z^{2}}{q_{\Phi}^{2}}\right)+\text { constant, }
\end{equation}
\noindent
where $q_{\Phi}$ is the axis ratio of equipotential surfaces, $q_{\Phi} = 1$ implies a circular shape. The circular velocity obtained from this potential is
$$v_{\mathrm{c}}=\frac{v_{0} R}{\sqrt{R_{\mathrm{c}}^{2}+R^{2}}},$$
which is indeed constant for $R>>R_{c}$.

%%%%%%%%%%%%%%% GENEREAL POTENTIALS %%%%%%%%%%%%%%%
\subsection{General potentials for flattened systems} 
\begin{figure}[ht]
\begin{subfigure}{.7\textwidth}
  \centering
  \includegraphics[width=1\linewidth]{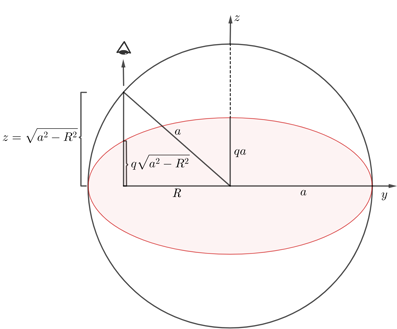}
  %\caption{}
  \label{homoeoid_a}
\end{subfigure}%
\begin{subfigure}{.25\textwidth}
  \centering
  \includegraphics[width=1\linewidth]{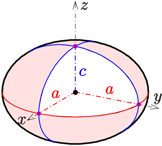}
  %\caption{}
  \label{homoeoid_b}
\end{subfigure}
\vspace{-1\baselineskip}
\caption[Ellipsoid of axis ratio $q$]{Left, an ellipsoid of semi-major axis $a$ and axis ratio $q$, viewed along a line of sight that cuts the spheroid's equatorial plane perpendicularly at radius R, based on the figure at \citealt{binney2011galactic}, p 96. In the right side is a 3D view of an oblate spheroid generated by rotating an ellipse around its minor axis. Licensed under \href{https://creativecommons.org/licenses/by-sa/4.0/}{CC BY-SA 4.0}}
\label{fig:homoeoid}
\end{figure}

One way to obtain a more general expression for the potential of arbitrary flattened systems is to consider a flattened spheroidal shell, with constant density $\rho$, semi-axes $a$ and $c$ and axial ratio $q=c/a$. Such object is called an \textbf{homoeoid}. Now, the volume of a triaxial spheroid with axis $a, b$ and $c$ is $V = \frac{4}{3} \pi abc$, in this case we have have a projected oblate spheroid, therefore $V = \frac{4}{3} \pi qa^{3}$, so the mass is $M(a) =\frac{4}{3} \pi q \rho a^{3} $. The projected density along the line of sight is $\Sigma = L\rho = 2q\sqrt{a^{2}-R^{2}}\rho$ (see \figref{homoeoid}). We can compute the mass and the density of a thin shell by differentiating $M(a) \, \text{and} \, \Sigma$ to get:
\begin{subequations}
\begin{gather}
    \delta M(a)=4 \pi \rho q a^{2} \delta a \\
    \delta \Sigma(a, R)=\frac{2 \rho q a}{\sqrt{a^{2}-R^{2}}} \delta a,
\end{gather}
\end{subequations}
\noindent
now let's define $2 \rho q a \equiv \Sigma_{0}(a)$ as the surface density at $R=0$, so we can rewrite both expressions as
\begin{subequations}
\begin{gather}
    \delta M(a)=2 \pi \Sigma_{0}(a) a \delta a \label{eq:deltaM_h} \\
    \delta \Sigma(a, R)=\frac{\Sigma_{0}(a)}{\sqrt{a^{2}-R^{2}}} \delta a. \label{eq:deltaSigma_h}
\end{gather}
\end{subequations}

\begin{figure}[ht]
    \centering
    \includegraphics[width=1\textwidth]{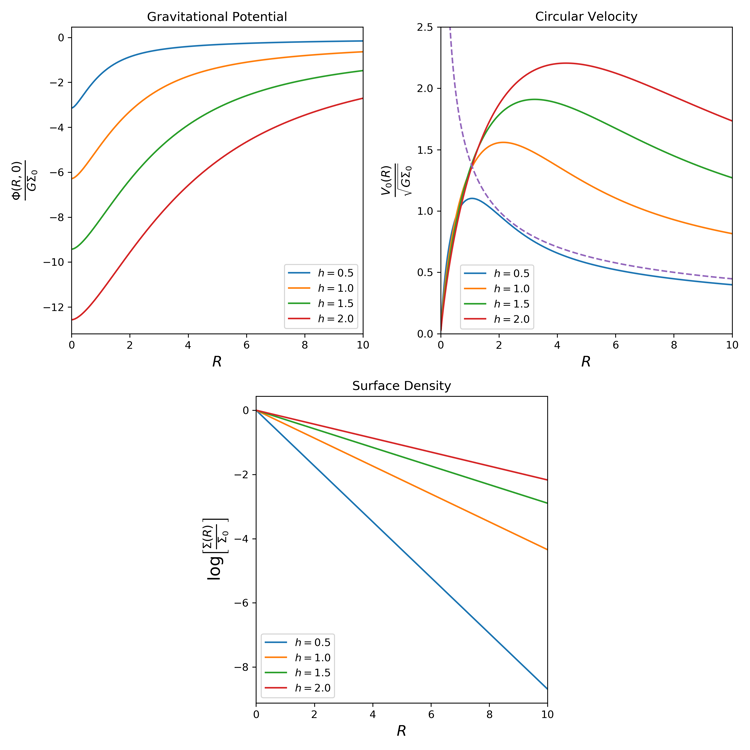}
    \caption[Potential, circular velocity and surface density of an exponential disk.]{Behaviour of the gravitational potential, circular velocity curve and surface density of an exponential disk.}
    \label{fig:plots_exponential}
\end{figure}

\noindent
We can now build $\Sigma(R)$, the density of a razor-thin disk, by adding all the homoeoids with density $\Sigma_{0}(a)$ from $a=R$ to $a \rightarrow \infty$, which is equivalent to integrate \autoref{eq:deltaSigma_h}:
\begin{equation}
\label{eq:sigma_integral}
    \Sigma(R)=\int_{R}^{\infty} \mathrm{d} a \frac{\Sigma_{0}(a)}{\sqrt{a^{2}-R^{2}}},
\end{equation}

\noindent
and find $\Sigma_{0}(a)$ by direct application of the Abel integral equation which reads as follow

\begin{equation}
\label{eq:abel_integral}
    f(x)=\int_{x}^{\infty} \frac{\mathrm{d}t \, g(t)}{(t-x)^{\alpha}} \\
    \Longrightarrow g(t)=-\frac{\sin \pi \alpha}{\pi} \frac{\mathrm{d}}{\mathrm{d}t} \int_{t}^{\infty} \frac{\mathrm{d}x \, f(x)}{(x-t)^{1-\alpha}}.
\end{equation}
By comparing \autoref{eq:sigma_integral} and \autoref{eq:abel_integral} we see that $t\rightarrow a^{2}, x\rightarrow R^{2} \, \text{and} \, \alpha = 1/2$, so we get

\begin{equation}
\label{eq:sigma_0_integral}
    \Sigma_{0}(a)=-\frac{2}{\pi} \frac{\mathrm{d}}{\mathrm{d} a} \int_{a}^{\infty} \mathrm{d} R \frac{R \Sigma(R)}{\sqrt{R^{2}-a^{2}}}.
\end{equation}
\noindent
We may calculate the potential of this razor-thin disk by adding the potentials that correspond to each one of the thin homoeoids that compose it. The potential of a thin homoeoid can be found by solving Laplace's equation $\nabla^{2} \Phi=0$ in oblate spheroidal coordinates $(u,v,\phi)$, more specifically, by assuming that $\Phi = \Phi(u)$. Such solution is of the form (see \citealt{binney2011galactic} §2.5.1 for further details): 
\begin{equation}
\label{eq:potential_homoeoid}
    \Phi=-\frac{G \delta M}{a e} \times\left\{\begin{array}{ll}
    \sin ^{-1}(e) & \left(u<u_{0}\right) \\
    \sin ^{-1}(\operatorname{sech} u) & \left(u \geq u_{0}\right)
\end{array}\right.,
\end{equation}
where $u = u_0 = \text{const}$ is the surface (in oblate spheroidal coordinates) that describes the thin homoeoid. Since Newton's third theorem states that a mass inside a homoeoid experiences no net gravitational force, the potential must be constant for $u<u_{0}$, as \autoref{eq:potential_homoeoid} states. We are interested, however, in the case when we are \textit{just outside} the thin homoeoid, $u>u_{0}$; for a completely flattened spheroid, $e=1-c/a=1$, because $c<<a$, replacing this and \autoref{eq:deltaM_h} into \autoref{eq:potential_homoeoid} we get

$$\delta \Phi(R, z)=-2 \pi G \Sigma_{0} \delta a \sin ^{-1}(\operatorname{sech} u).$$
\noindent
By use of the identity $1=\cos ^{2} v+\sin ^{2} v$ and the transformation equations $(u,v,\phi) \rightarrow (R,z,\phi)$ we can rewrite the expression like 
\begin{equation}
\label{eq:dphi_thin_homoeoid}
    \delta \Phi=-2 \pi G \Sigma_{0} \delta a \sin ^{-1}\left(\frac{2 a}{\sqrt{+} \, + \, \sqrt{-}}\right),
\end{equation}
where $\sqrt{\pm} \equiv \sqrt{z^{2}+(a \pm R)^{2}}$. Finally, we integrate this equation over all possible values of $a>0$ and substitute \autoref{eq:sigma_0_integral} to get the potential of an axisymmetric disk of arbitrary surface density profile: 

\begin{equation}
\label{eq:potential_flat_1}
    \Phi(R, z)=4 G \int_{0}^{\infty} \mathrm{d} a \sin ^{-1}\left(\frac{2 a}{\sqrt{+} \,+\sqrt{-}}\right) \frac{\mathrm{d}}{\mathrm{d} a} \int_{a}^{\infty} \mathrm{d} R^{\prime} \frac{R^{\prime} \Sigma\left(R^{\prime}\right)}{\sqrt{R^{\prime 2}-a^{2}}}.
\end{equation}
To get the potential in the disk we simply set $z=0$:
\begin{equation}
\label{eq:potential_flat_1_z0}
    \Phi(R, 0)=4 G \int_{0}^{\infty} \mathrm{d} a \sin ^{-1}\left(\frac{2 a}{(a+R)+|a-R|}\right) \frac{\mathrm{d}}{\mathrm{d} a} \int_{a}^{\infty} \mathrm{d} R^{\prime} \frac{R^{\prime} \Sigma\left(R^{\prime}\right)}{\sqrt{R^{\prime 2}-a^{2}}}.
\end{equation}
An alternative form of \autoref{eq:potential_flat_1}, which can be easier to evaluate numerically, can be obtained integrating by parts:
\begin{equation}
\label{eq:potential_flat_2}
\begin{aligned}
    \Phi(R, z)=-2 \sqrt{2} G \int_{0}^{\infty} \mathrm{d} a \frac{ \frac{a+R}{\sqrt{+}} - \frac{a-R}{\sqrt{-}} }{\sqrt{R^{2}-z^{2}-a^{2}+\sqrt{+} \sqrt{-}}} \int_{a}^{\infty} \mathrm{d} R^{\prime} \frac{R^{\prime} \Sigma\left(R^{\prime}\right)}{\sqrt{R^{\prime 2}-a^{2}}},
\end{aligned}
\end{equation}
\noindent
and get an analogous of \autoref{eq:potential_flat_1_z0} for the potential in $z=0$:
\begin{equation}
\label{eq:potential_flat_2_z0}
    \Phi(R, 0)=-4 G \int_{0}^{R} \frac{\mathrm{d} a}{\sqrt{R^{2}-a^{2}}} \int_{a}^{\infty} \mathrm{d} R^{\prime} \frac{R^{\prime} \Sigma\left(R^{\prime}\right)}{\sqrt{R^{2}-a^{2}}},
\end{equation}
and finally, to get the circular velocity we use the definition \autoref{eq:circular_velocity} applied to \autoref{eq:potential_flat_1_z0}
\begin{equation}
\label{eq:circular_velocity_flat}
v_{\mathrm{c}}^{2}(R)=-4 G \int_{0}^{R} \mathrm{d} a \frac{a}{\sqrt{R^{2}-a^{2}}} \frac{\mathrm{d}}{\mathrm{d} a} \int_{a}^{\infty} \mathrm{d} R^{\prime} \frac{R^{\prime} \Sigma\left(R^{\prime}\right)}{\sqrt{R^{\prime 2}-a^{2}}}.
\end{equation}
\noindent
\autoref{eq:potential_flat_1_z0}, \autoref{eq:potential_flat_2_z0} and \autoref{eq:circular_velocity_flat} provide the gravitational potential and circular velocity curve for any given surface density distribution. 
\\[12pt]
%%%%%%%%%%%%%%% THE EXPONENTIAL DISK %%%%%%%%%%%%%%%
\textbf{(a) The exponential disk}. One particularly useful application of the results in the previous section is the exponential disk. We talked about the brightness profiles of the different components of spiral galaxies in \cref{section2.1.1}, and mentioned that radial brightness profiles are exponential in nature, like \autoref{eq:brightness_disk_radial}. One could argue that the surface density scales in the same way:
\begin{equation}
\label{eq:surface_exponential}
    \Sigma(R)=\Sigma_{0} \mathrm{e}^{-R / h},
\end{equation}
so it is interesting to find out the potential and circular velocity associated to such surface density. For that purpose we star by using \autoref{eq:potential_flat_2_z0}, the innermost integral can be solved in terms of the modified Bessel function $K_{1}$; the complete integral yields:
\begin{equation}
\begin{aligned}
\label{eq:potential_exponential}
    \Phi(R, 0) &=-\pi G \Sigma_{0} R\left[I_{0}(y) K_{1}(y)-I_{1}(y) K_{0}(y)\right],
\end{aligned}
\end{equation}
\noindent
where $y=\frac{R}{2h}$, differentiating this we get the circular velocity squared:
\begin{equation}
\label{eq:circular_exponential}
    v_{c}^{2}(R)=R \frac{\partial \Phi}{\partial R}=4 \pi G \Sigma_{0} h y^{2}\left[I_{0}(y) K_{0}(y)-I_{1}(y) K_{1}(y)\right].
\end{equation}
In \figref{plots_exponential} are plotted \autoref{eq:surface_exponential}, \autoref{eq:potential_exponential} and \autoref{eq:circular_exponential} for different values of the radial scale, i.e, for disks with increasing extension.
\noindent
\\[12pt]
\textbf{(b) Thick Disks}. Vertical scales are in general small compared to radial scales, as we also mentioned in \cref{section2.1.1}, that means that the density decreases faster in directions perpendicular to the disk plane, and the orbits are mainly determined by the potential generated by the surface density at the equatorial plane of the galaxy. However, this is not a realistic description of galaxies. Observations of disks with $z_{0} \lesssim 4$ kpc show that there are thick disks for which the razor-thin assumption is inaccurate. A simple model of thick disks supposes a separable density, i.e, that for any value of $z=z'$ the density is only a function of $R$:
\begin{equation}
\label{eq:density_thick}
\rho(R,z) = \Sigma(R) \mathrm{Z}(z).
\end{equation}
The potential for a density distribution of this form is
\begin{equation}
\label{potential_thick_general}
    \Phi(R, z)=\int_{-\infty}^{\infty} \mathrm{d} z^{\prime} \, \mathrm{Z}\left(z^{\prime}\right) \Phi_{0}\left(R, z-z^{\prime}\right),
\end{equation}
\noindent
where $\Phi_{0}\left(R, z-z^{\prime}\right)$ is the potential at $z^{\prime}$ generated by a density distribution $\Sigma(R)$ at $z=0$; $\Phi_{0}$ can be computed using \autoref{eq:potential_flat_1} or \autoref{eq:potential_flat_2}. For the case of a exponential disk with the surface density of \autoref{eq:surface_exponential} we have that 
\begin{equation}
\label{potential_thick_exp}
    \Phi(R, z)=-\frac{4 G \Sigma_{0}}{h} \int_{-\infty}^{\infty} \mathrm{d} z^{\prime} \, \mathrm{Z}\left(z^{\prime}\right) \int_{0}^{\infty} \mathrm{d} a \sin ^{-1}\left(\frac{2 a}{\sqrt{+}+\sqrt{-}}\right) a K_{0}\left(a / R_{\mathrm{d}}\right),
\end{equation}
\autoref{potential_thick_exp} can be used to calculate the potential $\Phi(R,z)$ given a vertical distribution $\mathrm{Z}(z)$ that complements the surface density distribution $\Sigma(R)$, as in \autoref{eq:density_thick}. For example, the disk of the Milky Way galaxy is often modelled like 
\begin{equation}
\label{density_MW}
    \rho_{\mathrm{d}}(R, z)=\Sigma(R) \mathrm{Z}(z)=\Sigma_{d} \, \mathrm{e}^{-R / R_{h}}\left(\frac{\alpha_{0}}{2 z_{0}} \mathrm{e}^{-|z| / z_{0}}+\frac{\alpha_{1}}{2 z_{1}} \mathrm{e}^{-|z| / z_{1}}\right),
\end{equation}
\noindent
where the factor containing the exponentials in $|z|$ is the vertical density distribution $\mathrm{Z}(z)$; the parameters $z_{0}$ and $z_{1}$ are the vertical scales for the thin and thick components of the galactic disk. 
%%%%%%%%%%%%%%% SPIRAL STRUCTURE %%%%%%%%%%%%%%%
  
\section{Spiral structure}
Spiral arms are perhaps one of the must stunning views in the universe. These structures are present in a wide variety of shapes, from the elegant and easily traced arms in \enquote{grand-design} spiral galaxies, to the clumpy and fragmented configurations of \enquote{Flocculent} galaxies. Spiral arms play an important role in the secular evolution of galaxies. Their origin and dynamics has been a matter of discussion since the first observations, and there exists a robust theoretical framework to study their properties from a phenomenological standpoint. In this chapter we will review some of their principal features, and also give a brief overview of density wave theory. 
%%%%%%%%%%%%%%% FUNDAMENTALS %%%%%%%%%%%%%%%
\subsection{Observational Properties}
Observations have shown that spiral arms are the epicenter of star formation, therefore, the chemical, dynamical and thermodynamical properties of the disk are, in great proportion, regulated by them. Photometry of spiral galaxies performed in different bands shows a variety of morphological features that are related to physical phenomena occurring in the spiral arms. \par
\noindent
We can say that arms are outlined by different tracers, therefore, it is possible to classify different types of arms: mass, potential, gas and bright-stars arms. Each 
type of arm is traced where the property is maximum (or a minimum in the case of the potential), the mass arm is marked by the regions where the stellar surface density is maximum, the gas arm is marked by the regions where the surface density of gas is maximum, the potential arm by the regions of minimum potential and bright-stars arm where the surface luminosity density due to young stars is maximum. 

\subsubsection{Neutral Hydrogen ($\mathrm{HI}$)} The spiral structure has been observed through HI maps. The detection of neutral non-emitting hydrogen gas is made by measuring the 21-cm emission line. This photons are emitted due to a change in the orientation of the spin of the electron with respect to that in the proton. Those HI observations serve to measure the rotation velocity of spiral galaxies and therefore give estimates of their luminosity through the Tully-Fisher relations (\citealt{tully1977new}), with the luminosity, the distance modulus can be computed. The presence of HI arms is believed to be a product of rapid stellar formation, since those processes generate great amounts of UV radiation that dissociate molecular hydrogen. This conclusion derives from the fact that often times the HI arm coincides with the bright-stars arm. 
\subsubsection{Molecular Hydrogen ($ \mathrm{H}_{2}$) } 
\begin{figure}[H]
\begin{subfigure}{0.5\textwidth}
  \centering
  \includegraphics[width=0.9\linewidth]{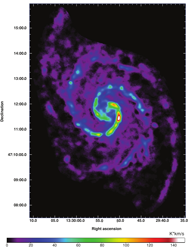}
  \label{fig:M51_koda}
\end{subfigure}%
\begin{subfigure}{.5\textwidth}
  \centering
  \includegraphics[width=0.9\linewidth]{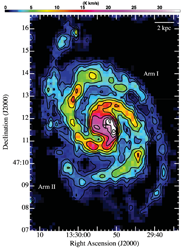}
  \label{fig:M51_vlahakis}
\end{subfigure}
\caption[M51 seen with CO(J=1-0) and CO(J=3-2) emission.]{Left: CO(J=1-0) emission map of M51 using combined data from CARMA (Combined Array for Research in Millimeter Astronomy) and NRO45 (Nobeyama 45 m telescope); taken from \citealt{koda2011co}. Right: CO(J=3-2) emission map of M51 using data from the HARP-B instrument on the JCMT (James Clerk Maxwell Telescope); taken from \citealt{vlahakis2013co}.}
\label{fig:CO-koda-vlahakis}
\end{figure}

It is not possible to directly observe $\mathrm{H}_{2}$, since in molecular clouds the temperatures are on the order of $T = 10 \mathrm{K}$ and none of the $ \mathrm{H}_{2}$ molecules are in states capable of emitting any photons. However CO luminosity is believed to trace the molecular gas content in a galaxy, in which $ \mathrm{H}_{2}$ is the majority. The CO molecule is detected through its rotational transitions $J = 1\rightarrow0,\ 2\rightarrow1,\ 3\rightarrow2 $; carbon monoxide rotational emission traces high density regions with low temperatures in the interstellar medium (ISM). Spiral arms have been observed in this type of low-energy emissions, see for example a map of the spiral galaxy M51 in \figref{CO-koda-vlahakis}, using CO(J=1-0) and CO(J=3-2) emission.

\subsubsection{$\mathrm{HII}$ regions} 
\begin{figure}[ht]
\begin{subfigure}{0.5\textwidth}
  \centering
  \includegraphics[width=0.9\linewidth]{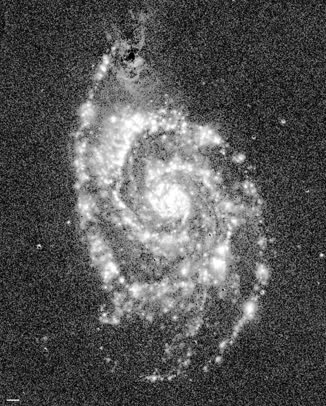}
  \label{fig:M51_greenawalt}
\end{subfigure}%
\begin{subfigure}{.5\textwidth}
  \centering
  \includegraphics[width=0.9\linewidth]{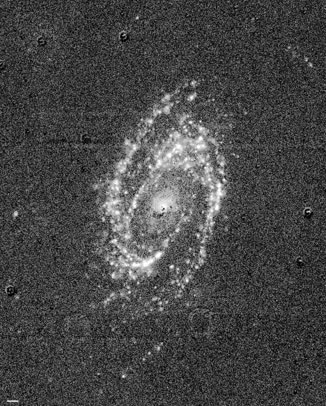}
  \label{fig:M81_greenawalt}
\end{subfigure}
\caption[M51 and M81 spiral galaxies in H$\alpha$.]{M51 (left) and M81 (right) spiral galaxies. Both images are continuum-subtracted H$\alpha$ + $[\mathrm{N} \text { II }]$. Taken from \citealt{greenawalt1998diffuse}.}
\label{fig:H_alpha-greenawalt}
\end{figure}

$\mathrm{H}_{II}$ regions trace young OB stars and sectors of star formation activity. They can be observed through different indicators, in particular, the $\mathrm{H}_{\alpha}$ emissions of the Balmer series is commonly used to observe this type of regions. The origin of this emission line is the combination of protons with free electrons and the posterior decaying of this atomic hydrogen, that is in an excited state, to the ground level. See for example \figref{H_alpha-greenawalt}, taken from \citealt{greenawalt1998diffuse}, where $\mathrm{H}_{\alpha}$ emission-line images are presented for grand-design spiral galaxies M51 and M81. 
%%%%%%%%%%%%%%% Geometry of the spiral arms %%%%%%%%%%%%%%%
\subsection{Geometry and morphology of the spiral arms}
Let's define $m$ as the number of spiral arms. A galaxy has $m$-fold rotational symmetry if its brightness distribution is the same when we perform a rotation of $2\pi/m$, i.e, $I(R, \phi+2 \pi / m)=I(R, \phi)$. Spiral arms shapes are though as a 2D mathematical curve in the plane of the galaxy, such curve defines the location of the arm at any given time and radius; in a galaxy with $m$-fold rotational symmetry, a curve that defines the location of all $m$ arms is
\begin{equation}
\label{eq:shape_function}
m \phi+f(R, t)=\text { constant },
\end{equation}
where $f(R,t)$ is defined as the \textbf{shape function}. One commonly used function to describe the shape of the spiral arms, is a logarithmic spiral function:
\begin{equation}
\label{eq:logarithmic_spirals}
    r=r_{0} e^{\theta \tan (\phi)},
\end{equation}
where $\phi$ is the polar angle, $\theta$ is the central angle,and $r_{0}$ is the radius at which the central angle is equal to zero. The brightness distribution $I(R, \phi)$ can be represented as a Fourier series
\begin{equation}
\label{eq:fourier_spiral}
    \frac{I(R, \phi)}{\bar{I}(R)}=1+\sum_{m=1}^{\infty} A_{m}(R) \cos m\left[\phi-\phi_{m}(R)\right],
\end{equation}
where the coefficients of the expansion $A_m$ define the amplitude of the spiral structure's components and $\phi_{m}$ the corresponding phase. $\bar{I}(R)$ is the brightness averaged over $0 < \phi < 2\pi$:
$$\bar{I}(R) \equiv \frac{1}{2\pi} \int_{0}^{2 \pi} \mathrm{d} \phi I(R, \phi).$$

\begin{figure}[ht]
    \centering
    \includegraphics[width=0.6\textwidth]{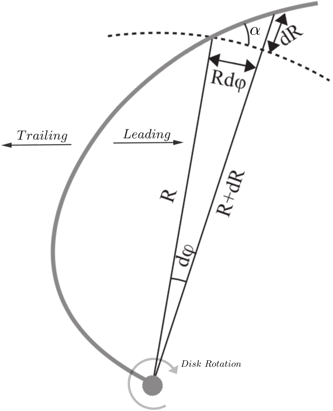}
    \caption[Schematic representation of the pitch angle]{Definition of the pitch angle as the angle between a tangent to the arm at $R$ and a circumference of radius $R$. Adapted from \citealt{ferreras2019fundamentals}, licensed under \href{https://creativecommons.org/licenses/by-sa/4.0/}{CC BY-SA 4.0}}
    \label{fig:pitch_angle}
\end{figure}

Other two properties of spiral arms are the pitch angle, that plays an important role in Hubble classification as previously discussed in \cref{section_hubble_classification}, and whether or not the spiral structure trails or leads. The pitch angle is a measure of how wound are the spiral arms, from \figref{pitch_angle} we can easily see that
\begin{equation}
\label{eq:pitch_angle}
    \cot \alpha=\left|R \frac{d \phi}{d R}\right|.
\end{equation}
Another important property that describes the dynamic of spiral arms is the \textbf{radial wavenumber}, which is defined as
\begin{equation}
\label{eq:radial_wavenumber}
k(R, t) \equiv \frac{\partial f(R, t)}{\partial R}.
\end{equation}
The radial wavenumber sign defines if the arms are trailing ($k>0$) or leading ($k<0$); in \figref{pitch_angle} we indicate what those two terms mean: the tip of a \textit{trailing} arm points opposite to the direction of rotation, while the tip of a \textit{leading} arm points in the same direction of rotation.  

Differentiating \autoref{eq:shape_function} with respect to $R$ and replacing \autoref{eq:radial_wavenumber} we can write the pitch angle as: 
\begin{equation}
\label{eq:pitch_angle_2}
    \cot \alpha=\left|\frac{k R}{m}\right|.
\end{equation}
Furthermore, if the azimuthal position of the arm is given by $\phi(R, t)=\phi_{0}+\Omega(R) t$, the pitch angle is written as
\begin{equation}
\label{eq:pitch_angle_3}
\cot \alpha=R t\left|\frac{\mathrm{d} \Omega}{\mathrm{d} R}\right|.
\end{equation}
Let us consider the evolution of the position of an arm, $\phi(R,t)$, such that $\phi(R,0)=\phi_{0}$, where $\phi_{0}$ is the azimuthal angle. If the disk has differential rotation, the angular speed depends on $R$, $\Omega = \Omega(R)$. We can write the angular position of the arm as
\begin{equation}
\label{eq:arm-position}
    \phi(R, t)=\phi_{0}+\Omega(R) t.
\end{equation}
The pitch angle can now be rewritten using \autoref{eq:pitch_angle} and \autoref{eq:arm-position} to give
\begin{equation}
\label{eq:pitch_angle4}
    \cot \alpha=R t\left|\frac{\mathrm{d} \Omega}{\mathrm{d} R}\right|.
\end{equation}
For a disk with a flat circular-velocity curve, $v_{c}=200 km/s$, at $R = 5 kpc$, after $t = 10 \text{Gyr}$ the pitch angle is $\alpha \approx 0.14$. Such angle is way smaller that observed pitch angles, which are around $\alpha=6-18 \deg$.
The disparity is known as the \textbf{winding problem}. The differential rotation of the disk causes the spiral patterns to wind up in short timescales. 
\subsection{Density wave theory}
So far we have only described some aspects of the morphology of the spiral arms that are present in disk galaxies. Those results are mainly observational, and try to describe what we see as is, i.e, independent of the origin of these structures. \par
The first attempt to address the problem on the origin of the spiral structure in disk galaxies were made by Bertil Lindblad who devoted a great part of his work to this issue. He suggested that such spiral patterns occurred due to instabilities at the edge of Maclaurin ellipsoids, and his work had a big focus on the orbits of individual particles. He modelled the spiral pattern as a perturbation of the orbits in the disk, more specifically, perturbations at the edges of Maclaurin spheroids which generated orbits with high-ellipticity, and consequently originated the arms. \par
Lindblad did not exploit the ideas behind wave-mechanics, although he established an analogy between spiral arms and the harmonic waves in an unstable Maclaurinn spheroid. He is considered a pioneer in the field and \enquote{all problems that in later developments turned out to be important in the theory of spiral structure had, in one way or another, already been touched upon or even studied by Lindblad} (\citealt{dekker1976spiral}). \par 
There were contributions in this field from authors such as Alar Toomre, Donald Lynden-Bell, Agris Kalnajs, Leonid Marochnik, C.C Lin and Frank Shu which constitute a great part of the foundation of what we know today as the density-wave spiral theory. A complete review of the contributions and a historical perspective on the origin of the theory was made by \citet{pasha2002I}, \citet{pasha2004II} and \citet{marochnik2005westside}. \par 
It was in the decade of 1960 with the publication of \citet{lin1964spiral} that a general interest in the field of spiral structure arouse among the scientific community. In this section we will try to outline the main aspects of this theory.

\subsection{Quasi-stationary density wave theory.}
The quasi-stationary density wave theory tries to explain the nature and dynamics of the spiral structures that are present in disk galaxies. It states that the spiral arms are features that rotate with a constant pattern speed in the disc, and they do it slowly. This theory portraits long-living structures that evolve slowly and remain in large scale-times. \par

This theory was proposed by \citet{lin1964spiral} as we have previously mentioned. They assumed that the spiral arms are an over-density pattern, and gas (or stars) travel \textit{through} this pattern, i.e, they go in and out of it, as they rotate. This spiral pattern rotates as well, and the theory proposes that it has a constant angular velocity rotation which is suggestively called the \textit{pattern speed}. \par

Indeed, this pictures the arms as \textit{density waves} that rotate around the galaxy. As these waves move, matter accumulates in the form of spiral arms. This accumulation causes localized over-densities that, at the same time, induce a perturbation in the gravitational potential of the disk. \par

A common analogy to explain this theory, that is often found in the literature, is that of a traffic jam in a highway. The cars in the traffic jam travel at a speed of $v_{jam}$, vehicles coming from behind it, travel at a faster speed of $v_{fast}$, such that $v_{fast}>v_{jam}$; those cars coming from behind the traffic jam, have to reduce their speed as they go into the it, eventually reaching the velocity of $v_{jam}$. As they move out the jam, they gradually increase their speed to reach $v_{fast}$ once again. We can think of the cars as the stars and gas rotating in the galaxy disk, and the traffic jam as the density waves in which matter accumulates. \par

\subsubsection{Dispersion relations}
Dispersion relations are equations that characterize the dynamic of waves (or \textit{modes}) that propagate in a medium. One classical example is the wave equation 
\begin{equation}
\label{eq:wave-equation-light}
    \frac{\partial^{2} \psi}{\partial t^{2}}=c^{2} \frac{\partial^{2} \psi}{\partial x^{2}},
\end{equation}
whose solution (in an homogeneous and borderless medium) has the form of traveling plane waves, 
\begin{equation}
\label{eq:traveling-wave}   
    \psi(x, t)=A e^{i(k x-\omega t)}.
\end{equation}
When substituting this solution in \autoref{eq:wave-equation-light} we get the relation
\begin{equation}
\label{eq:dispersion-relation-light}
    \omega^{2}=c^{2} k^{2},
\end{equation}
which is a trivial dispersion relation of the form $\omega = \omega(k)$. The phase velocity of the wave is defined as $\omega/k$, and in the case of \autoref{eq:dispersion-relation-light} it is constant and equal to $\pm c$. The phase velocity in this case coincides with the group velocity, defined as 
\begin{equation}
\label{eq:group-velocity}
    v_{g} = \frac{d\omega(k)}{dk} = \pm c = v_{phase}.
\end{equation}
The group velocity describes the speed of a wave packet, i.e, the speed of a superposition of traveling waves with different wavenumbers. In this case, the two velocities are the same. However this is not always true. In a \textit{dispersive} system, the phase velocity is not constant, this a consequence of the fact that waves with different values of wavenumber move at different speeds. \par

\subsubsection{Dispersion relations for the density waves}
Dispersion relations define the dynamics of the waves, so finding those for the case of density waves is of huge importance. \citet{lin1964spiral} derived dispersion for both fluid and stellar disk, following this method:
\begin{enumerate}
    \item They linearized the continuity and momentum equations (Euler equations) and the Poisson equation. They assumed that the unperturbed disk has axial symmetry and \textit{no radial motion}. They assumed a disk with zero thickness.
    \item They used what is called the \enquote{Tight-winding approximation}. Since gravity is a long-range force, the perturbations in all the disk are coupled. If we consider tightly-wound density waves, the long-range coupling is negligible, since the radial wavelength (\autoref{eq:radial_wavenumber}) is small compared to the radius, and the perturbations are only important locally. 
    \item The spiral arms rotate as a rigid-body with a determined angular velocity and pitch angle. This angular velocity is the pattern speed $\Omega_{p}$. This implies that there are regions where the stars and gas rotate faster than the pattern speed and other where they rotate slower. The radius at which $\Omega_{p} = \Omega(R)$ is called the corotation radius, where $\Omega(R)$ is the circular angular speed of the material in the disk.
\end{enumerate}

With this method (and the assumptions in it), they derived the dispersion relation
\begin{equation}
\label{eq:dispersion-relation-fluid}
    (\omega-m \Omega)^{2}=c_{s}^{2} k^{2}+\kappa^{2}-2 \pi \mathrm{G} \Sigma_{0}|k|,
\end{equation}
where $m$ is the number of spiral arms, $\Omega$ is the angular frequency at any given radius, $\Sigma_{0}$ is the surface density, $\kappa$ is the epicyclic frequency. This dispersion relation gives a criteria to determine the local stability of the disk against perturbations. \par
In a rotating frame, $\omega-m\Omega$ is the angular frequency of the density wave at radius $R$. A perturbation has the form of $exp[-i(\omega-m\Omega)t]$ so if $(\omega-m\Omega)^{2}<0$, it turns into a real-valued exponential $exp[\pm(\omega-m\Omega)t]$, which will eventually increase its value without limit and disrupt the disk's stability.  \par
From this we can establish a stability limit:
\begin{equation}
\label{eq:stability-limit}
  c_{s}^{2} k^{2}+\kappa^{2}-2 \pi \mathrm{G} \Sigma_{0}|k| = 0.  
\end{equation} 
 This equation is quadratic in $k$, so we can solve it to get:
$$ k = \frac{1}{c_{s}^{2}} \left(\pi \mathrm{G} \Sigma_{0} \pm \sqrt{\pi^{2} \mathrm{G}^{2} \Sigma_{0}^{2} -c_{s}^{2} \kappa^{2} } \right), $$ 
and for stability we require that there are no solutions for \autoref{eq:stability-limit}, so we end up with the Toomre stability criteria, first derived by \citealt{toomre1964gravitational}:
\begin{equation}
\label{eq:toomre-parameter}
    Q \equiv \frac{\kappa c_{s}}{\pi G \Sigma_{0}} > 1,
\end{equation}
if $Q>1$ the disk is stable to tight-winding perturbations, if $Q<1$ it is not. We can also define, for cold disks ($c_{s}=0$), a critical unstable wavelength as $\lambda_{\text {crit }}=2 \pi / k_{\text {crit }}$ where \autoref{eq:stability-limit} holds. The perturbations with wavelenghts $\lambda > \lambda_{crit}$ are unstable, or, equivalently, perturbations with $|k|<k_{crit}$, where $k_{\text {crit }}=\kappa^{2} /\left(2 \pi G \Sigma_{0}\right)$ as per \autoref{eq:stability-limit} (with $c_{s}=0$). Waves with $|k| \ll k_{\text {crit }}$ are called \textit{long waves} ($\lambda \gg \lambda_{crit} $), while those with $|k| \gg k_{\text {crit }}$ are called \textit{short waves} ($\lambda \ll \lambda_{crit} $). 

The physical meaning of \autoref{eq:stability-limit} is the following. The first term represents thermal pressure from the proper motion of the gas, which stabilizes the disk; the second term accounts for rotation-supported stability. The third term, with a negative sign, accounts for the effects of self-gravity and it is responsible for the instabilities. When those effects surpass the thermal and rotation-provided stability support, the existence of a stable density wave becomes impossible. \par

\subsubsection{Propagation of density waves}
The density waves propagate radially like wave packets with a defined group velocity (\autoref{eq:group-velocity}), and it carries energy and angular momentum. Following the dispersion relation of \autoref{eq:dispersion-relation-fluid} we get the group velocity of the density wave:
\begin{equation}
\label{eq:groupvel-fluid}
    v_{g}=\frac{\partial \omega(k, R)}{\partial k}=\pm \frac{|k| c_{s}^{2}-\pi G \Sigma_{0}}{\omega-m \Omega},
\end{equation}
where the $\pm$ indicates leading $k<0$ or trailing $k>0$ arms, as in \figref{pitch_angle}. If $v_{g}>0$, the wave packet propagates radially outwards, if $v_{g}<0$ it propagates radially inwards. \par 
The numerator of \autoref{eq:groupvel-fluid}  is negative for short waves ($|k|\gg k_{crit}$), and it is positive for long waves ($|k|\ll k_{crit}$). On the other hand, the denominator can also be positive or negative, depending of we are inside or outside the co-rotation radius. A spiral pattern with $\Omega(R) > \Omega_{p}$ is inside the corotation radius $R<R_{CR}$. A pattern with $\Omega(R) > \Omega_{p}$ lies outside the corotation radius, $R>R_{CR}$. The denominator in \autoref{eq:groupvel-fluid} is positive if $R>R_{CR}$ and negative if $R<R_{CR}$.
\begin{itemize}
    \item Short waves + trailing arms $\rightarrow$ $v_{g}>0$, propagate away from $R_{CR}$.
    \item Short waves + leading arms $\rightarrow$ $v_{g}<0$, propagate approaching to $R_{CR}$.
    \item Long waves + trailing arms $\rightarrow$ $v_{g}<0$, propagate approaching to $R_{CR}$.
    \item Long wave + leading arms $\rightarrow$ $v_{g}>0$, propagate away from $R_{CR}$.
\end{itemize} 
\chapter{Simulations and methods} % Main chapter title

\label{Chapter3} % For referencing the chapter elsewhere, use \ref{Chapter3} 

\section{Description of the simulations}
\label{Description-of-the-simulations}

In this chapter we give a description of the simulations that were used in this work, the galaxy model, the parameters used for it and a brief descriptions of the physics involved in the dynamics of the gas.

\subsection{The GIZMO code}
Gizmo is a muti-physics simulations code that solves the fluid equations and self-gravity, the code implements meshless Lagrangian methods for the resolution of the hydrodynamics in which the mesh elements follow the fluid. The code includes ideal MHD (Magnetohydrodynamics). A full description of the numerical methods used in the code can be found in \citet{hopkins2015new}, for the specific details regarding the MHD part see \citet{hopkins2016accurate}. The GIZMO code is built on top of the cosmological simulation code GADGET-2 by \citet{springel2005cosmological}. \par
The way in which the gravitational interactions are computed in GIZMO is the same as in GADGET-2, where a \enquote{Tree-PM} algorithm is implemented. The main difference between the GIZMO and GADGET-2 lies in how the hydrodynamic equations are discretized. GADGET uses traditional SPH (Smoothed Particle Hydrodynamics), while GIZMO implements various methods, which include MFM (Meshless Finite-Mass) as the default and most accurate method.

\subsubsection{Ideal MHD}
In this work we mainly study the gas component of the simulated galaxies, therefore it is necessary that we review the physics involved in its dynamics. MHD studies the interactions between fluid flows and magnetic fields. The ideal approximation lies in considering the fluid as infinitely conducting. The main equations that rule the interplay between the plasma and the magnetic field are the equations of fluid dynamics (Euler equations) and the Maxwell equations. \par
The Euler equations are a set of partial hyperbolic differential equations that describe conservation laws for mass, momentum and energy. The form of these equations in a rest reference frame, with no source terms, is:
\begin{equation}
\label{eq:general_euler}
    \frac{\partial \mathbf{U}}{\partial t}+\nabla \cdot \mathbf{F}=0,
\end{equation}
and the conserved quantities are contained in the vector $\mathbf{U}$:
\begin{equation}
\label{eq:conserved_vector}
    \mathbf{U}=\begin{bmatrix}
    \rho \\
    \rho \mathbf{v} \\
    \rho e \\
    \end{bmatrix}.
\end{equation}
The tensor $\mathbf{F}$ contains the \textit{flux} of the corresponding conserved quantities:
\begin{equation}
\label{eq:flux_tensor}
    \mathbf{F}=\begin{bmatrix}
    \rho \mathbf{v} \\
    \rho \mathbf{v} \otimes \mathbf{v}+P_{T} \mathcal{I}-\mathbf{B} \otimes \mathbf{B} \\
    \left(\rho e+P_{T}\right) \mathbf{v}-(\mathbf{v} \cdot \mathbf{B}) \mathbf{B} \\
    \end{bmatrix},
\end{equation}
where $\otimes$ is the outer product of two vectors, $\mathcal{I}$ is the identity tensor, $\rho$ is the mass density, $P_{T}=P+|\mathbf{B}|^{2} / 2$ is the total pressure in the fluid (thermal and magnetic contributions), $e=u+|\mathbf{B}|^{2} / 2 \rho+|\mathbf{v}|^{2} / 2$ is the total specific energy (internal, magnetic and kinetic contributions). \par
The Maxwell equations for ideal MHD are:
\begin{align}
\begin{split}
\label{eq:maxwell_1}
    \nabla \cdot \mathbf{E} &=\frac{\rho}{\varepsilon_{0}}
\end{split} \\
\begin{split}
\label{eq:maxwell_2}
    \nabla \cdot \mathbf{B} &=0
\end{split} \\
\begin{split}
\label{eq:maxwell_3}
    \nabla \times \mathbf{E} &=-\frac{\partial \mathbf{B}}{\partial t}
\end{split} \\
\begin{split}
\label{eq:maxwell_4}
    \nabla \times \mathbf{B} &=\mu_{0} \mathbf{J},
\end{split}
\end{align}
where displacement current is neglected in \autoref{eq:maxwell_4} as it is considered negligible by comparison with the current density $\mathbf{J}$. Consider Ohm's Law for a reference frame in which the particle is at rest
\begin{equation}
\label{eq:ohms-law}
\mathbf{J}=\sigma (\mathbf{E}+\mathbf{u} \times \mathbf{B}).
\end{equation}
If we replace \autoref{eq:ohms-law} and \autoref{eq:maxwell_4} into \autoref{eq:maxwell_3} and consider that $\mathbf{B}$ is solenoidal, we get the induction equation:
\begin{equation}
\label{eq:induction_equation}
    \frac{\partial \mathbf{B}}{\partial t}=\nabla \times(\mathbf{v} \times \mathbf{B})+\lambda \nabla^{2} \mathbf{B},
\end{equation}
where $\lambda$ is the magnetic diffusivity. This equation is a transport equation for the magnetic field, it determines its temporal and spatial evolution given a field of velocities $\mathbf{v}$ and appropriate initial conditions. \autoref{eq:induction_equation} can be written in the form of \autoref{eq:general_euler} and completes the set of equation that govern the fluid evolution in the ideal MHD scenario. \par
It is important to mention that in GIZMO, \autoref{eq:general_euler} is not homogeneous, i.e, there are source terms in the right hand side of the equations. Those \textit{source} terms are called divergence-cleaning terms, their goal is to maintain the divergence of the magnetic field as close to zero as possible, in order to respect Maxwell's \autoref{eq:maxwell_2}. The divergence cleaning schemes used are explained in \citet{hopkins2016accurate} and references therein.  

\subsection{Parameters of the galaxy model}
\label{Parameters-of-the-galaxy-model}
\begin{figure}[ht]
    \centering
    \includegraphics[width=0.4\textwidth]{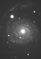}
    \caption[Image of the galaxy AM2322A in the r'-band.]{A zoom to the main component of the AM2322-821 system, a Sab spiral galaxy. This image was taken with the GMOS-S instrument at an effective wavelength of 630nm. This image file was extracted from a fits in the Gemini Observatory Archive.}
    \label{fig:AM2322A}
\end{figure}

The galaxy model that we are studying in this work comes from a previous study made by \citet{arboleda2019}. The system AM2322-321 is a minor merger located at the Octans constellation. The image that appears in \figref{AM2322A} is part of the GMOS-S r' acquisition images,  this image was taken with the GMOS-S instrument at the Gemini observatory. The image does not show the interacting galaxy, but zooms only on the main component of the system, which we shall call AM2322A from now on. Some key parameters of AM2322A were measured by \citet{krabbe2011effects}, in Table~\ref{tab:props-galaxy} we show their results.

\begin{table}[ht]
\centering
\caption[Observational properties of the galaxy AM2322A]{Measured properties for the main component of the AM2322-821 interacting system obtained by \citet{krabbe2011effects}}
\label{tab:props-galaxy}
\begin{tabular}{@{}lllll@{}}
\toprule
 \hfil                                                              &   &\hfil \textbf{AM2322A}              \\ 
 \midrule
 \hfil Distance [Mpc]  &   &\hfil$49.6$   \\
 %\midrule
 \hfil $M_{B}$ [mag]                     &   &\hfil $ -20.98 $                  \\ 
 %\midrule
 \hfil Mass $[M_{\odot}]$                                           &   &\hfil$1.7\times10^{11}$        \\ 
 %\midrule
 \hfil $R_{25}$ [Kpc]                                           &   &\hfil$13.5$        \\ 
 \bottomrule
 
\end{tabular}
\end{table}

By performing surface photometry to this galaxy, \citet{arboleda2019} fitted a exponential luminosity profile of the form \autoref{eq:brightness_disk_radial}, and found that the radial scale lenght is $h=4.35$ kpc. 
\subsubsection{Dark Matter Halo}
The dark matter halo parameters are the virial radius $r_{200}$, and the mass inside this radius defined as $M_{200}$. Inside $r_{200}$ the halo is considered in virial equilibrium, so we call $r_{200}$ and $M_{200}$ its virial radius and virial mass, respectively. A way to estimate these parameters is by using virial relations. We have that the maximum rotational velocity and the radius of dark matter halos are both related with the mass:
\begin{equation}
\label{eq:virial_velmax}
    \log _{10}\left(\frac{V_{\mathrm{cmax}}}{[k m / s]}\right)=0.332285 \log _{10}\left(\frac{M_{200}}{\left[M_{\odot} / h\right]}\right)-1.755139
\end{equation}

\begin{equation}
\label{eq:virial_radius}
    \log _{10}\left(\frac{R_{200}}{[k p c / h]}\right)=0.330886 \log _{10}\left(\frac{M_{200}}{\left[M_{\odot} / h\right]}\right)-1.75495.
\end{equation}

This relations are obtained from cosmological simulations of structure formation performed by \citet{arboleda2019}. To obtain $V_{cmax}$, \citet{arboleda2019} used Tully-Fisher relations constructed by \citet{bell2001stellar}, and the value of the absolute magnitude in the B-band of $M_{B} = -16,99 \pm 0.30$ measured by \citet{ferreiro2008sample}. With this value of the maximum circular velocity, $M_{200}$, and $r_{200}$ are easily computed from \autoref{eq:virial_velmax} and \autoref{eq:virial_radius}. \par

Other important parameters that describe the morphology of dark matter halos are the \textit{NFW scale length}, the \textit{Hernquist scale length} and the \textit{concentration parameter}. The first two are characteristic lengths for the NFW and Hernquist density profiles:
\begin{equation}
\label{eq:NFW_profile}
    \rho_{\mathrm{NFW}}(r) = \frac{\rho_{0}}{\left(r / r_{\mathrm{s}}\right)\left(1+r / r_{\mathrm{s}}\right)^{2}}
\end{equation}
\begin{equation}
\label{eq:hernquist_profile}
    \rho_{\mathrm{H}}(r)=\frac{M_{\mathrm{h}}}{2 \pi} \frac{r_{\mathrm{h}}}{r\left(r+r_{\mathrm{h}}\right)^{3},}
\end{equation}
while the concentration parameter is
\begin{equation}
\label{eq:concentration_parameteer}
    c=\frac{R_{200}}{r_{s}}. 
\end{equation}
The concentration parameter was set to $c=12$ and from this value both $r_{\mathrm{s}}$ and $r_{\mathrm{h}}$ were computed.
\subsubsection{The Disk}
The composition of the gas disk is assumed to be mainly of neutral hydrogen. The ratio between the stellar mass and the gas mass used to determine the gas content is the one from \citet{catinella2010galex}, who found that $M_{HI}/M_{\star,d} \approx 0.1$. An estimate of the total stellar mass of the disk can be computed from the mass of the dark halo, known as the stellar-to-halo mass ratio. The model of \citet{moster2010constraints} was used, getting a value for the stellar mass of the disk of $M_{\star,d} \sim 1.6 \times 10^{10} M_{\odot}$ and therefore a gas mass of $M_{HI} \sim 1.6 \times 10^{9}$

\subsubsection{The star formation model}
The model of star formation used in these simulations is the one proposed by \citet{springel2003cosmological}. This model is an statistical formulation of the process of stellar formation, i.e , it uses spatially averaged properties of the ISM in an attempt to account for the behaviour at sub-resolved scales. \par
The ISM is two-phased: cold giant molecular clouds in pressure equilibrium and a hot ambient gas. Of course, another component of the model are the stars that result from this process. Both phases of the gas interact, there is mass and energy exchange processes. The mass exchange takes place through star formation, cloud evaporation (caused by supernovae) and cloud growth. \par 
The characteristic time-scale in which the star formation process takes place is denoted as $t_{\star}$ and the fraction of stars that rapidly die out as supernovae is $\beta$, so the effective rate at which stars are formed is 
\begin{equation}
\label{eq:sfr_s&h2003}
    \frac{\mathrm{d} \rho_{\star}}{\mathrm{d} t}=\frac{\rho_{\mathrm{c}}}{t_{\star}}-\beta \frac{\rho_{\mathrm{c}}}{t_{\star}}=(1-\beta) \frac{\rho_{\mathrm{c}}}{t_{\star}},
\end{equation}
where $\rho_{\star}$ and $\rho_{\mathrm{c}}$ are the density of stars and the cold clouds, respectively. All the ejected gas from supernovae explosions is considered as hot gas, so it contributes to $\rho_{\mathrm{h}}$ (hot ambient gas density). The parameter $\beta$ accounts for the fraction of the stars that eventually explodes as supernovae, i.e, the fractions of stars with $\mathrm{M} > 8\mathrm{M}_{\odot}$. This fraction is highly dependent of the IMF. For a \citet{salpeter1955luminosity} IMF function and a range of mass between $0.1\mathrm{M}_{\odot} < M < 40\mathrm{M}_{\odot}$, the fraction is $\beta = 0.106$. The returned gas from supernovae is hot, metal rich gas that mixes with the ISM. \par
In addition to the mass exchange, there is also an energy input that happens through supernovae. This \enquote{feedback energy} heats the ambient hot phase of the gas; the rate at which this heating caused by supernovae happens is given by
\begin{equation}
\label{eq:heating_s&h2003}
    \left.\frac{\mathrm{d}}{\mathrm{d} t}\left(\rho_{\mathrm{h}} u_{\mathrm{h}}\right)\right|_{\mathrm{SN}}=\epsilon_{\mathrm{SN}} \frac{\mathrm{d} \rho_{\star}}{\mathrm{d} t}=\beta u_{\mathrm{SN}} \frac{\rho_{\mathrm{c}}}{t_{\star}},
\end{equation}
where $u_{\mathrm{h}}$ is the internal energy of the hot gas, $\epsilon_{\mathrm{SN}}=4 \times 10^{48} \mathrm{erg} \mathrm{M}_{\odot}^{-1}$ and $u_{\mathrm{SN}} \equiv(1-\beta) \beta^{-1} \epsilon_{\mathrm{SN}}$. Another way in which mass is exchanged between the cold clouds and the hot ambient gas is through cloud evaporation, which is caused by the heat input described by \autoref{eq:heating_s&h2003}:
\begin{equation}
\label{eq:cloudevap_s&h2003}
    \left.\frac{\mathrm{d} \rho_{\mathrm{c}}}{\mathrm{d} t}\right|_{\mathrm{EV}}=A \beta \frac{\rho_{\mathrm{c}}}{t_{\star}},
\end{equation}
where $A \propto \rho^{-4/5}$ is the efficiency of the evaporation process, that depends on the local properties of the ISM (\citealt{mckee1977theory}). Another process to take into consideration is the one of cold cloud creation and growth. The model assumes a mass exchange between ambient gas and clouds through thermal instability. Furthermore, as the hot ambient gas radiates energy away, the cold clouds' mass grows at the same rate. Such mass flux between phases can be expressed as      
\begin{equation}
\label{eq:cloudgrowth_s&h2003}
    \left.\frac{\mathrm{d} \rho_{\mathrm{c}}}{\mathrm{d} t}\right|_{\mathrm{TI}}=-\left.\frac{\mathrm{d} \rho_{\mathrm{h}}}{\mathrm{d} t}\right|_{\mathrm{TI}}=\frac{1-f}{u_{\mathrm{h}}-u_{\mathrm{c}}} \Lambda_{\mathrm{net}}\left(\rho_{\mathrm{h}}, u_{\mathrm{h}}\right),
\end{equation}
where $u_{\mathrm{c}}$ is the internal energy of cold clouds and $\Lambda_{\mathrm{net}}$ is the cooling function that corresponds to the radiative process of a H/He-composed plasma (\citet{katz1996numerical}). The parameter $f$ can be either $f=0$ when thermal instability operates, or $f=1$ when an ordinary cooling process take place, $\frac{\mathrm{d} \rho_{\mathrm{c}}}{\mathrm{d} t}=-\frac{\mathrm{d} \rho_{\mathrm{h}}}{\mathrm{d} t}$. Finally, the rates of cold clouds and hot gas growth are

\begin{equation}
\label{eq:coldgasgrowth_s&h2003}
    \frac{\mathrm{d} \rho_{\mathrm{c}}}{\mathrm{d} t}=-\frac{\rho_{\mathrm{c}}}{t_{\star}}-A \beta \frac{\rho_{\mathrm{c}}}{t_{\star}}+\frac{1-f}{u_{\mathrm{h}}-u_{\mathrm{c}}} \Lambda_{\mathrm{net}}\left(\rho_{\mathrm{h}}, u_{\mathrm{h}}\right)
\end{equation}

\begin{equation}
\label{eq:hotgasgrowth_s&h2003}
    \frac{\mathrm{d} \rho_{\mathrm{h}}}{\mathrm{d} t}=\beta \frac{\rho_{\mathrm{c}}}{t_{\star}}+A \beta \frac{\rho_{\mathrm{c}}}{t_{\star}}-\frac{1-f}{u_{\mathrm{h}}-u_{\mathrm{c}}} \Lambda_{\mathrm{net}}\left(\rho_{\mathrm{h}}, u_{\mathrm{h}}\right),
\end{equation}
where the first term accounts for star formation, the second for cloud evaporation and the third for radiative cooling (or heating) of the gas. For the energy rates of change, we have
\begin{equation}
\label{eq:energycoldchange_s&h2003}
    \frac{\mathrm{d}}{\mathrm{d} t}\left(\rho_{\mathrm{c}} u_{\mathrm{c}}\right)=-\frac{\rho_{\mathrm{c}}}{t_{\star}} u_{\mathrm{c}}-A \beta \frac{\rho_{\mathrm{c}}}{t_{\star}} u_{\mathrm{c}}+\frac{(1-f) u_{\mathrm{c}}}{u_{\mathrm{h}}-u_{\mathrm{c}}} \Lambda_{\mathrm{net}}
\end{equation}

\begin{equation}
\label{eq:energyhotchange_s&h2003}
    \frac{\mathrm{d}}{\mathrm{d} t}\left(\rho_{\mathrm{h}} u_{\mathrm{h}}\right)=\beta \frac{\rho_{\mathrm{c}}}{t_{\star}}\left(u_{\mathrm{SN}}+u_{\mathrm{c}}\right)+A \beta \frac{\rho_{\mathrm{c}}}{t_{\star}} u_{\mathrm{c}}-\frac{u_{\mathrm{h}}-f u_{\mathrm{c}}}{u_{\mathrm{h}}-u_{\mathrm{c}}} \Lambda_{\mathrm{net}},
\end{equation}

where \autoref{eq:energycoldchange_s&h2003} is obtained by doing \autoref{eq:coldgasgrowth_s&h2003}$\times u_{\mathrm{c}}$, similarly for \autoref{eq:energyhotchange_s&h2003} but with an additional term that contains the supernovae energy input $u_{\mathrm{SN}}$. The model assumes a fixed temperature for the clouds, so $u_{\mathrm{c}}$ is treated as a constant. \par

A key characteristic of this model is that it does not consumes and turns all the gas into stars in an unrealistic way. It is, in this sense, self-regulated. This self-regulation comes from two things: cloud evaporation and hot ambient gas cooling. The evaporation process (caused by supernovae) reduces the density of the clouds and therefore the SFR. This evaporation process causes an increase in the density of the hot gas, which leads to an increase in its cooling rate that replenishes the clouds and increases SFR. \par

\subsubsection{Stability-related parameters}

When simulating isolated galaxies, one of the main goals is to ensure that the system is stable. We want to make sure that the galaxy won't be violently disrupted due to unphysical initial configurations.  \par
In a simple model of the origin of disk galaxies, the disk inherits properties of its surrounding dark matter halo. The relation between the dark matter halo's angular momentum and that of the disk, plays an important role in the stability of the latter. The \textit{spin parameter} quantifies the importance of rotation for self gravitating systems, the value of this parameter was set based in the criterium obtained by \citet{quiroga2018polar}. \par
In Table~\ref{tab:final-parameters} there is a summary of the relevant parameters of the galaxy for every component, including the bulge, whose mass was determined as $M_{b} = 0.3M_{d}$ for stability purposes as well. These are the values upon which the initial conditions of the simulation are constructed. \par
\begin{table}[ht]
\centering
\caption[Physical parameters of the galaxy]{Physical parameter for each component of the system. }
\label{tab:final-parameters}
\tabcolsep=0.11cm
\begin{tabular}{@{}lllll@{}}
\toprule
 \hfil \textbf{Component}   &\hfil \textbf{Parameters}                       &\hfil \textbf{AM2322A}     \\ 
 \midrule
 \hfil                      &\hfil Virial mass ($M_{200}[M_{\odot}]$)        &\hfil$1.443\times10^{12}$  \\
 \hfil                      &\hfil Virial radius ($r_{200}[kpc]$)            &\hfil$185.401$             \\
 \hfil                      &\hfil Velocity ($V_{200}[km/s]$)                &\hfil$194.535$             \\
 \hfil \textbf{Halo}        &\hfil NFW scale length ($r_{s}[kpc]$)           &\hfil$37.966$              \\
 \hfil                      &\hfil Hernquist scale length ($r_{h}[kpc]$)     &\hfil$77.871$              \\
 \hfil                      &\hfil Concentration ($c$)                       &\hfil$12.0$                \\
 \hfil                      &\hfil Spin parameter ($\lambda$)                &\hfil$0.4$                 \\ 
 \midrule
 \hfil                      &\hfil Stellar Mass ($M_{\star,d}[M_{\odot}]$)   &\hfil$1.6\times10^{10}$    \\
 \hfil                      &\hfil Disk mass fraction ($m_{d}$)              &\hfil$0.01$                \\
 \hfil \textbf{Disk}        &\hfil Gas mass fraction (of $M_{\star,d}$)      &\hfil$0.1$                 \\
 \hfil                      &\hfil Angular momentum ($j_{d}$)                &\hfil$0.01$                \\
 \hfil                      &\hfil Radial scale ($h[kpc]$)                   &\hfil$4.35$                \\
 \midrule
 \hfil                      &\hfil Total mass ($M_{\star,d}[kg]$)            &\hfil$4.8\times10^{9}$     \\
 \hfil \textbf{Bulge}       &\hfil Mass fraction ($m_{b}$)                   &\hfil$0.003$               \\
 \hfil                      &\hfil Scale length in units of h                &\hfil$0.2$                 \\
 \bottomrule
\end{tabular}
\end{table}

Regarding the numerical parameters of the simulation, it is worth mentioning the softening length, which is necessary to avoid numerical errors in close encounters as $r \rightarrow 0 $. A way in which it can be defined, takes into consideration the amount of particles of the component and its characteristic length:
\begin{equation}
\label{eq:softening}
    \epsilon=\frac{r}{\sqrt{N_{\mathrm{p}}}},
\end{equation}
where $r$ is the scale length of the component and $N_{p}$ is the number of particles of the corresponding component. \autoref{eq:softening} is from \citet{dehnen2011n}. The different softening lengths are presented in  Table~\ref{tab:final-parameters-sim}, along with the number of particles of each component and the mass resolution, i.e, the mass of each particle for every component of the galaxy. 
\subsubsection{Visualization of the gas disk evolution.}
\begin{figure}[H]
    \centering
    \includegraphics[height=9.4cm]{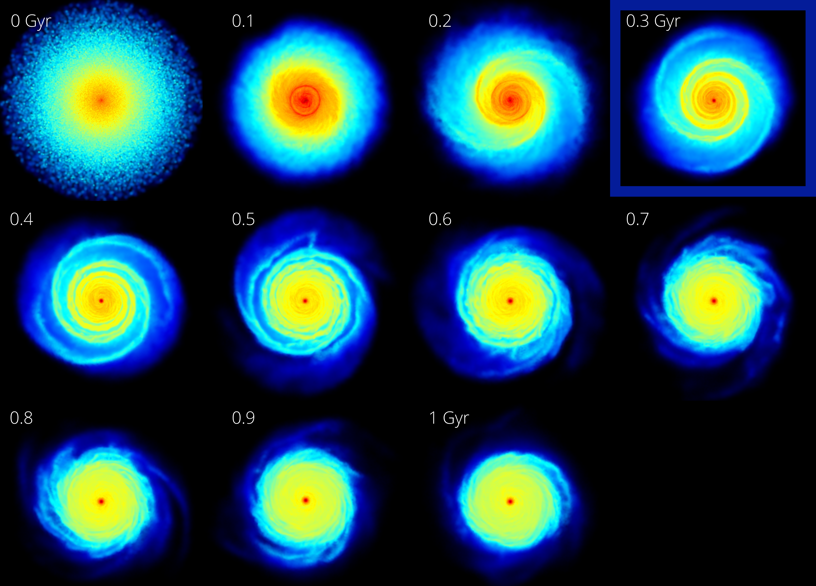}
    \caption[A organized sequence showing the gas disk of the galaxy during the simulation.]{From top to bottom and left to right, we show an organized sequence of the gas disk during the whole simulation. The time step between snapshots is $\Delta t = 0.1 \ \text{Gyr}$, the top left image shows the initial configuration of the gas particles at $t=0$. The last image (bottom right) shows the end of the simulation at $t = 1 \ \text{Gyr}$. The image in the blue box, at $t = 0.3 \text{Gyr}$, is the the snapshot we are studying in this work. The visualization tool used is Glnemo2, an open source software released under the terms of the \href{https://spdx.org/licenses/CECILL-2.0.html}{CeCILL2 Licence} }
    \label{fig:snapshot_collage}
\end{figure}

In \figref{snapshot_collage} we see a sequence of the gas disk of the galaxy we are studying. The whole simulations runs during $1 \ \text{Gyr}$ (in code units). However, we do not study the time evolution of the disk, we choose only a snapshot where spiral structure is present. It is important to mention this, we emphasize that this work deals with a stationary configuration of the disk, any other subjects such as the evolution of the spiral pattern in time are beyond the scope of this work. The election of snapshot 3 at $t = 0.3 \ \text{Gyr}$ is based on how identifiable the spiral pattern is, we considered that at this moment of the evolution the spiral arms are best formed and therefore this configuration is ideal for the study we will perform. \par
The color-map in \figref{snapshot_collage} indicates the values of density of the gas particles being the coolest color the less dense particles. The opposite for the more dense particles (hotter is denser).
\vspace{-0.8cm}
\begin{table}[H]
\centering
\caption[Numerical parameters of the galaxy.]{Numerical parameters of the galaxy simulations, including the resolution and the values of the softenings for the different components.  }
\label{tab:final-parameters-sim}
\tabcolsep=0.11cm
\begin{tabular}{@{}lllll@{}}
\toprule
 \hfil \textbf{Component}     &\hfil \textbf{Parameters}                    &\hfil \textbf{AM2322A}     \\ 
 \midrule
 \hfil                        &\hfil Number of particles                    &\hfil$2000000$             \\
 \hfil \textbf{Halo}          &\hfil Total Mass ($M_{\odot}$)               &\hfil$1.9\times10^{12}$    \\
 \hfil                        &\hfil Mass resolution (($M_{\odot}$)         &\hfil$9.5\times10^{5}$     \\
 \hfil                        &\hfil Softening length ($\epsilon_{h}[kpc]$) &\hfil$0.1$                 \\
 \midrule
 \hfil                        &\hfil Number of particles (stars)            &\hfil$1000000$             \\
 \hfil \textbf{Stellar Disk}  &\hfil Total Mass ($M_{\odot}$)               &\hfil$1.5\times10^{10}$    \\
 \hfil                        &\hfil Mass resolution (($M_{\odot}$)         &\hfil$1.5\times10^{4}$     \\
 \hfil                        &\hfil Softening length ($\epsilon_{d}[kpc]$) &\hfil$0.015$               \\
 \midrule
 \hfil                        &\hfil Number of particles (stars)            &\hfil$375000$              \\
 \hfil \textbf{Gas Disk}      &\hfil Total Mass ($M_{\odot}$)               &\hfil$1.5\times10^{9}$     \\
 \hfil                        &\hfil Mass resolution (($M_{\odot}$)         &\hfil$4\times10^{3}$       \\
 \hfil                        &\hfil Softening length ($\epsilon_{d}[kpc]$) &\hfil$0.015$               \\
 \midrule
 \hfil                        &\hfil Number of particles                    &\hfil$250000$              \\
 \hfil \textbf{Bulge}         &\hfil Total Mass ($M_{\odot}$)               &\hfil$4.7\times10^{9}$     \\
 \hfil                        &\hfil Mass resolution (($M_{\odot}$)         &\hfil$1.88\times10^{4}$    \\
 \hfil                        &\hfil Softening length ($\epsilon_{b}[kpc]$) &\hfil$0.015$               \\
 \bottomrule
\end{tabular}
\end{table}

\section{Data extraction from the simulations}
\label{Data-extraction}

Now we describe all the data extraction and preparation process. The first thing to do is to extract all the necessary data from the snapshots. The output of the simulations are binary files in blocks. This type of file stores the information of each particle in blocks of data, so we extracted all the necessary information using a simple C code that reads through each block and outputs the information in standard ascii format. The information related to the star formation is our main focus in this work, we will target the gas component of the disk, since it contains the information of the SF of the particles. The SFR is not the only parameter that describes the gas particles of the simulation, hence, we also study the mean free-electrons number per proton, neutral Hydrogen fraction, gas density and gas' internal energy. 

\subsection{Method for the extraction of the spiral structure}
\label{extraction}
\begin{figure}[H]
\begin{subfigure}{0.49\textwidth}
  %\captionsetup[subfigure]{aboveskip=-1pt,belowskip=-1pt}
  \centering
  \includegraphics[height=7cm]{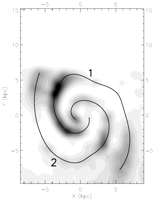}
  \caption{Arm tracing made by \citet{scheepmaker2009spatial}}
  \label{fig:tracing-arms-a}
\end{subfigure}
\begin{subfigure}{0.49\textwidth}
  \centering
  \includegraphics[height=7cm]{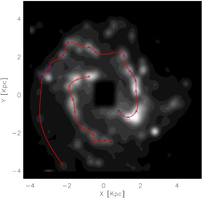}
  \caption{Arm tracing made by \citet{silva2012relation}}
  \label{fig:tracing-arms-b}
\end{subfigure}
\caption[Photometric arm tracing on M51 and M83.]{\textbf{(a)} Shows the spiral galaxy M51 with a gaussian blur that enhances pronounced emissions, as well as the interpolation that marks the arm path. The emission is near infrared. \textbf{(b)} Same but for the galaxy M83, and with H$\alpha$ emission.}
\label{fig:tracing-arms}
\end{figure}

Now we proceed to explain the process we followed to extract the spiral structure from the simulations. There are a number of different ways to identify the regions of over-density in spiral galaxies. From an observational standpoint, various indicators have been used to trace the spiral structure of galaxies, for example \citet{scheepmaker2009spatial} used near infrared to trace the mass spiral arms on the disk of the spiral galaxy M51; \citet{silva2012relation} used H$\alpha$ emissions. Both authors used similar methodology, which includes a masking of the central region $\lesssim 1$ kpc, detection of pronounced emission points using surface photometry techniques, and finally a process of interpolation between these points. \figref{tracing-arms} shows the results the authors got when performing the procedure. 

We want to achieve a similar outcome for the gas component of the galaxy, i.e, we want to detect and extract these regions of over-density in a self-consistent fashion. Moreover, we want to find the arm structure using every property of the gas particles. The goal is to \enquote{see} the arms through different indicators, as it is similarly done in the observational works previously mentioned.  \par
The methodology can be broken down into the following steps, and applies to every property that we are going to study: 
\begin{enumerate}
    \item Narrow down the data set in both R and z to isolate the disk (cylindrical coordinates).
    \item Find the mean property profile using a uniform radial binning.
    \item Perform a cubic-spline interpolation to the data points obtained in the previous step in order to get a continuous function that can be evaluated at any radius. 
    \item Determine the gas particles that belong to the background, i.e, particles that are below the mean density profile. To determine that we simply use a contrast parameter 
    \begin{equation}
    \label{eq:deltax}
        \Delta x=\frac{x(R,\theta)}{\overline{x}(R)} - 1,
    \end{equation}
    if $\Delta x < 0$ the particle's property is below the mean profile and hence belongs to the background of this particular property.
    \item The over-density regions are given by all the gas particles for which $\Delta x > 0$. 
\end{enumerate}

We will illustrate the procedure using $x \equiv \rho$, the density of the gas. First off, we filter the data based on the radial and vertical density distributions. For the $R$ range, we ignore all the gas inside $R<1$ kpc. The reason to do this is because the structure that dominates in this region is the bulge, and we are interested only in the disk. This fact manifests itself when we plot the radial profile and observe the behaviour of the mean density in the inner region of the disk. 

\begin{figure}[H]
    \centering
    \captionsetup{aboveskip=3pt,belowskip=3pt}
    \includegraphics[height=7.25cm]{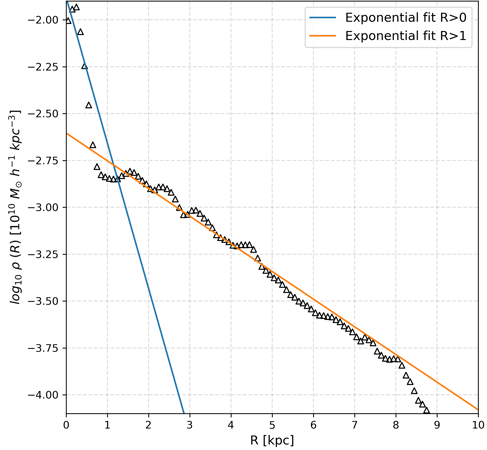}
    \caption[Exponential profile of the gas disk.]{Exponential fit for the gas density, taking the whole data for the fit (blue line) and restricting the fit at $R>1$ kpc (orange line). The black triangles are the radially-averaged density of the disk.}
    \label{fig:radial-density}
\end{figure}

In \figref{radial-density} we show an exponential fit of the form of \autoref{eq:surface_exponential} following the two approaches: the blue line, fits the complete data, i.e, for $R \geq 0 \text{kpc}$; the orange line, on the other hand, excludes the density particles inside $R \leq 1 \text{kpc}$. This clearly indicates that the density distribution in the inner region acts differently than the rest of the gas disk, i.e, its behaviour is not an exponential like \autoref{eq:surface_exponential}. 

\begin{figure}[H]
    \centering
    \captionsetup{aboveskip=3pt,belowskip=3pt}
    \includegraphics[height=7.25cm]{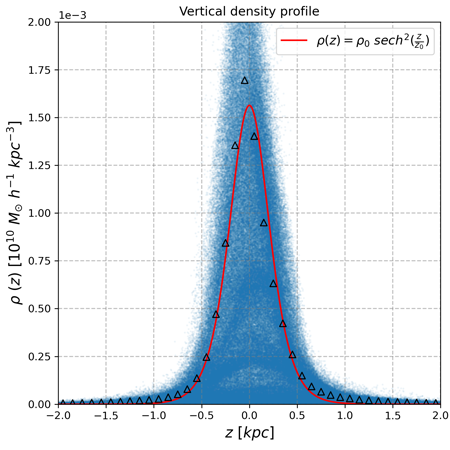}
    \caption[Vertical profile of the gas disk.]{Average vertical density distribution obtained by averagind the density in $\Delta z = 0.1 $ kpc bins. The fit of the data is an equation of the form of \autoref{eq:brightness_disk_vertical}, with vertical scale $z_{0}=0.28$ kpc}
    \label{fig:vertical-density}
\end{figure}

The criteria to limit the $z$ range is based on the vertical density profile of the disk. In \figref{vertical-density} we plot the mean vertical density, averaged in uniform $\Delta z$ bins, similarly as the radial profile. From this we can safely choose $z\lesssim1.5$ kpc to ensure that we get the regions of disk that are of our interest. We only want to study the most dense parts of the disk. Notice that beyond $z\sim 1.5 \text{kpc}$ the disk is made out of low-density and hot gas that escapes from the galaxy plane. They are part of the disk, but we are not interested in them.  
\begin{figure}[ht]
    \centering
    \captionsetup{aboveskip=0pt,belowskip=0pt}
    \includegraphics[height=9cm]{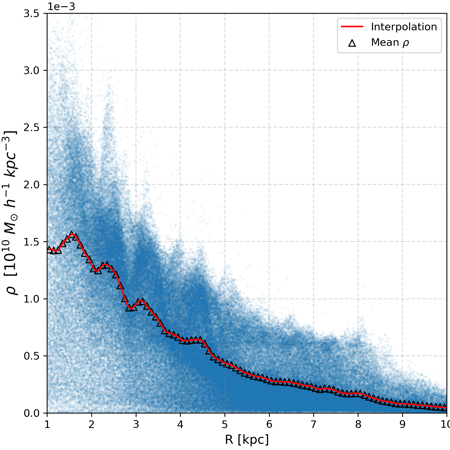}
    \caption[Interpolation of the radially averaged density.]{Interpolation of the radially averaged density for $R>1$ kpc (red line). The interpolation is made using a cubic spline method. The black triangles are the data points of $\bar{\rho}$.}
    \label{fig:interp-density}
\end{figure}

The second step is simply an interpolation of the radially-averaged data for $R>1$ kpc using a cubic spline, as shown in \figref{interp-density}. We also plot the whole distribution of points with an adequate value of transparency in order to visualize more features of the density distribution. The chosen binning is $R=0.1 $ kpc in order to account for the overdense values and avoid noise related to oversampling of the data. \par

Now we determine the background threshold using \autoref{eq:deltax}, and determine the background. This is made using the interpolating function $f(R) = \overline{\rho}(R) $, which returns the mean density at any given radius. \autoref{eq:deltax} then turns into

\begin{equation}
\label{eq:delta_rho}
    \Delta \rho= \frac{\rho(R,\theta,z)}{f(R)} - 1,
\end{equation}

so we are comparing every gas particle with density $\rho(R,\theta,z)$ to the the mean density profile using $f(R)$. \par
\begin{figure}[H]
    \centering
    \captionsetup{aboveskip=3pt,belowskip=3pt}
    \includegraphics[height=8cm]{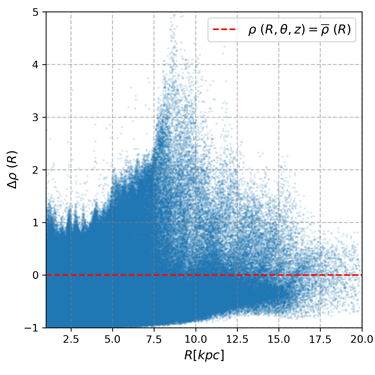}
    \caption[Density threshold.]{Plot of \autoref{eq:deltax} vs R. The red dotted line at $\Delta \rho = 0$ determines the density threshold. }
    \label{fig:density-threshold}
\end{figure}

\figref{density-threshold} shows the results of this calculation, as a function of the galactic radius. This procedure basically divides our data set in two, depending if the data points are above or below this threshold. This is important because we now know which particles of gas belong to the over-density regions and hence to the spiral arm structure. This way of filtering the data lets some outliers in, i.e, particles that not necessarily belong to the spiral structure, but still have high values of density. \par

\begin{figure}[H]
    \centering
    \captionsetup{aboveskip=3pt,belowskip=3pt}
    \includegraphics[width=\linewidth]{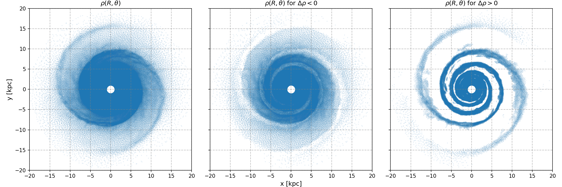}
    \caption[Comparison of the disk before and after the density filtering.]{The first picture is the data once we do the filters in $R$ and $z$ (step 1), the second one corresponds to the background of the property (step 4), and the final plot is only the data above threshold (step 5).}
    \label{fig:density-compared}
\end{figure}

The last step is to pick only the elements above the property threshold. In \figref{density-compared} we show a plot of the gas distribution of the initial data set, its background (particles below the density threshold) and the particles above threshold. 

\subsection{Analysis of the arm's local properties: measuring the arm widths.}
\label{analysis-local}
In the previous section, we explained in detail how the spiral structure was extracted from the gas disk. We ended up with two sets of data: a background and the over-density, that happens to contain the spiral structure in which we are interested, as shown in \figref{density-compared}.\par
We want to make a local study of the behaviour of the gas properties inside the spiral arms and outside of them (on which is usually referred to in the literature as the inter-arm regions). Particularly, we want to assess the morphology of the arms in all of the available properties, with a particular interest in the SFR.  \par
In this section, we will explain how we performed the analysis on the arm's morphology. Specifically, we will show the procedure through which we evaluated the width of the arms in a consistent way. We want to stress the fact that this procedure applies not only to the density (which we are simply using to illustrate the process and the reasoning behind it) but to all the different properties of the gas.  \par

\begin{figure}[ht]
    \centering
    \captionsetup{aboveskip=3pt,belowskip=3pt}
    \includegraphics[height=10cm]{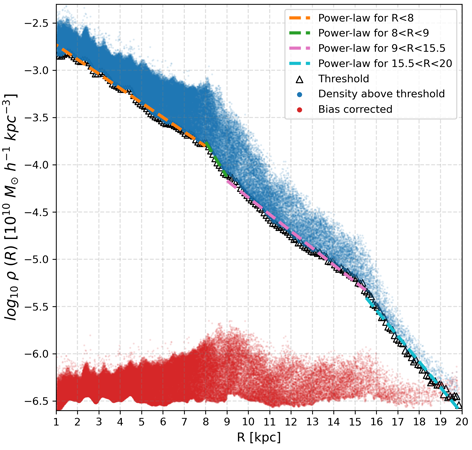}
    \caption[Flattened radial density profile.]{We plot the log of the density distribution above threshold (this is the same data of \figref{interp-density} but in log scale and only the points above threshold), the power laws fitted to the threshold data points (hollow triangles), and the resulting distributions after doing the subtraction (red points). }
    \label{fig:uptrend-bias}
\end{figure}

We started by looking at the density distribution of the data above the threshold, and noticing that along with the distribution there are density spikes, which are correlated with the arms positions in the disk. This can be seen on both \figref{interp-density} and \figref{density-threshold}. \par

Since the density distribution decreases exponentially with radius, the height of these density spikes also decrease with radius. This downward tendency of the distribution interferes in the correct modelling of the spikes, so we need to remove it. The correction is made by subtracting a fit to the threshold. \par
In \figref{uptrend-bias} we plot the logarithm of the density distribution \textit{above threshold}, before (blue) and after (red) the correction. As mentioned in the previous paragraph, the correction is a simple subtraction of power-laws fits to the threshold data in order to remove the downward tendency of the distribution. The result is that, now the corrected distribution (red) is much more \enquote{flat} than the original one (blue). \par
It is important to say that although it is true that the values of density are changing (we are re-scaling the points in the y axis), the properties of the density distribution remain unchanged, i.e, the width of the spikes is the same. \par

 By doing this correction to the data we can better analyze the over-density structures (the arms) without the interference of the uptrend tendency of the distribution. \par
 
 Now we are ready to take a closer look to the arms. Since we want to evaluate the extension and width of the arms, it is necessary to analyse the $\rho \, \text{vs} \, x$ plane (or $\rho \, \text{vs} \, y$, which would be as valid).\par
 
\begin{figure}[hbt!]
    \centering
    \captionsetup{aboveskip=3pt,belowskip=3pt}
    \includegraphics[width=\linewidth]{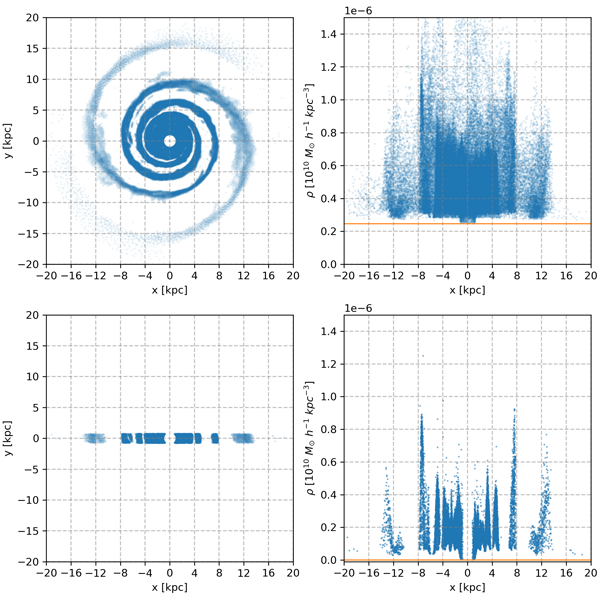}
    \caption[Band selection of width $y=1.5$ kpc for the density distribution.]{The first row of the figure shows the filtered gas disk ($\Delta \rho > 0$), and its corresponding density plot in the x coordinate. The second row of the figure, shows only a horizontal band of the disk of width $y = 1.5 $ kpc, and its density distribution on the right. The orange horizontal line in plots 2 and 4 mark the minimum value of the density. The data set is shifted so $\rho_{min} = 0$}
    \label{fig:band-selection}
\end{figure}

 \figref{band-selection} shows what we want to illustrate. The top row of plots shows the density distribution of the whole disk (with $\Delta \rho > 0$), we notice that this approach makes a localized characterization of the arms impossible. The second row of plots shows the distribution when we only take a \enquote{strip} of the galaxy disk, with a width of $y = 1.5 $ kpc. We choose this value to have enough data to perform the analysis and account for the actual width of the arms. \par
Once this is done, we can say that the density spikes that appear in the density distribution of \figref{band-selection}, describe the horizontal density distribution of the arm at a particular radius. It is worth noting that, in the inner radius, the curvature of the arms can introduce some noise in the $x$ vs $\rho$ distribution. To deal with this effect, we use a narrower strip to evaluate the width of the arms located close to the center, at $ R < 2$ kpc. We found that a strip with a width of  $y \approx 0.8$ kpc reduces the noise associated to the curvature while keeping a good amount of data of the arm.  \par
Now that we have selected this strip of disk, we can more easily fit the gaussians to the spikes present in the $x$ vs $\rho$ distributions.

\subsubsection{Locating the arms}
Before going ahead and do all the fits to the overdensities, we need to identify where the arms are located in the $x$ position, at least visually. The idea is to avoid fitting every single spike in the distribution of $x$ vs $\rho$, since some of then may not be part of an arm, i.e, there might be inter-arm scattered particles that represent a density spike in the $x$ vs $\rho$ but do not belong to the structure of any particular arm and hence do not contribute to its width. \par 

\begin{figure}[ht]
    \centering
    \captionsetup{aboveskip=3pt,belowskip=3pt}
    \includegraphics[height=12.5cm]{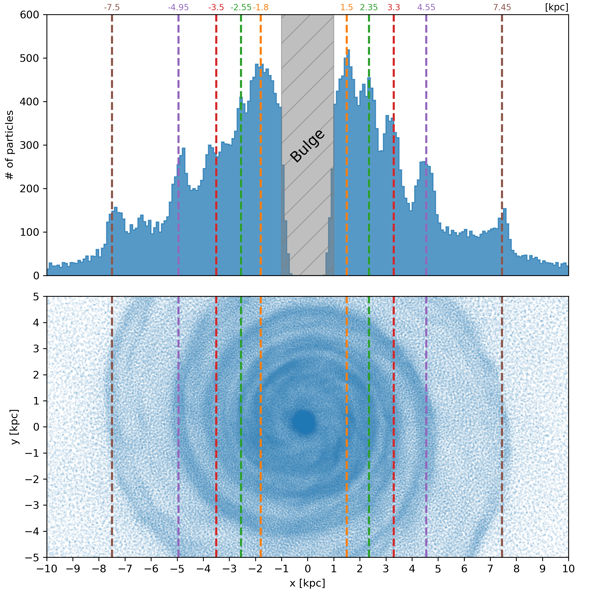}
    \caption[Location of the arms with a histogram.]{In the top plot, we show a histogram of numbers of particles and vertical lines that indicate the approximate position of the spiral arms in the $x$ axis. In the bottom plot, we show the spatial distribution of the particles in the disk. This particle distribution corresponds to the unaltered gas disk (without any type of threshold filtering). The $x$ axis is shared between the two plots so the vertical lines are in the same positions. }
    \label{fig:arm-position}
\end{figure}

In order to estimate the arms positions, we made a histogram that shows the distributions of the particles inside the horizontal band of width $1.5$ kpc (see \figref{band-selection}). The bin size is $ \Delta R = 0.1$ kpc. In  \figref{arm-position} we show both histogram and disk. The spiral arms clearly show up in the histogram since the number of particles increases as we move horizontally through the arms. \par
We indicate with vertical lines the position of this overdensities in the histogram and observe how they match with the position of the arms. In the upper part of the plot we indicate the $x$ coordinate of each one of the arms. This plot only goes to $R=10$ kpc, however both arms extend up to around $R=14$ kpc (see \figref{density-compared}. We do not show the entire disk because this would make it harder to visualize the arms in the inner regions of the disk, where the majority of the arms are.\par

\subsubsection{Modelling the arm's horizontal profile}
\label{gaussian-model}
Now we are left with the task of modelling the density spikes in order to give a measure of the width of the spiral arms. It is important that this measure is consistent, so we can compare the results of the different properties, i.e, we would like to see if the width of the arms is the same (or if it behaves similarly as a function of galactic radius) in density as it is in SFR, for instance. \par
For this purpose, we have taken a rather simple approach. We have chosen a Gaussian-type distribution of three parameters that has the form
\begin{equation}
\label{eq:gaussian}
   F(x;A,\mu,\sigma) = \frac{A}{\sqrt{2\pi} \sigma} e^{{ \frac{(x-\mu)^{2}}{2\sigma^{2}} }},
\end{equation}

where A is the amplitude (and the reason we are calling the function Gaussian-type and not simply Gaussian), $\mu$ is the mean value (the center of the arm, in this case), and the parameter $\sigma$ will be regarded as the width of the arms. It is worth mentioning that the election of $\sigma$ as the width of the arm is as good as any multiple of $\sigma$; for instance we could have chosen the FWHM (Full Width at Half Maximum) 
\begin{equation}
\label{eq:FWHM}
    \mathrm{FWHM}=2 \sqrt{2 \ln 2} \sigma,
\end{equation}
and it will still be a valid way to give a measure of the arm's width. Any reasonable choice of a multiple of sigma, $n\sigma$, could be in principle used to provide an estimate of the width. It is important that we stay consistent on how we measure the width, independent of the analysed property. For the density and all the other properties, our benchmark of the width will always be $\sigma$, this way, we achieve the desired consistency, i.e, we will be able to assert if any particular arm is wider that any other, without ambiguity. \par 

To fit \autoref{eq:gaussian} to the arm's density distribution, we need to extract the \enquote{envelope} of the density spikes that appear in \figref{band-selection}. To do this, we simply take the maximum value of $\rho$, with a step of $0.01 <\Delta x < 0.1 $ kpc, depending on the particular arm. The chosen value is such that the envelope is extracted in a precise way. This process requires some careful manipulation of the data, since there are points that do not belong to the distributions (to the spikes) and may interfere in the envelope-extraction procedure. 

For example, take a look at the lower right plot of \figref{band-selection}. We can see that there are data points scattered above the spikes in the range $1 \text{kpc} < R < 4 \text{kpc}$, if we take the maximum value of $\rho$ in that range, it is clear that those scattered values will be the ones returned. This detail has to be taken into consideration when doing the procedure, and hence, the evaluation of the maximum value needs an upper bound in density: take the maximum value between $R_{min} < R < R_{max}$ and $\rho_{min} < \rho <\rho_{max}$. \par

\begin{figure}[hbt!]
    \centering
    \captionsetup{aboveskip=3pt,belowskip=3pt}
    \includegraphics[width=\linewidth]{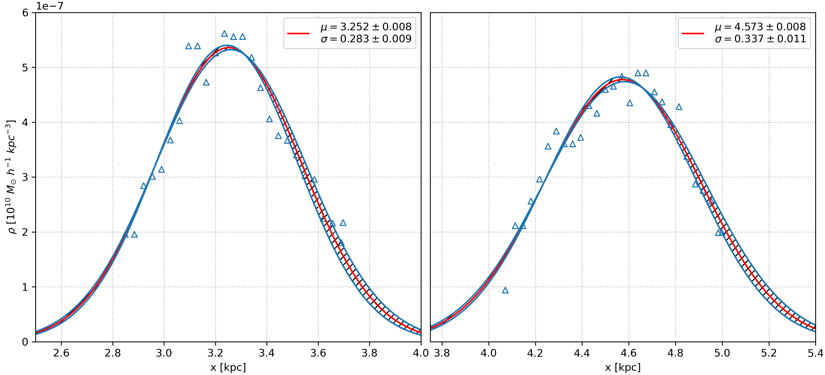}
    \caption[Two examples of arm envelopes in density.]{Two of the \enquote{envelopes} that wrap around the density spikes in \figref{band-selection}. The step is $\Delta x = 0.05$ kpc, and the fitted function,\autoref{eq:gaussian}, is plotted over the data, as well as the obtained numerical values for $\mu$ and $\sigma$.}
    \label{fig:arm-profiles}
\end{figure}

In \figref{arm-profiles} we show two examples of envelopes for two arms located at $x \approx 3,2$ and $x \approx 4.6$, in the right side of the disk. We indicate the obtained fitted values of $\mu$ and $\sigma$ and their corresponding errors. 

\subsubsection{Dealing with fractured arms}
\begin{figure}[hbt!]
    \centering
    \captionsetup{aboveskip=3pt,belowskip=3pt}
    \includegraphics[height=8cm]{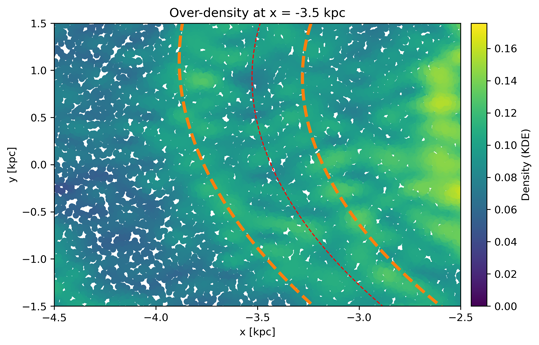}
    \caption[KDE in a region where the arm is fractured.]{Zoom in the arm at $x=-3.5$ kpc. The color map corresponds to a 2D Kernel Density Estimate of this region. Higher values (more yellow in this case) means more crowded regions.}
    \label{fig:divided-arm-KDE}
\end{figure}

The arms' shape is not uniform. This is both a reality in the actual observed galaxies and the one we are studying in this work. There are points where the arm brokes apart, or presents a clumpy, irregular shape. There are also cases where the arm gets divided in two, like the case of the arm at $x=-3.5$ kpc. If we look closely at the histogram in \figref{arm-position}, we see that the red line at $x=-3.5$ kpc is located between to spikes in number of particles despite being at the center of the spiral arm. This is how a fractured arm looks like. 
In \figref{divided-arm-KDE} we make a zoom in the region where this portion of the arm is. We performed a Kernel Density Estimate in 2D to make a color map of the particles based on how close they are. In this plot, there are two orange lines that limit the internal and external parts of the arm and a red line that goes through the center of it. In the middle, we can see that there is a hole in the density of particles. This fact is showed by the colouring of the particles in there, which appear more dark. 

To estimate the width of such a divided arm, we modelled the two parts individually with separate gaussians. From the fitted parameters of these two, we create a third one that is the sum of them. If we call $G_{out}(A_{out},\mu_{out},\sigma_{out})$ the gaussian that fits the outer part of the arm, and $G_{in}(A_{in},\mu_{in},\sigma_{out})$ to the one that fits the inner part, then $G_{T}(A_{T},\mu_{T},\sigma_{T})$
is the gaussian that fits the entire arm, where
$$ A_{T} = \sqrt{A_{in}^{2} + A_{out}^{2} }  $$
$$ \mu_{T} = \sqrt{\mu_{in}^{2} + \mu_{out}^{2} }  $$
$$ \sigma_{T} = \sqrt{\sigma_{in}^{2} + \sigma_{out}^{2} }.  $$
\begin{figure}[hbt!]
    \centering
    \captionsetup{aboveskip=3pt,belowskip=3pt}
    \includegraphics[height=8cm]{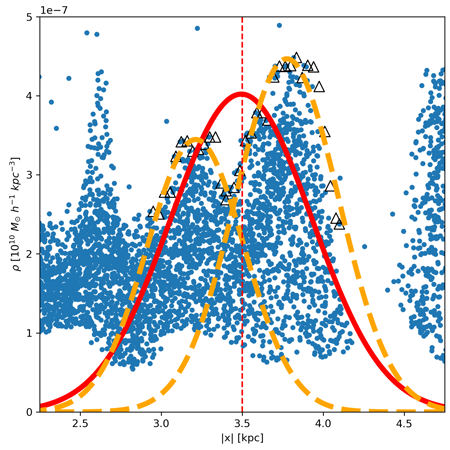}
    \caption[An example of a divided of fractured arm and its fit.]{Overdensity located at $x=-3.5$ kpc that presents a fractured shape. The black hollow triangles show the points of the envelope upon which the fit is done. The orange dashed lines show both $G_{in}$ and $G_{out}$. The solid red line is the sum of those two, $G_{T}$.}
    \label{fig:divided-arm-gaussians}
\end{figure}
In \figref{divided-arm-gaussians} we show the three gaussian fits in the case of the divided arm at $x=-3.5$ kpc. The dashed orange lines correspond to $G_{in}$ and $G_{out}$, the solid red line is $G_{T}$. Taking this approach allows us to still get a good estimate of the arm width.\par

\begin{figure}[H]
    \centering
    \captionsetup{aboveskip=3pt,belowskip=3pt}
    \includegraphics[width=\linewidth]{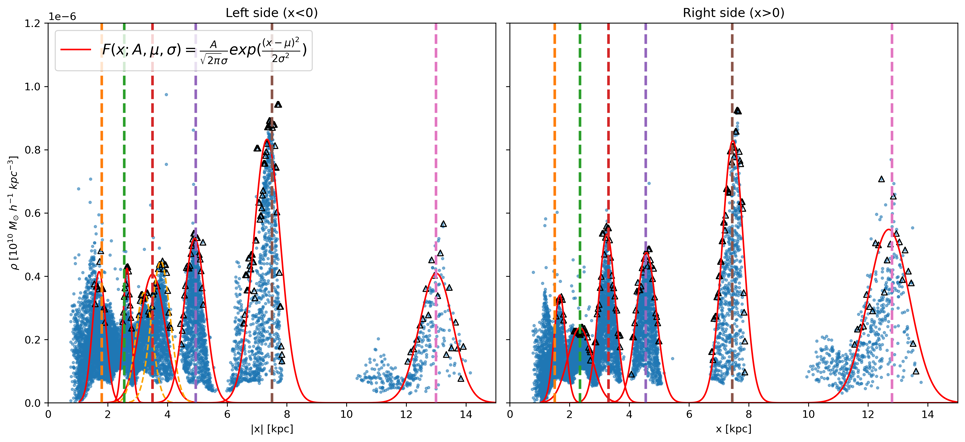}
    \caption[Final fitted arm profiles for the density.]{Plot of all the envelopes that wrap the density spikes in \figref{band-selection} and the corresponding fits of \autoref{eq:gaussian}. }
    \label{fig:left-right-arms}
\end{figure}

Finally, \figref{left-right-arms} shows the result for both sides of the disk. We indicate with vertical lines the positions of the arms that we identified with \figref{arm-position}, we have 6 overdensities at each side of the disk. In the left side, notice that we are plotting against the absolute value of the $x$ coordinate. This is the final outcome of the process and it is applicable to the rest of the gas properties (with some elements of the procedure that may differ slightly, specially in the case of the Free Electrons and Neutral Hydrogen abundances, as we shall see in the next Chapter). The specific details for the other properties will be reviewed in the next chapter as well.\par 

\chapter{Results} % Main chapter title

\label{Chapter4} % For referencing the chapter elsewhere, use \ref{Chapter3} 

In this chapter we present the structure of spiral arms obtained for the remaining properties. In Sections \ref{extraction} and \ref{analysis-local} we described a methodology for extracting the spiral structure (and in general, the over-density regions) from the gas disk, and gave a method to get an estimation of the width of the spiral arms. We described the process using the density of the gas, now we will present the results when applied to the other properties of the gas.  

\section{Star Formation Rate (SFR)}
\label{SFR-results}
\begin{figure}[!htbp]
\captionsetup[subfigure]{labelformat=empty}
\begin{subfigure}{0.49\textwidth}
  \centering
  \includegraphics[width=0.99\linewidth]{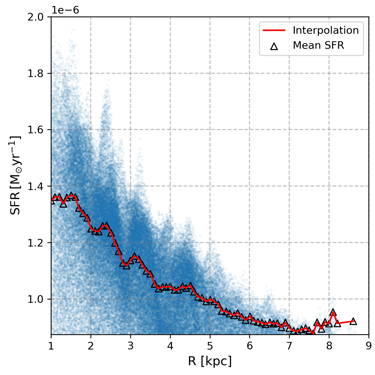}
  \caption{}
  \label{fig:interp-sfr}
\end{subfigure}%
\begin{subfigure}{.5\textwidth}
  \centering
  \includegraphics[width=0.99\linewidth]{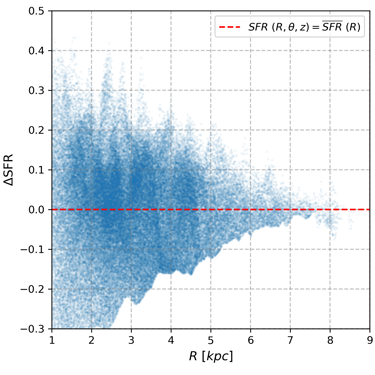}
  \caption{}
  \label{fig:sfr-threshold}
\end{subfigure}
\vspace{-1\baselineskip}
\caption[Radial profile and threshold determination for SFR.]{(a) Radial profile for the SFR, including its radially-averaged mean values (hollow rectangles) and the interpolation. (b) Threshold determination, computed with \autoref{eq:deltax}.}
\label{fig:SFR-interp-threshold}
\end{figure}

We found that the Star Formation Rate behaves in a similar way as the density when it comes to its radial distribution, as \figref{SFR-interp-threshold} shows. The star formation stops at around $R \approx 8.6 $ kpc, i.e, all gas particles beyond this radius have $\mathrm{SFR} = 0$. This is a direct consequence of the density not being enough to allow for star formation processes to occur at large radii. 
In \figref{sfr-threshold} we see the \enquote{SFR threshold}, similarly as the \enquote{density threshold} that we previously discussed and showed in \figref{density-threshold}. \par
\begin{figure}[!htbp]
    \centering
    \captionsetup{aboveskip=3pt,belowskip=3pt}
    \includegraphics[height=5.cm]{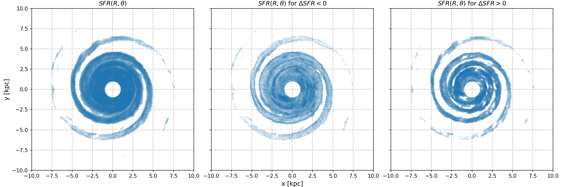}
    \caption[Comparison of the disk before and after the SFR filtering.]{The first figure shows the gas particles that have non-zero SFR, the second and third figures show the gas particles below and above threshold, respectively.}
    \label{fig:sfr-compared}
\end{figure}

\begin{figure}[!htbp]
    \centering
    \captionsetup{aboveskip=3pt,belowskip=3pt}
    \includegraphics[height=13.2cm]{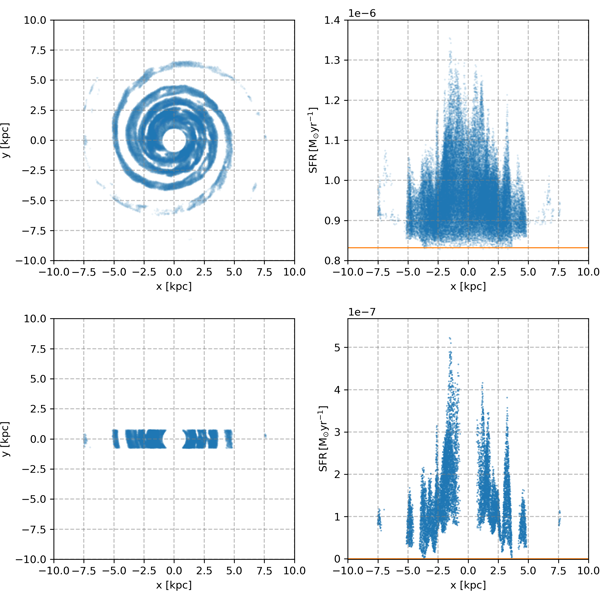}
    \caption[Band selection for the SFR distribution.]{The top row of the figure shows the filtered gas disk ($\Delta \mathrm{SFR} > 0$), and the corresponding SFR vs x plot in the right. The bottom row of the figure, shows only a horizontal band of the disk of width $y = 1.5 $ kpc, and its SFR distribution on the right. The orange horizontal line in plots 2 and 4 mark the minimum value of the SFR. The data-set is shifted so $\mathrm{SFR}_{min} = 0$}
    \label{fig:band-selection-sfr}
\end{figure}
For the Star Formation Rate, the disk looks like \figref{sfr-compared}, where the first plot shows all the gas particles with non-zero SFR, the second shows all the particles below threshold and the third shows all the particles with SFR above the threshold. This behaviour is different from the density plotted in \figref{density-compared}, because all the gas particles have a defined density while not all of them have non-zero SFR. \par 
Only imposing that $\mathrm{SFR} > 0$ is a good filter that exposes the regions of over-density. However we further reduce the sample by applying the same criteria we did for the density, the effect of this filtering is most noticeable in the inner regions where a separation between the arms appears. \par
\begin{figure}[!htbp]
    \centering
    \captionsetup{aboveskip=3pt,belowskip=3pt}
    \includegraphics[height=6.75cm]{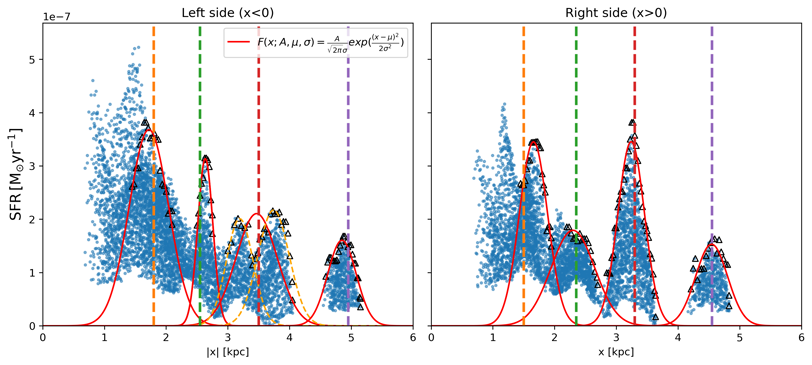}
    \caption[Final fitted arm profiles for the SFR.]{Plot of all the envelopes that wrap the density spikes in \figref{band-selection-sfr} and the corresponding fits of \autoref{eq:gaussian}. }
    \label{fig:left-right-arms-sfr}
\end{figure}

The top row of \figref{band-selection-sfr} shows the $x$ vs SFR plot for the whole disk ($\Delta \mathrm{SFR} > 0$) and the bottom row shows when we only take the horizontal strip, as described in Section \ref{analysis-local}. Following the procedure, we now take the SFR spikes and locally fit \autoref{eq:gaussian} to the envelopes. The result of this is shown in \figref{left-right-arms-sfr}.

Notice that we find a double spike and $x=-3.5$ kpc. This also appears in the density distribution of the particles, indicating a fractured arm that is divided in two parts. We follow the procedure described in the previous chapter to make the fit of this arm. \par

Other thing to see, is that the vertical lines do not precisely coincide with the mean of each fit, this is because the arm-location procedure is not exact and its role is only to be a visual aid to locate the spikes. \par 

\section{Internal Energy}
\label{Us-results}
\begin{figure}[!htbp]
\captionsetup[subfigure]{labelformat=empty}
\begin{subfigure}{0.49\textwidth}
  \centering
  \includegraphics[width=0.99\linewidth]{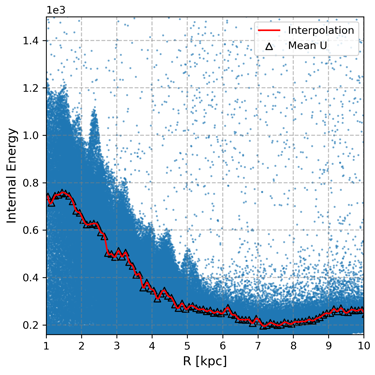}
  \caption{}
  \label{fig:interp-U}
\end{subfigure}%
\begin{subfigure}{.5\textwidth}
  \centering
  \includegraphics[width=0.99\linewidth]{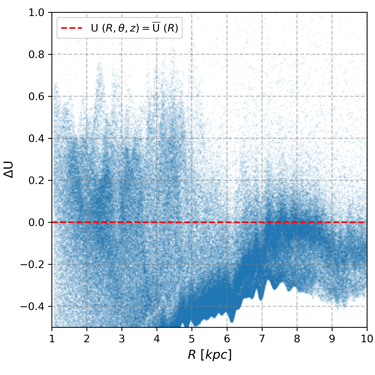}
  \caption{}
  \label{fig:U-threshold}
\end{subfigure}
\vspace{-1\baselineskip}
\caption[Radial profile and threshold determination for the Internal Energy.]{(a) Radial profile for the Internal Energy, including its radially-averaged mean values (hollow rectangles) and the interpolation. (b) Threshold determination, computed with \autoref{eq:deltax}.}
\label{fig:U-interp-threshold}
\end{figure}

We begin by showing the radial distribution of the internal energy and the corresponding internal energy threshold in \figref{interp-U} and \figref{U-threshold}. Again the behaviour is similar to that of they density and the SFR. The internal energy is expected to trace regions of the ISM with strong magnetic fields, which are present in star-forming regions. It also locates regions where heating and ionization processes such as wind and supernovae feedback take place.  \par

\begin{figure}[!htbp]
    \centering
    \captionsetup{aboveskip=3pt,belowskip=3pt}
    \includegraphics[height=5cm]{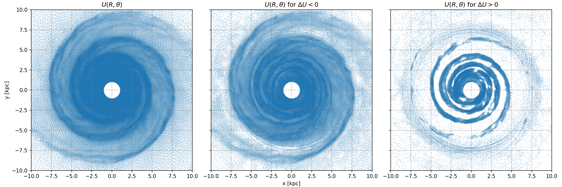}
    \caption[Comparison of the disk before and after the Internal Energy filtering.]{The first picture shows the gas particles that have non-zero Internal Energy, the second and third show the gas particles below and above threshold, respectively.}
    \vspace{-0.5em}
    \label{fig:U-compared}
\end{figure}

Next, we show in \figref{U-compared} the before and after of the filtering using the Internal Energy threshold determined by \figref{U-threshold}. One difference of the gas distribution when we apply this \enquote{Internal Energy filter} is that there appears to be a lot of energetic particles (at least with $\Delta U >0$) at large radii, as it can be seen in the last plot of \figref{U-compared}. We believe that this sort of \enquote{hot gaseous cloud} could originate due to gas escaping at high temperatures from the inner regions of the galaxy as a result of the galactic winds implemented in the feedback model. Another possible explanation is that this gas is a residue of the early stages of the galaxy evolution. Observational records of this type of diffuse thermal emission has been observed both in our own galaxy, \citet{sembach2003highly}, and in other star-forming disk galaxies (see \citealt{strickland2004high}). The problem of the origin of this type of phenomena is not the goal of our work, however it is worth mentioning in the present analysis as we would like to explain the particular results obtained in the case on the Internal Energy and how it differs from the other properties of the gas.\par

\begin{figure}[!htbp]
    \centering
    \captionsetup{aboveskip=3pt,belowskip=3pt}
    \includegraphics[height=13.2cm]{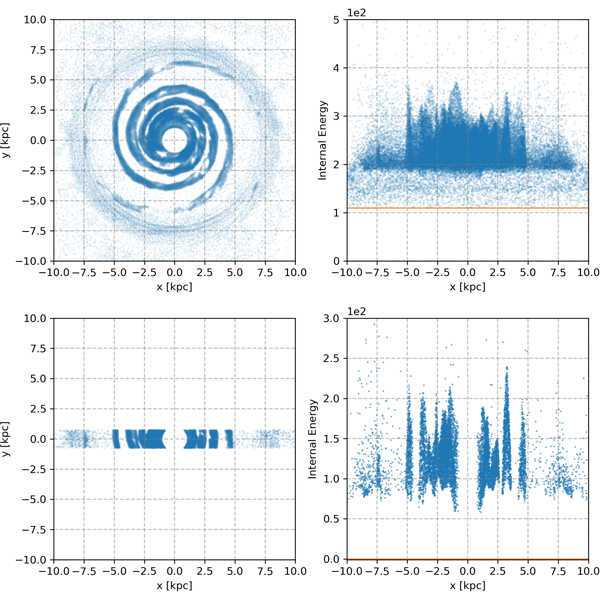}
    \caption[Band selection for the Internal Energy distribution.]{The first row shows the filtered gas disk ($\Delta \mathrm{U} > 0$), and the corresponding $\mathrm{U}$ vs x plot in the right. The second row, the horizontal band of the disk and its Internal Energy distribution on the right. The orange horizontal line marks the minimum of the Internal Energy. The data-set is shifted so $\mathrm{U}_{min} = 0$}
    \label{fig:band-selection-U}
\end{figure}

\begin{figure}[!htbp]
    \centering
    \captionsetup{aboveskip=3pt,belowskip=3pt}
    \includegraphics[height=6.75cm]{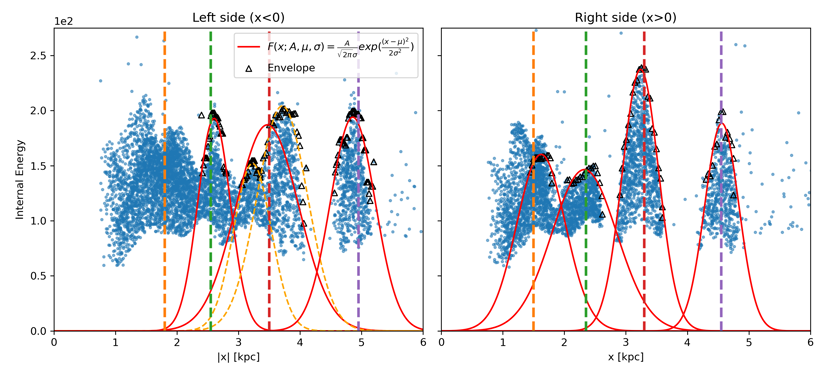}
    \caption[Final fitted arm profiles for the Internal Energy.]{Plot of all the envelopes that wrap the Internal Energy spikes in \figref{band-selection-U} and the corresponding fits of \autoref{eq:gaussian}. }
    \label{fig:left-right-arms-U}
\end{figure}

Moving on with the procedure, we now plot the horizontal distribution of the Internal Energy in \figref{band-selection-U} and the final results of the envelope extraction procedure and the corresponding Gaussian fits in \figref{left-right-arms-U}. \par

Notice that there is a \enquote{halo} of energetic particles at around $R=7.5$ kpc, beyond this radius the spiral structure once present (see left plot in \figref{U-compared}) disappears. In general, we notice that there are scattered particles across all the disk, with not a particular distribution, except from what appears to be a halo of particles at $R=7.5$ kpc. \par

In \figref{left-right-arms-U} we show the gaussian fits that we managed to perform on the overdensities. Notice that it is not always possible to make the fit, like in the first overdensity in the left side, indicated with the orange dashed line. We again see an arm that is divided in two, like we already discussed in Section \ref{analysis-local} and like we also saw in the case of the SFR. \par

The absence of a fit where the first arm is supposed to be is again present in the Internal Energy distribution, like it was in the SFR distribution. In general, finding a clean spike to perform a fit is more difficult in the inner parts of the disk, since the frontiers of the arms in there are more diffuse and harder to locate. \par

\section{Free Electrons}
\label{Nes-results}
We now analyse the case of the Free Electrons per number proton, this is the mean free-electron number per proton (hydrogen nucleon), averaged over the mass of the gas particle, as stated in the GIZMO documentation. When we tried to replicate the same procedure used with the other properties, we encountered that there were two regimes when it comes to the behaviour of this property, so it was not straightforward to do the radial binning and interpolation of the mean field.   
\begin{figure}[!htbp]
    \centering
    \captionsetup{aboveskip=1pt,belowskip=0pt} \includegraphics[height=6cm]{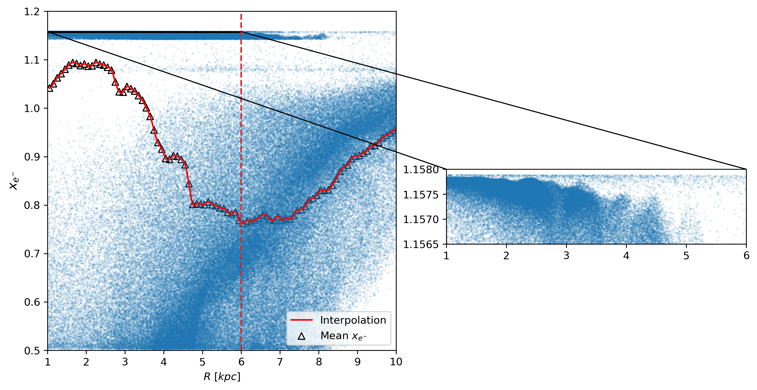}
    \caption[Radial profile for the Free Electrons.]{Radial distribution of the free electrons. The Figure shows an attempt of interpolation computed over the whole data. The sub-axes zoom in the inner regions of the disk, where the proportion of free electrons is $x_{e^{-}} \approx 1.15$. The red dotted line separates what we identify as two different regimes of ionization. For $R>6$ kpc, the behaviour of the Free Electrons is noticeably different.}
    \label{fig:Ne-interp-whole}
\end{figure}

\begin{figure}[!htbp]
    \centering
    \captionsetup{aboveskip=3pt,belowskip=3pt}
    \includegraphics[height=5.9cm]{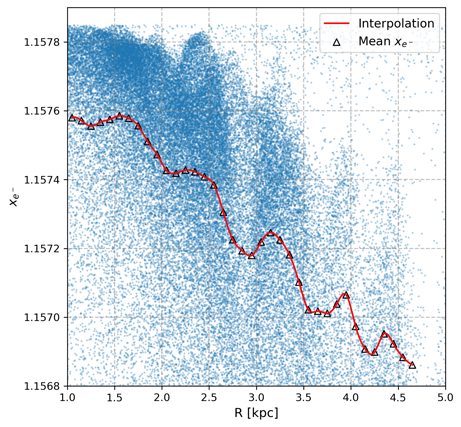}
    \caption[Radial profile for the Free Electrons, highly ionized region.]{Radial distribution of the highly-ionized gas particles (high proportion of free electrons). The Figure shows the interpolation once it is restricted to only this part of the data.}
    \label{fig:Ne-interp-lim}
\end{figure}

\begin{figure}[H]
\captionsetup[subfigure]{labelformat=empty}
\centering
 \vspace*{5pt}%
 \hspace*{\fill}% 
  \begin{subfigure}{0.75\textwidth}    
    \centering
    \includegraphics[width=0.75\linewidth]{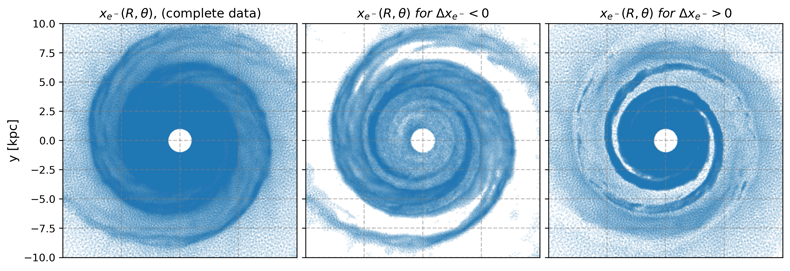}%
    \captionsetup{skip=-20pt}%
    \caption{}
    \label{fig:ionization-regimes-whole}
  \end{subfigure}%          
  \hspace*{\fill}%          

  \vspace*{8pt}%  

  \hspace*{\fill}%  
   \begin{subfigure}{0.75\textwidth}        
    \centering
    \includegraphics[width=0.75\linewidth]{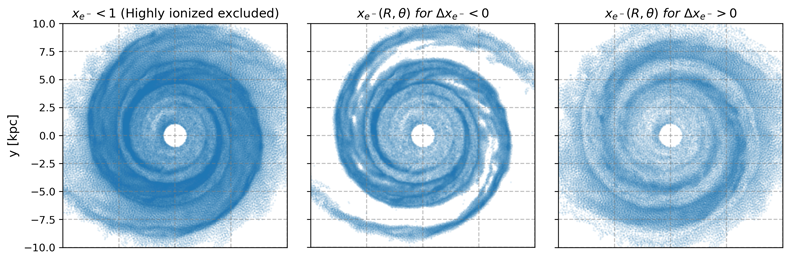}%
    \captionsetup{skip=-20pt}%
    \caption{}
    \label{fig:ionization-regimes-not}
  \end{subfigure}%            
  \hspace*{\fill}%          

  \vspace*{8pt}%

  \hspace*{\fill}%  
  \begin{subfigure}{0.75\textwidth}     
    \centering
    \includegraphics[width=0.75\linewidth]{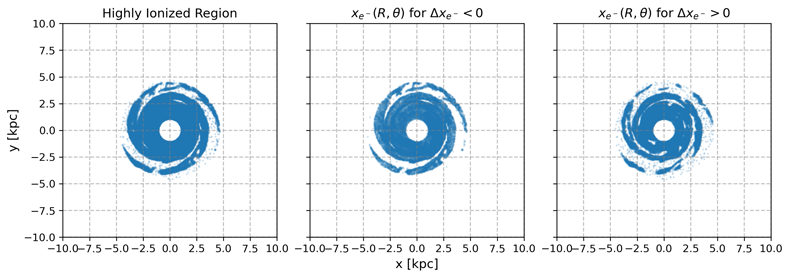}%
    \captionsetup{skip=-15pt}%
    \caption{}
    \label{fig:ionization-regimes-ionized}
  \end{subfigure}%              
  \hspace*{\fill}%
  
  \caption[Comparison between different ionization regimes.]{Comparison between all the gas particles with defined $x_{e^{-}}$, the particles with $x_{e^{-}}<1$ and the ones with $1.1565<x_{e^{-}}<1.1579$, which corresponds to the \enquote{highly ionized regime} showed in \figref{Ne-interp-whole} and \figref{Ne-interp-lim}. Notice that when we exclude this region (the highly ionized), the inner regions of the disk look considerably less populated, and the shape of the spiral structure is now almost empty (middle panel), whereas if we only consider this ionized region (third panel) we only get particles inside $R<5$ kpc, that trace the spiral arms in the inner regions. }
  \vspace{-0.5em}
  \label{fig:ionization-regimes-compared}
\end{figure}

Inspecting the distribution of the $R \  \mathrm{vs} \ x_{{e^-}}$ plot, we found that there is region from $1 \mathrm{kpc} <\mathrm{x}<5\mathrm{kpc} $ and $1.1565 < x_{e^{-}} < 1.1579$ that corresponds to highly-ionized gas particles, as \figref{Ne-interp-whole} shows. We now proceed to restrict our discussion to only this portion of the data, which are of interest in our study. When we only take data points for which $1.1565 < x_{e^{-}} < 1.1579$, we get the interpolation shown in \figref{Ne-interp-lim}. \par

In \figref{ionization-regimes-compared} we show the gas distributions and the filtering process with three different data-sets. The first one is the whole data, without any restriction in the free electron proportions of the gas particles; in the second one, we exclude the more ionized particles and in the third one we restrict the data to only the ionized particles. \par

\begin{figure}[H]
    \centering
    \captionsetup{aboveskip=3pt,belowskip=3pt}
    \includegraphics[height=11cm]{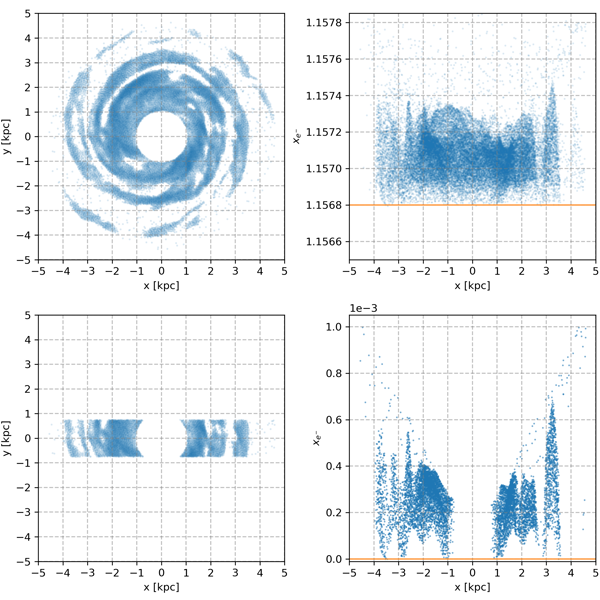}
    \caption[Band selection for the Free Electrons distribution.]{The first row shows the filtered gas disk ($\Delta x_{e^{-}} > 0$), and the corresponding $x_{e^{-}}$ vs x plot in the right. The second row, shows the horizontal band of the disk and its Free Electrons distribution on the right. The orange horizontal line marks the minimum of $x_{e^{-}}$. The data-set is shifted so $x_{e^{-},{min}} = 0$}
    \label{fig:band-selection-Ne}
\end{figure}

We will restrict out analysis to the region corresponding to \figref{ionization-regimes-ionized}, since this represents the parts of the gas disk that we are interested in, i.e, the overdense, hihgly ionized regions where star formation takes place. In \figref{band-selection-Ne} and \figref{left-right-arms-Ne} we show the final results of the procedure.

We only recover three envelopes from each side of the disk. This is a consequence of picking only the more ionized part of the disk and hence having only the interior part of the spiral pattern. \par

\begin{figure}[H]
    \centering
    \captionsetup{aboveskip=3pt,belowskip=3pt}
    \includegraphics[height=8.5cm]{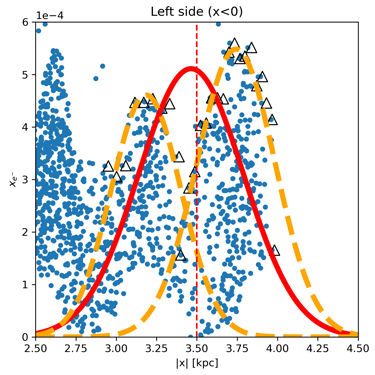}
    \caption[Fractured arm in the Free Electrons distribution.]{Fractured arm in the Free Electrons distribution.}
    \label{fig:divided-arm-gaussians-Ne}
\end{figure}

\begin{figure}[H]
    \centering
    \captionsetup{aboveskip=0pt,belowskip=3pt}
    \includegraphics[height=6.75cm]{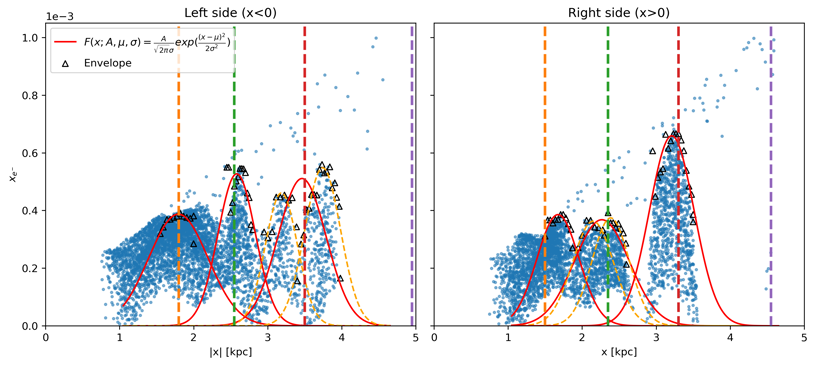}
    \caption[Final fitted arm profiles for the Free Electrons.]{Plot of all the envelopes that wrap the Free Electrons spikes in \figref{band-selection-Ne} and the corresponding fits of \autoref{eq:gaussian}. }
    \label{fig:left-right-arms-Ne}
\end{figure}

In \figref{left-right-arms-Ne}, we pick the fractured arm that has also been present in the previous properties. Notice that the division is, in this case, much more noticeable at $x=-3.5$ kpc. \figref{divided-arm-gaussians-Ne} shows with more detail how this arm looks like and the gaussian fits made.

\section{Neutral Hydrogen}
\label{H-results}
Finally, we will now review the results for the Neutral Hydrogen fraction, defined as the percentage of a gas particles that is made out of $\mathrm{HI}$. All of the previous properties of the gas presented roughly the same behaviour: the numerical values increase in the more dense regions: gas particles become more dense, their SFR increases, they have more Internal Energy and their Free Electrons per number proton increases as well. As they increase considerably inside the arms, they decrease in the inter-arm regions, as \figref{band-selection}, \figref{band-selection-sfr}, \figref{band-selection-U} and \figref{band-selection-Ne} show. \par

The case of the Neutral Hydrogen fraction is somewhat different than the other properties. In the most dense regions, the neutral hydrogen fraction actually decreases, so instead of having spikes pointing upwards, we have them pointing downwards. This makes sense since in this parts of the disk, the feedback mechanisms and star formation processes are more active, and the interstellar medium becomes hostile for the presence of non-ionized gas, i.e, for Neutral Hydrogen. \par 

\begin{figure}[H]
    \centering
    \captionsetup{aboveskip=1pt,belowskip=0pt}
    \includegraphics[height=7.5cm]{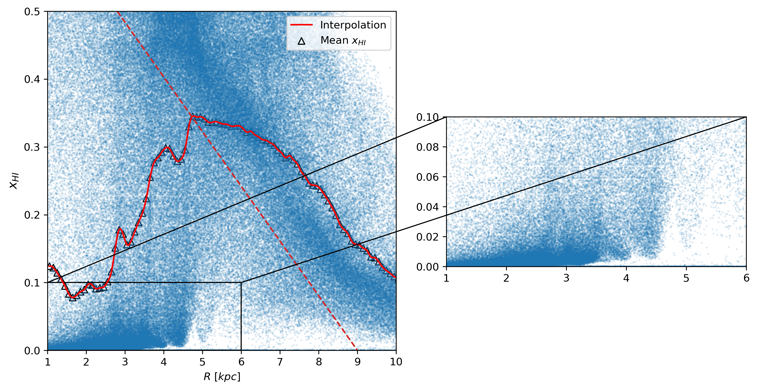}
    \caption[Radial profile for the Neutral Hydrogen fraction.]{Radial distribution of the Neutral Hydrogen fraction of the gas particles. The sub-axes zoom in the inner regions of the disk, where the fraction of Neutral Hydrogen is around $x_{HI} \lesssim 10 \% $. The red oblique dotted line separates what we identify as two different regimes of of the Neutral Hydrogen distribution.}
    \label{fig:Hn-interp-whole}
\end{figure}

In \figref{Hn-interp-whole} we show the radial distribution of $x_{HI}$ and a zoom in the inner regions where the fraction of $x_{HI}$ is less than $10 \%$. This portrays what we recently mentioned, the Neutral Hydrogen fraction drops to very low values in the inner part of the disk.  \par 

We also see that there are two separate behaviours for the distribution, we indicate this in \figref{Hn-interp-whole} with a diagonal red dashed line that separates what we identify as two different regimes for the Neutral Hydrogen distribution. The one we are most interested in is the one below that line.  \par

\begin{figure}[!htbp]
    \centering
    \captionsetup{aboveskip=1pt,belowskip=0pt}
    \includegraphics[height=7.5cm]{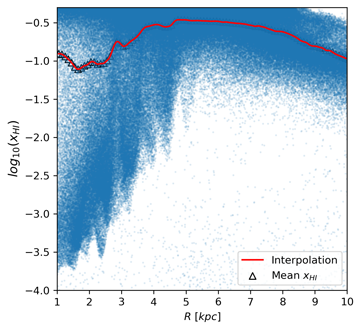}
    \caption[Radial profile for the Neutral Hydrogen fraction.]{Radial distribution of the Neutral Hydrogen fraction of the gas particles in logarithmic scale.}
    \label{fig:Hn-interp-log}
\end{figure}

We notice that the overdensities show up but are very hard to see for $R<4$ kpc. For that reason, we decide to plot the same distribution in logarithmic scale. The result is shown in the \figref{Hn-interp-log}. We notice that in this case, the overdensities are more easy to identify. We will now use the log of the Neutral Hydrogen fraction to perform all the analysis of this property. \par

\begin{figure}[H]
    \centering
    \captionsetup{aboveskip=3pt,belowskip=3pt}
    \includegraphics[height=5.2cm]{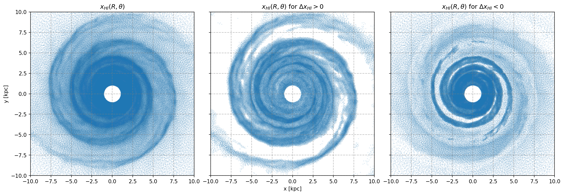}
    \caption[Comparison of the disk before and after the Neutral Hydrogen filtering.]{The first picture shows all the gas particles with defined $0<x_{HI}<1$, the second and third show the gas particles below and above threshold, respectively.}
    \label{fig:Hn-compared}
\end{figure}

Previously, we defined the \enquote{background} with the criteria that $\Delta x <0$. The overdensity regions were determined by $\Delta x >0$, as it is shown in \figref{density-compared}, \figref{sfr-compared}, \figref{U-compared} and \figref{ionization-regimes-compared}, where we compare how the disk looks like when this \enquote{filter} is applied. In the case of the Neutral hydrogen the criteria is the other way around, the background is set by all the gas particles with $\Delta x_{HI} >0$ and the overdensity regions where $\Delta x_{HI} <0$, as \figref{Hn-compared} shows. The reason behind this, is that we want to investigate the more ionized region, where the SFR is more concentrated at. For that reason we are picking gas particles with less Neutral Hydrogen fractions, which is equivalent to picking the particles that are more ionized. \par

\begin{figure}[H]
\captionsetup[subfigure]{labelformat=empty}
\begin{subfigure}{0.49\textwidth}
  \centering
  \includegraphics[height=7cm]{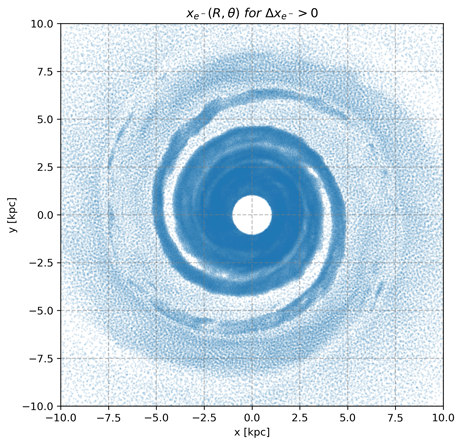}
  \caption{}
  \label{fig:dNe>0}
\end{subfigure}%
\begin{subfigure}{.5\textwidth}
  \centering
  \includegraphics[height=7cm]{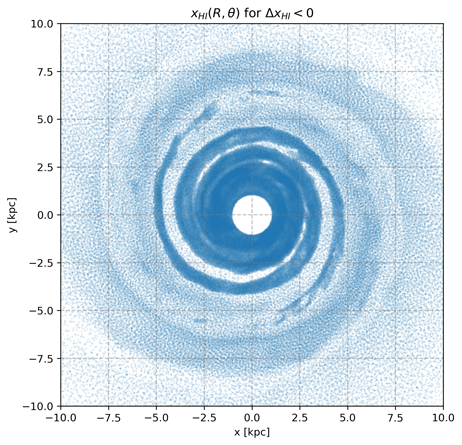}
  \caption{}
  \label{fig:dH<0}
\end{subfigure}
\vspace{-1\baselineskip}
\caption[High ionization region through Free Electrons and Neutral Hydrogen.]{Left: high ionization regions in the particle distribution of the Free Electrons, by restricting the data-set to particles with $\Delta x_{e^{-}}>0$ only. Righ: high ionization regions in the particle distributions of the Neutral Hydrogen, by restricting the data-set to particles with $\Delta x_{HI}<0$) only.}
\label{fig:dNe-dH}
\end{figure}

To further illustrate this point, in \figref{dNe-dH} we plot side by side the particle distributions for the Free Electrons with $\Delta x_{e^{-}}>0$ (this is the same as the top right plot of \figref{ionization-regimes-compared}), and the particle distribution for the Neutral Hydrogen fraction with $\Delta x_{HI}<0$. Notice the similarity in the regions that are picked up by the two different thresholds ($\Delta x_{e^{-}}>0$ and $\Delta x_{HI}<0$).  \par 

We only recover three envelopes from each side of the disk. This is a consequence of picking only the more ionized part of the disk and hence having only the interior part of the spiral pattern. \par

In \figref{left-right-arms-Ne}, we pick the fractured arm that has also been present in the previous properties. Notice that the division is, in this case, much more noticeable at $x=-3.5$ kpc. \figref{divided-arm-gaussians-Ne} shows with more detail how this arm looks like and the gaussian fits made.

\begin{figure}[H]
    \centering
    \captionsetup{aboveskip=3pt,belowskip=3pt}
    \includegraphics[height=11cm]{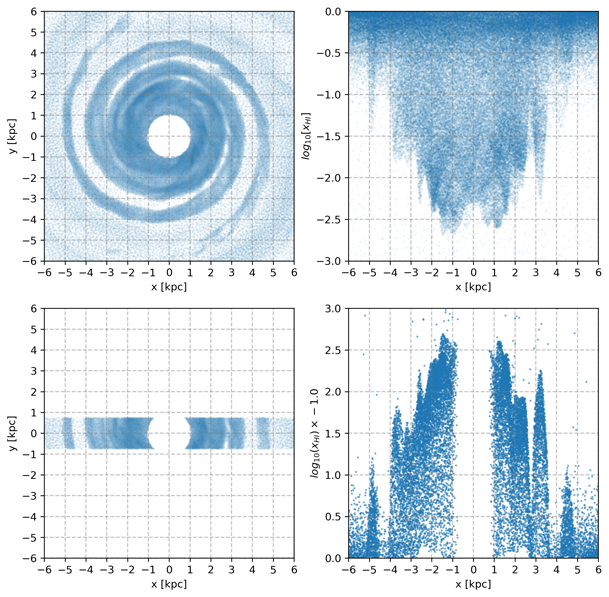}
    \caption[Band selection for the Neutral Hydrogen distribution.]{The first row shows the filtered gas disk ($\Delta \mathrm{x_{HI}} < 0$), and the corresponding $\mathrm{x_{HI}}$ vs x plot in the right. The second row, the horizontal band of the disk and its Neutral Hydrogen distribution on the right. Notice that the data is turned upside-down from the first row to the second, this manipulation is necessary in order to perform the fit of \eqref{eq:gaussian}.}
    \label{fig:band-selection-Hn-log}
\end{figure}
\vspace{-1cm}
\begin{figure}[H]
    \centering
    \captionsetup{aboveskip=3pt,belowskip=3pt}
    \includegraphics[height=6cm]{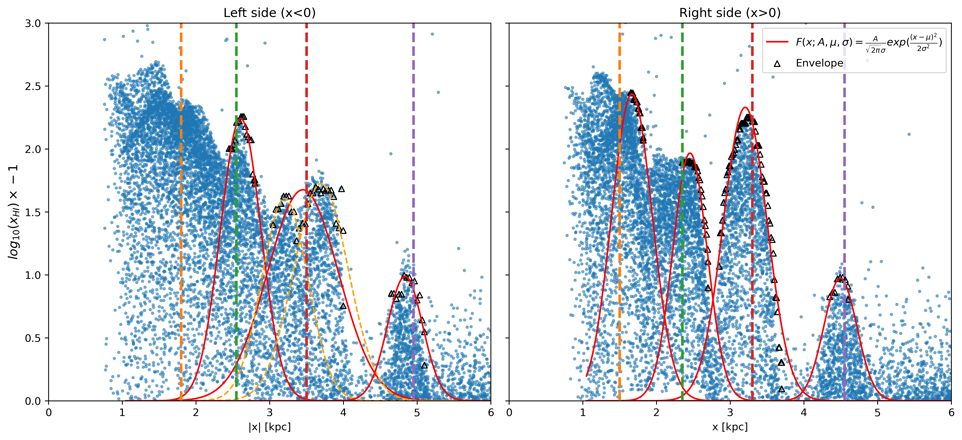}
    \caption[Final fitted arm profiles for the Neutral Hydrogen.]{Plot of all the envelopes that wrap the Neutral Hydrogen spikes in \figref{band-selection-Hn-log} and the corresponding fits of \autoref{eq:gaussian}.}
    \label{fig:left-right-arms-Hn}
\end{figure}

We now can fit the gaussian model over the spikes in the $x$ vs $\log(x_{HI}) \times -1$ distribution, as shown in the bottom right plot of \figref{band-selection-Hn-log}. The results are in \figref{left-right-arms-Hn}, where we were able to fit seven gaussians. Notice that the fractured arm at $x=-3.5$ kpc is clearly present, as it has been in all the previous properties. \par

\section{Getting more data through rotations}
Up to this points, we have reviewed the procedure for each one of the properties. However, we only have between 7 and 12 data points (widths) for each one of them, which is a very low number to conclude confidently. We need to increase the number of data points. \par

\begin{figure}[H]
    \centering
    \captionsetup{aboveskip=3pt,belowskip=3pt}
    \includegraphics[height=8cm]{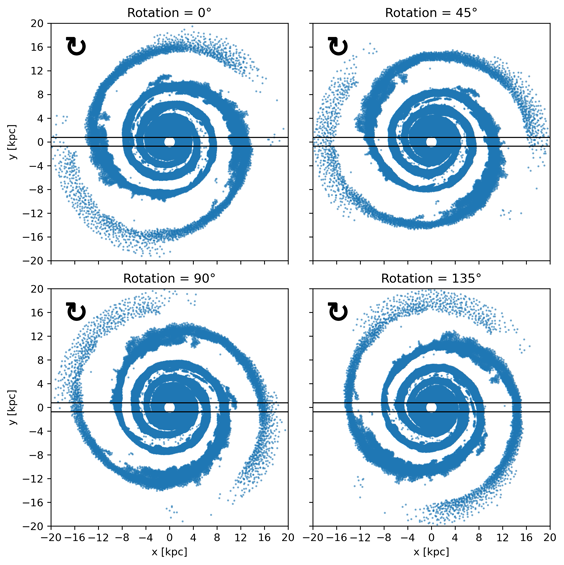}
    \caption[Gas disk in different orientations]{Gas disk in different orientations. In this case, we show as an example the density particle distribution obtained in \figref{band-selection}. We indicate the sense of rotation with a curved clock-wise arrow. We also plot the horizontal \enquote{strip} of the disk in which we evaluate the width. }
    \label{fig:rotations-density}
\end{figure}

Fortunately, this whole procedure can be repeated for any orientation of the disk. We can rotate the particle distribution and evaluate the arm width across another horizontal strip. This way we can increase the data-set considerably.\par

In \figref{rotations-density} we show the four rotations that we have made for this work. The first one, $\alpha=0^{\circ}$, is what have been using so far to describe the procedure for each one of the properties. The another three rotations are evenly spaced by $45^{\circ}$: $\alpha= [ 45^{\circ},90^{\circ},135^{\circ}]$. This spacing is to avoid evaluating the same part of the arm twice. \par

We have now 4 times the amount of the data that we would normally have, had we only perform the analysis on a single disk orientation. We will now show the final results. \par

\newpage
\section{Results \& Discussion}
We now present a summary of all the results that we have gotten for the four disk orientations and all the gas properties. \par

\begin{figure}[H]
    \centering
    \captionsetup{aboveskip=3pt,belowskip=-3pt}
    \includegraphics[height=17cm]{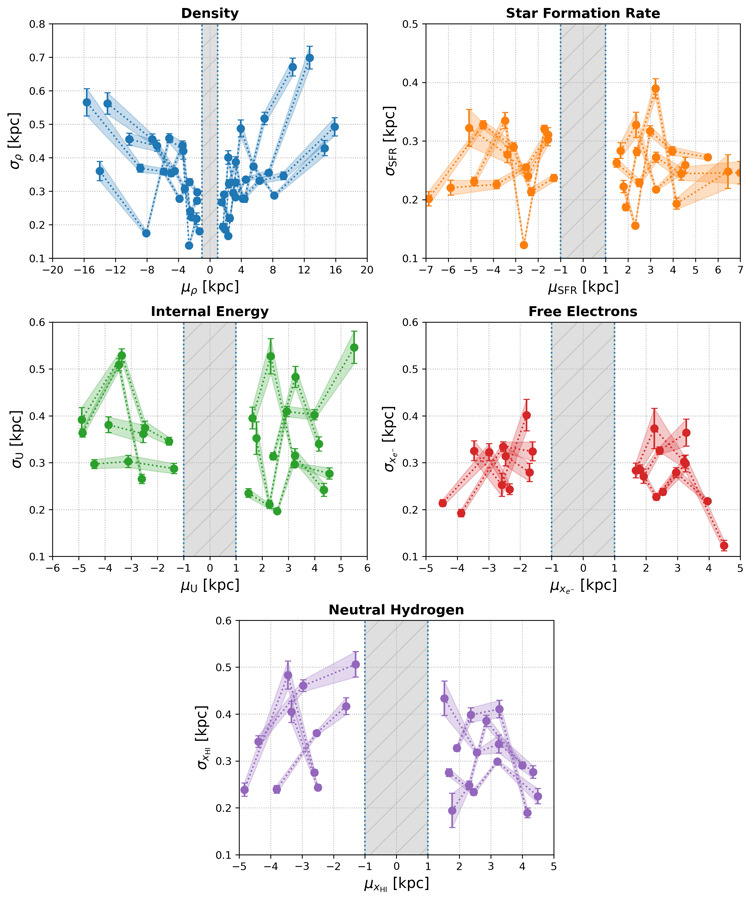}
    \caption[Summary of the $\sigma$ vs $\mu$ plot for all properties.]{Summary of the $\sigma$ vs $\mu$ plot for all properties. We indicate the excluded zone that corresponds to the bulge. Error bars are the standard error of the fitted parameters. Shaded areas show possible values of the parameters, inside the uncertainty determined by the errors.}
    \label{fig:results-grid}
\end{figure}

\begin{figure}[H]
    \centering
    \captionsetup{aboveskip=3pt,belowskip=-3pt}
    \includegraphics[height=18.2cm]{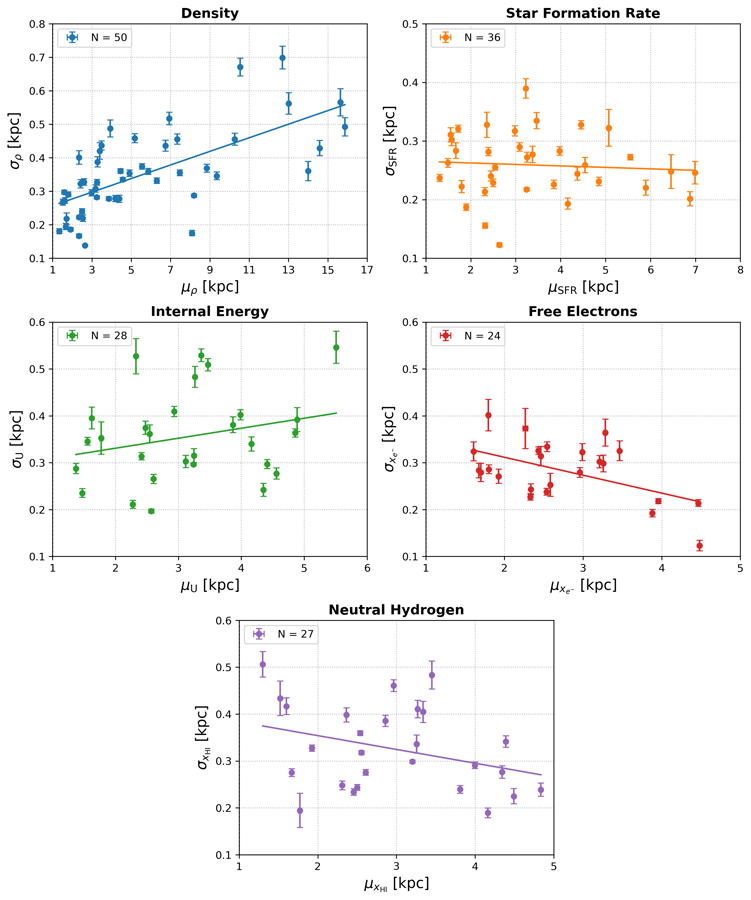}
    \caption[Summary of the $\sigma$ vs $\mu$ plot for all properties (radially).]{Summary of the $\sigma$ vs $\mu$ plot for all properties, but we take the absolute value of the $\mu_{x}$ values in the left side of the disk (negative $\mu_{x}$).}
    \label{fig:results-grid-radial}
\end{figure}

In \figref{results-grid} we plot the parameters $\mu$ and $\sigma$ that the fit process returns. The distinction between the arms in the left and right sides is made, and the region $|R|<1$ kpc is coloured in gray to indicate the position of the bulge. We also indicate the errors, which are the standard errors associated to the fitted function in \autoref{eq:gaussian}. We used the open source library \enquote{lmfit} to perform the fit of \autoref{eq:gaussian} to the envelopes. To have a better idea of the behaviour of the widths as a functions of galactic radius, we turn the negative values of $\mu_{x}$ into positive, the results are displayed in \figref{results-grid-radial}.

We performed a linear fit to that data in each case, $\sigma_{x} = mR + b$, the obtained functions in each case are:
\begin{equation}
    \begin{aligned}
    \label{eq:linear-fits-band}
     \sigma_{\rho} &= (0.020 \pm 0.003)R + (0.236 \pm 0.022)  \\
     \sigma_{\mathrm{SFR}} &= (-0.003 \pm 0.006)R + (0.268 \pm 0.022)  \\
     \sigma_{\mathrm{U}} &= (0.021 \pm 0.016)R + (0.288 \pm 0.054) \\
     \sigma_{x_{e^{-}}} &= (-0.038 \pm 0.013)R + (0.388 \pm 0.038)  \\
     \sigma_{x_{\mathrm{HI}}} &= (-0.029 \pm 0.017)R + (0.413 \pm 0.053) 
    \end{aligned}
\end{equation}

We also show all the values of the parameters in \autoref{tab:parameters-band}. 

\begin{table}[H]
\caption[Parameters of the linear fit for the Horizontal Band method.]{Parameters of the linear fit for the Horizontal Band method.}
\label{tab:parameters-band}
\centering
\begin{tabular}{|c|c|c|c|c|}
\hline
                     & \bm{$m$} & \bm{$\delta m$} & \bm{$b$} & \bm{$\delta b$} \\ \hline
\bm{$\rho$}      & 0.020        & 0.003               & 0.236        & 0.022               \\ \hline
\textbf{SFR}         & -0.003       & 0.006               & 0.268        & 0.022               \\ \hline
\textbf{U}           & 0.021        & 0.016               & 0.288        & 0.054               \\ \hline
\bm{$x_{e^{-}}$} & -0.038       & 0.013               & 0.388        & 0.038               \\ \hline
\bm{$x_{HI}$}    & -0.029       & 0.017               & 0.413        & 0.053               \\ \hline
\end{tabular}
\end{table}

From \figref{results-grid} and \figref{results-grid-radial} we can extract some information about the widths in each property:
\begin{itemize}
    \item \textbf{Density ($\rho$)}: the density is the property of the gas for which more data we could collect, using this method ($N=50$). It is perhaps the only one that presents a clear widening in the arm's horizontal profile as the galactic radius increments. This behaviour, as it should be expected, can be seen in both sides of the disk. This widening is, however, not uniform, there are a few parts where we encounter thinner arms, however the overall trend is clear: the gas particles in the spiral arms are more spread in the outer parts of the galactic disk. This could be explained by the fact the gravitational potential is less steep as $R$ increases, allowing for the overdensities to spread out more in the outer regions. In the inner parts of the disk, the gravitational potential is stronger, which compresses the particles in the spiral arms. \par
    
    \item \textbf{Star Formation Rate ($\mathrm{SFR}$)}: we can see that when studying the particles with non-zero SFR, the spiral arms don't go as far, in fact they only go up to $R=7$ kpc. The scatter in the widths distribution is clearly greater than that of the density. We are uncertain, from this results, if the arms widen systematically with galactic radius. We do note that the values of the width are not as big as the ones in the density. This becomes clear when we see the $y$ axis range and note that all values (except for one) are between $0.1 < \sigma_{SFR} < 0.35$. This is an indicator that the SFR only occurs in the innermost parts of the spiral arms. With this results, we are motivated to test if there is an actual systematic decrease in arm width that is not clear due to the mixing of results from the two different arms. In the next section we will investigate further. \par
    
    \item \textbf{Internal Energy ($\mathrm{U}$)}: the Internal Energy shows the more scattering between all the properties.  We observe that for some disk orientations, the width of the arms remains approximately constant. We are unable to determine if there exists any particular behaviour, up or down, in this case.  \par
    
    \item \textbf{Free Electrons ($x_{e^{-}}$)}: we got the least amount of data points in this case, $N=24$, however, from the available data, we can see that the widths distribution favours an anti-correlation between the arm width and the galactic radius. This result makes sense if we consider that Free Electrons are located in the more ionized regions of the arms, and that such ionization depends on whether or not feedback mechanisms are taking place. As we move out of the disk, the new-born stars and supernovae explosions become scarce, that could explain the observed decrease in the width of the ionized regions in the outer parts of the disk. \par
    
    \item \textbf{Neutral Hydrogen ($x_{\mathrm{HI}}$)}: last but not least, we have the case of the Neutral Hydrogen, which, presents a similar behaviour to the Free Electrons distributions, but with more scatter. This similarity between the two properties can be explained with they way the threshold was chosen, as explained in \figref{dNe-dH}. Both particle distributions were obtained for the more ionized regions only. \par 

\end{itemize}

Overall, we could draw some conclusions for the behaviour of the widths, specially for the density of the gas (which is the more evident). For the Free Electrons and the Neutral Hydrogen we encountered some similarities, which in general favour an anti-correlation between the widths and the galactic radius. For the star formation rate, we got the hint of a possible anti-correlation, but we are unable to conclude that due to the high scatter of the data. For the Internal Energy, we found the most erratic behaviour from the 5 properties, and also the more dispersion. Any possible trend or tendency is impossible to determine, in this particular case. \par

We believe that the scatter in the widths distributions might be caused by the method itself. Notice that when we take the horizontal strip, we evaluate the widths of the two arms every other point, i.e, we are mixing the widths from the two arms at different radii. Such mixing can introduce noise and scatter when trying to visualize the results. For this reason, we will now take a different approach. We will evaluate the widths of each arm separately, an see if the  results are any different. \par

\subsection{A different approach: Arm Tracing}
\label{A-different-approach}

We observed a great dispersion in the results obtained, for almost all properties. Perhaps the only exception was the widths of the arms when we looked at the density. This scatter could be caused, as mentioned in the previous section, by the fact that we are mixing the widths of the two arms at different galactic radii.\par

When we take the horizontal strip (see \figref{band-selection}, \figref{band-selection-sfr}, \figref{band-selection-U}, \figref{band-selection-Ne}, \figref{band-selection-Hn-log}), the two arms pass through the middle of it, interspersing each other as we move towards the edge of the band. This results in a data-set that contains the widths of the two arms mixed at different radii, and might be causing the scattering observed in \figref{results-grid} and \figref{results-grid-radial}. \par

\begin{figure}[t!]
    \centering
    \captionsetup{aboveskip=3pt,belowskip=-3pt}
    \includegraphics[height=18.2cm]{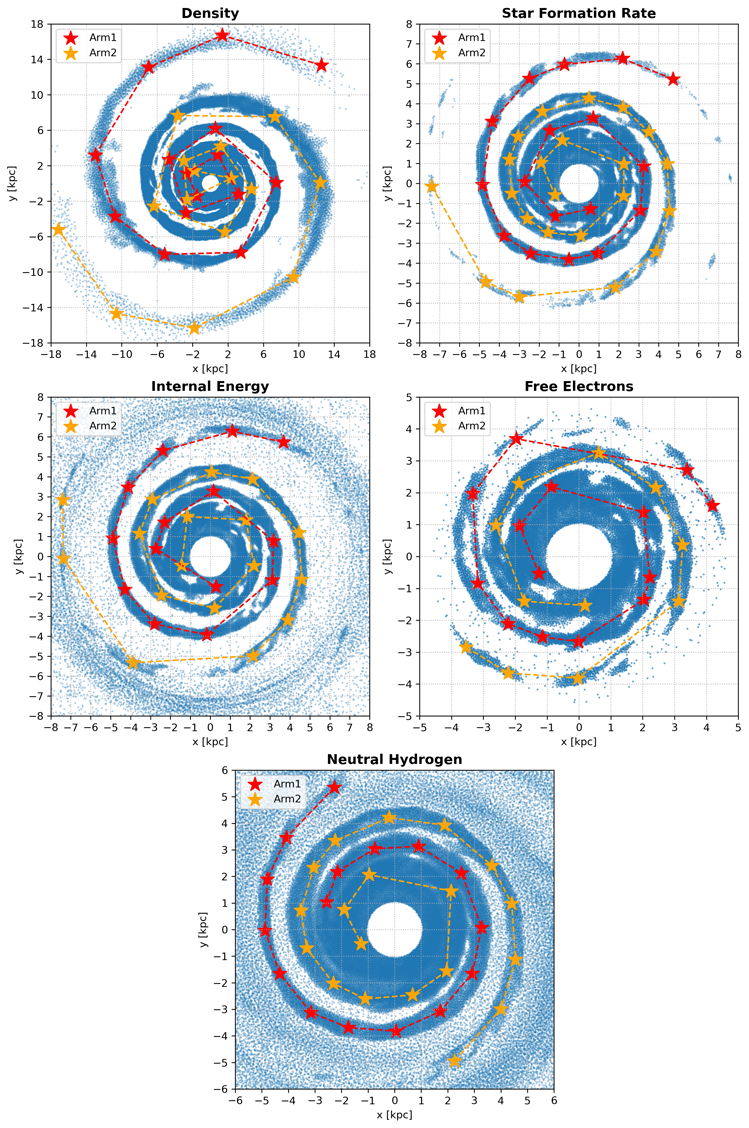}
    \caption[Tracing points to evaluate the arm width in each property.]{Tracing points used for each property. Red stars trace what we labeled as Arm1, orange stars trace Arm2.}
    \label{fig:arm-tracing-disks-grid}
\end{figure}

Now we will take a different approach. We will trace each of the arms individually and evaluate the arm widths at different radii, but keeping the data for each arm separate. In \figref{arm-tracing-disks-grid} we show the points we picked for each arm in every property. These data corresponds to the threshold-filtered and final data-sets for every property (see \figref{density-compared}, \figref{sfr-compared}, \figref{U-compared}, \figref{Hn-compared}, \figref{ionization-regimes-compared}).  \par

We considered that the easiest way to measure the width at every point, was to rotate the disk an angle of $\alpha = \tan^{-1}(y_{i}/x{_{i}})$, where $[x_{i},y_{i}]$ is the coordinate of each point that appears over the arms in \figref{arm-tracing-disks-grid}. Doing such rotation, places the point in the line $y=0$, where we can easily evaluate its horizontal profile, like we did in the band method. \par

We tried to get at least 10 or more data points in each spiral arm. Regions where the arms are fragmented or divided were ignored. The reason to ignore those parts, is that we want to make calculations more efficient and free of noise due to insufficient data. We tried, however, to distribute the points in a way that guarantees data across all galactic radii (at least where the arm is still visible). \par 

\begin{figure}[H]
    \centering
    \captionsetup{aboveskip=3pt,belowskip=-3pt}
    \includegraphics[height=18.2cm]{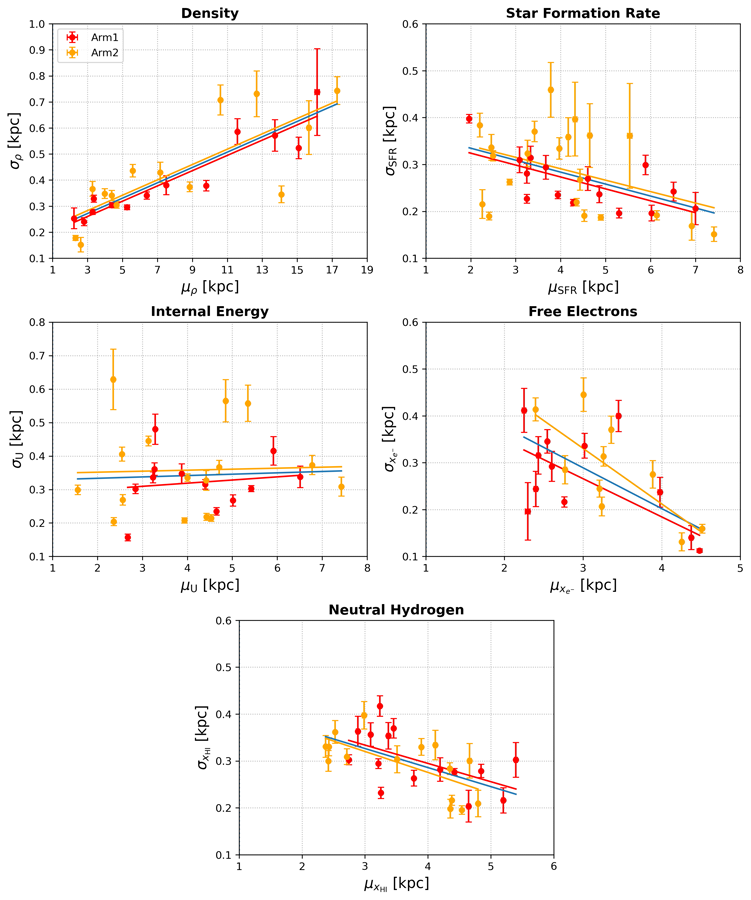}
    \caption[Final results for the Arm Tracing.]{Final results for the widths of the Arm Tracing procedure.}
    \label{fig:arm-tracing-results-grid}
\end{figure}

The results are showed in \figref{arm-tracing-results-grid}. As mentioned, we indicate which points belong to which spiral arm. Along with the scattered points, we plot, like we did in the previous method (which from now on we will call the Horizontal Band Method), linear best-fits to the data. In this case, there are three fits: the red line that fits data of Arm1, the orange line that fits data of Arm2, and a blue line that fits the complete data-set. The resulting linear equations (the blue fit for the complete data-set) are:

\begin{equation}
    \begin{aligned}
    \label{eq:linear-fits-trace}
     \sigma_{\rho} &= (0.029 \pm 0.004)R + (0.183 \pm 0.035)  \\
     \sigma_{\mathrm{SFR}} &= (-0.025 \pm 0.008)R + (0.386 \pm 0.036)  \\
     \sigma_{\mathrm{U}} &= (0.004 \pm 0.015)R + (0.325 \pm 0.067) \\
     \sigma_{x_{e^{-}}} &= (-0.087 \pm 0.023)R + (0.550 \pm 0.078)  \\
     \sigma_{x_{\mathrm{HI}}} &= (-0.041 \pm 0.010)R + (0.448 \pm 0.039) 
    \end{aligned}
    \vspace{0.3cm}
\end{equation}

In \autoref{tab:parameters-arm1-arm2}, we gather the parameters of the linear fit for Arm1 and Arm2. We observe that the linear fits for the Arm1 and Arm2 are very similar in every property, this can be seen directly on \figref{arm-tracing-results-grid} and in the numerical values of \autoref{tab:parameters-arm1-arm2}. \par

The slopes in the case of the density are exactly the same up to the third decimal number. For the SFR, up to the second decimal number. On the density, SFR and Neutral Hydrogen, the general behaviour of the width in both spiral arms is almost identical. \par

The slopes vary for the Free Electrons (ionization regions), and we see that Arm2's width has a steeper decrease than the Arm1. It shows a systematic decrease in the width as a function of $R$. The Neutral Hydrogen shows a similar behaviour than that of the Free Electrons but with less scatter. Both arms, in this case, get thinner at approximately the same rate. \par

For the Internal Energy distribution, we still have a high scatter in the distribution, the widths appear to oscillate around a constant value of $\sigma_{U} \approx 0.35$ kpc. \par 

\begin{table}[H]
\centering
\caption[Parameters of the linear for the Arm1 and Arm2.]{Parameters obtained from the linear fit for each property and for each arm with the Arm Tracing method.}
\label{tab:parameters-arm1-arm2}
\centering
\begin{tabular}{|c|c|c|c|c|c|c|c|c|}
\hline
\multirow{2}{*}{}    & \multicolumn{4}{c|}{\textbf{Arm1}}                                      & \multicolumn{4}{c|}{\textbf{Arm2}}                                      \\ \cline{2-9} 
                     & \bm{$m$} & \bm{$\delta m$} & \bm{$b$} & \bm{$\delta b$} & \bm{$m$} & \bm{$\delta m$} & \bm{$b$} & \bm{$\delta b$} \\ \hline
\bm{$\rho$}      & 0.029        & 0.003               & 0.172        & 0.030               & 0.029        & 0.007               & 0.194        & 0.063               \\ \hline
\textbf{SFR}         & -0.025       & 0.008               & 0.375        & 0.037               & -0.024       & 0.013               & 0.388        & 0.055               \\ \hline
\textbf{U}           & 0.009        & 0.020               & 0.281        & 0.090               & 0.003        & 0.022               & 0.346        & 0.096               \\ \hline
\bm{$x_{e^{-}}$} & -0.081       & 0.030               & 0.510        & 0.096               & -0.118       & 0.038               & 0.686        & 0.130               \\ \hline
\bm{$x_{HI}$}    & -0.039       & 0.016               & 0.451        & 0.064               & -0.045       & 0.014               & 0.454        & 0.051               \\ \hline
\end{tabular}
\end{table}

\subsection{Comparing the two approaches}
In \autoref{tab:parameters-two-methods} we present all the parameters (slope $m$, intercep $b$ and their associated errors $\delta m$ and $\delta b$) for both methods. \par
\begin{figure}[t!]
    \centering
    \captionsetup{aboveskip=3pt,belowskip=-3pt}
    \includegraphics[height=18.2cm]{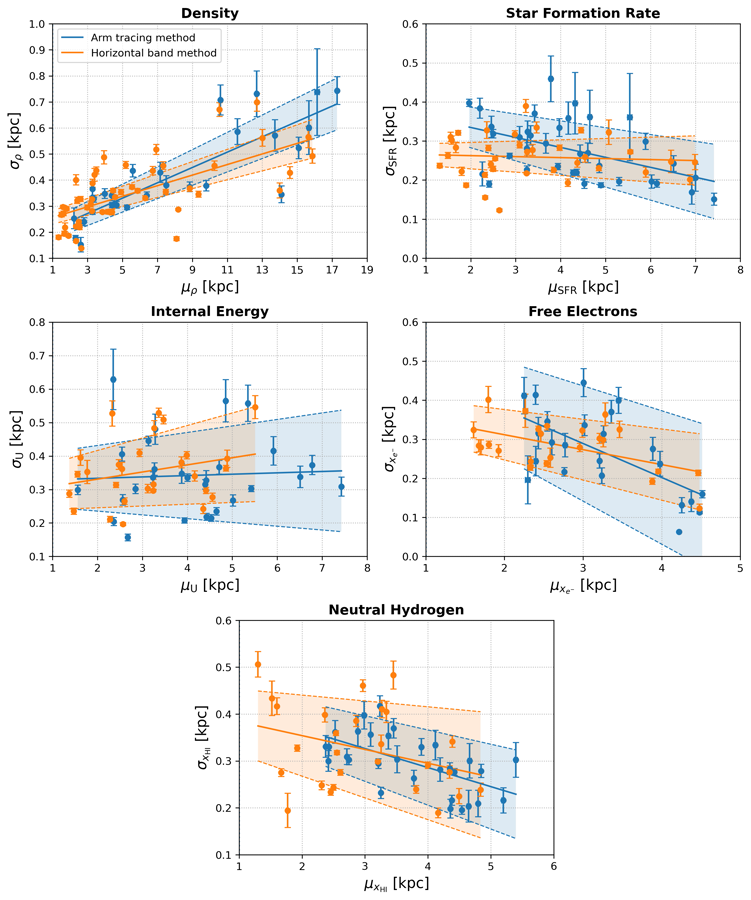}
    \caption[Comparison between the two methods: linear fits.]{A comparison between the two methods. We plot the best-fits and their $1-\sigma$ error.}
    \label{fig:band-vs-tracing-fits}
\end{figure}

 
To better grasp this information, we have plotted the linear-fits and their corresponding $1-\sigma$ error in \autoref{fig:band-vs-tracing-fits}. Given the amount of data and the differences between the two methods, we can better discuss the features of each property.

\begin{itemize}
    \item \textbf{Density ($\rho$)}: with the addition of the data from the Arm Tracing method, we further confirm the growing tendency of the widths of the spiral arms measured in density with galactic radii. We see that when we do the Arm Tracing we can reach farther regions of the spiral arms. This was perhaps the only property where we could conclude confidently, and the addition of new data complements what we already mentioned in the the first part of this section.  \par
    
    \item \textbf{Star Formation Rate ($\mathrm{SFR}$)}: we initially mentioned the great amount of scatter in the widths that made it difficult to determine if there exists (or not) a systematic widening. The slope obtained with the Horizontal Band method is $m=-0.003$, when using this Arm Tracing method, we got $m=-0.024$, showing a more clear decreasing tendency. By avoiding the mixing of data from the two spiral arms, we still see dispersion but the data shows a more clear anti correlation with $R$.   \par
    
    \item \textbf{Internal Energy ($\mathrm{U}$)}: while we managed to get more data at larger radii, the widths distribution appears even flatter than what we got with the Horizontal Band method. In fact, when we plot the $1-\sigma$ fits, the upper bound has a positive slope ($m=0.019$) and the lower bound a negative slope ($m=-0.011$). This is caused by how close to zero the slope of the best-fit is ($m=0.004$) and how big the dispersion is ($\delta m = 0.015$) in comparison with the best-fit value. If anything, we can say that the arm widths measured with the Internal Energy thresholds, appear to oscillate around a value of $\bar{\sigma}_{U} \approx 0.35 $ with a high dispersion. \par
    
    \item \textbf{Free Electrons ($x_{e^{-}}$)}: despite trying to pick points in the arms at large radii, these do not go beyond $4.5$ kpc in the case of the most ionized regions. With the Arm Tracing method the distribution keeps favouring an anti-correlation with galactic radii, however the dispersion of the data increased. This shows that the variability of the width is intrinsic to the arms themselves and is not caused by the Horizontal Band Method as we argued in \autoref{A-different-approach}. \par
    
    \item \textbf{Neutral Hydrogen ($x_{\mathrm{HI}}$)}: in this case the amount of disperion decreased, we were able to estimate the widths beyond $R=5$ kpc and the anti-correlation remains statistically valid. It maintains its similarity with the results obtained for the Free Electrons. \par 

\end{itemize}

\begin{table}[H]
\centering
\caption[Parameters of the linear fit for both methods.]{Parameters obtained from the linear fit for each property using the two methods.}
\label{tab:parameters-two-methods}
\begin{tabular}{|c|c|c|c|c|c|c|c|c|}
\hline
\multirow{2}{*}{}    & \multicolumn{4}{c|}{\textbf{Horizontal Band Method}}                    & \multicolumn{4}{c|}{\textbf{Arm Tracing Method}}                            \\ \cline{2-9} 
                     & \bm{$m$} & \bm{$\delta m$} & \bm{$b$} & \bm{$\delta b$} & \bm{$m$} & \bm{$\delta m$} & \bm{$b$} & \bm{$\delta b$} \\ \hline
\bm{$\rho$}      & 0.020        & 0.003               & 0.236        & 0.022               & 0.029        & 0.004               & 0.183        & 0.035                 \\ \hline
\textbf{SFR}         & -0.003       & 0.006               & 0.268        & 0.022               & -0.025       & 0.008               & 0.386        & 0.036               \\ \hline
\textbf{U}           & 0.021        & 0.016               & 0.288        & 0.054               & 0.004        & 0.015               & 0.325        & 0.067               \\ \hline
\bm{$x_{e^{-}}$} & -0.038       & 0.013               & 0.388        & 0.038               & -0.087       & 0.023               & 0.550        & 0.078               \\ \hline
\bm{$x_{HI}$}    & -0.029       & 0.017               & 0.413        & 0.053               & -0.041       & 0.010               & 0.448        & 0.039               \\ \hline
\end{tabular}
\end{table}

\begin{figure}[H]
    \centering
    \captionsetup{aboveskip=3pt,belowskip=3pt}
    \includegraphics[height=7cm]{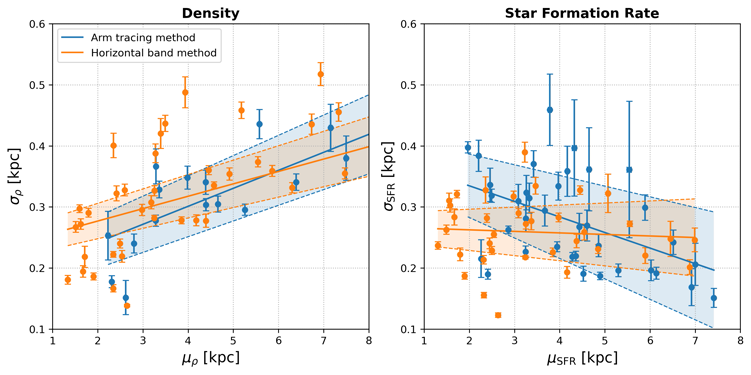}
    \caption[Arm widths in the same $R$ range for the density and the SFR.]{Arm widths in the same $R$ range for the density and the SFR.}
    \label{fig:density-vs-sfr}
\end{figure}

In \autoref{fig:density-vs-sfr} we plot, side-by-side, the widths of the arms measured in density and in star formation rate. The upper bound of the $x$ axis is limited to $R=8$ so we can compare the regions where data is available for both properties. The limit is imposed by the SFR, which arms reach up to $R \approx 7.3$ kpc. \par

From \autoref{fig:density-vs-sfr}, we see that despite the arms expand and enlarge their diameter as we move away from the center of the galaxy (in the density distribution), the star formation does not broaden with the arms, on the contrary, star formation regions shrink and only take place in the most inner parts of the spiral arms. \par

We can conclude that the abundance of star formation activity depends more on the global density distribution than on how width the spiral arms are. The density threshold to trigger star formation is more easily reached in the inner parts of the gas disk. When we compare the widths of the arms in SFR and the density, we see that they have similar values for $R < 5$ kpc:

\begin{equation}
    \begin{aligned}
    \bar{\sigma}_{\rho}[R<5 \ \text{kpc}] = (0.287 \pm 0.014) \ \text{kpc} \\
    \bar{\sigma}_{\text{SFR}}[R<5 \ \text{kpc}] = (0.278 \pm 0.017) \ \text{kpc}.
    \nonumber
    \end{aligned}
\end{equation}
This indicates that star formation is taking place all across the spiral arm, i.e, the density threshold for the star formation to take place is reached across all the arm. \par 

The opposite occurs when we look at the widths at larger radii:
\begin{equation}
    \begin{aligned}
    \bar{\sigma}_{\rho}[5\ \text{kpc}<R<8 \ \text{kpc}] = (0.426 \pm 0.022) \ \text{kpc} \\
    \bar{\sigma}_{\text{SFR}}[5\ \text{kpc}<R<8 \ \text{kpc}] = (0.238 \pm 0.024) \ \text{kpc}.
    \nonumber
    \end{aligned}
\end{equation}
We have that the arms are $\approx 1.8$ times bigger in density than in the SFR. This indicates that star formation is taking place only in the most interior parts of the spiral arms, where the density threshold for star formation rate is met. \par 

\begin{figure}[H]
    \centering
    \captionsetup{aboveskip=3pt,belowskip=3pt}
    \includegraphics[height=6cm]{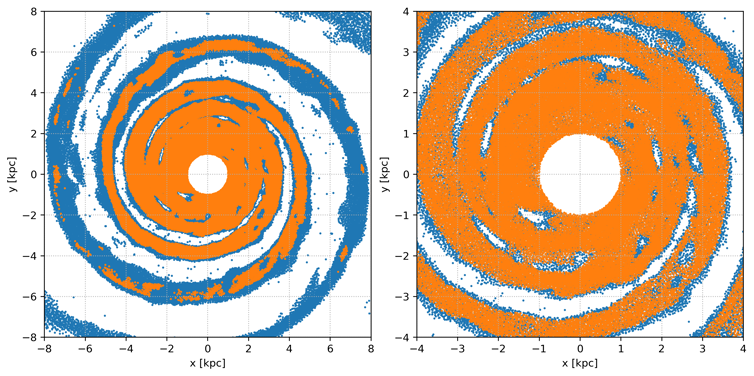}
    \caption[Superposition of the density and SFR spiral structure.]{Superposition of the density and SFR spiral structure. Blue points are the arms when using the density threshold. Orange points are the arms when using the SFR threshold. }
    \label{fig:density-vs-sfr-disk}
\end{figure}
A superposition of both spiral structures confirms what the data says. We can clearly observe in \autoref{fig:density-vs-sfr-disk} how the star formation regions only happen in the mos inner parts of the disk as the arms go away from the center. In the central region, the superposition of the SFR spiral structure almost completely hides the density spiral structure. We got slightly narrower arm widths in the case of the SFR for $R<5$ kpc, which supports what we see in the right plot of \autoref{fig:density-vs-sfr-disk}. \par
\begin{figure}[H]
    \centering
    \captionsetup{aboveskip=3pt,belowskip=3pt}
    \includegraphics[height=8cm]{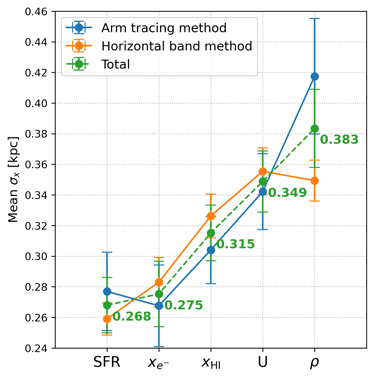}
    \caption[Mean arm width for every property.]{Mean arm width for every property, in increasing order. We see that the width of the arms measured in both $\mathrm{SFR}$ and $x_{e^{-}}$ have similar values of width. Blue points show the values for the Arm Tracing method, orange points show the results for the Horizontal Band method. The green dashed lines averages these two. The numerical values are plotted beside every point.}
    \label{fig:results-mean-sigmas}
\end{figure}

 
\figref{results-mean-sigmas} summarizes the results, we plot the mean values of the widths, the errors are the mean of all the individual error associated to each value of width. We arranged the mean widths in increasing order, being the SFR and Free Electrons arms the narrower, at $\sigma \approx 0.27$. The Neutral Hydrogen and Internal energy arms are in an intermediate point at $\sigma \approx 0.32$ and finally the widest arms are for the density at $\sigma \approx 0.38$.

\subsection{Displacement of the star formation in the spiral arms}

From the data of the Horizontal Band Method, we can evaluate if there is a separation between the position of the spiral arms when measured in density and when measured in SFR. We want to see if there is any systematic displacement between the zones where star formation takes place and the density-arm itself. \par 

\begin{figure}[H]
    \centering
    \captionsetup{aboveskip=3pt,belowskip=3pt}
    \includegraphics[width=1.0\linewidth ]{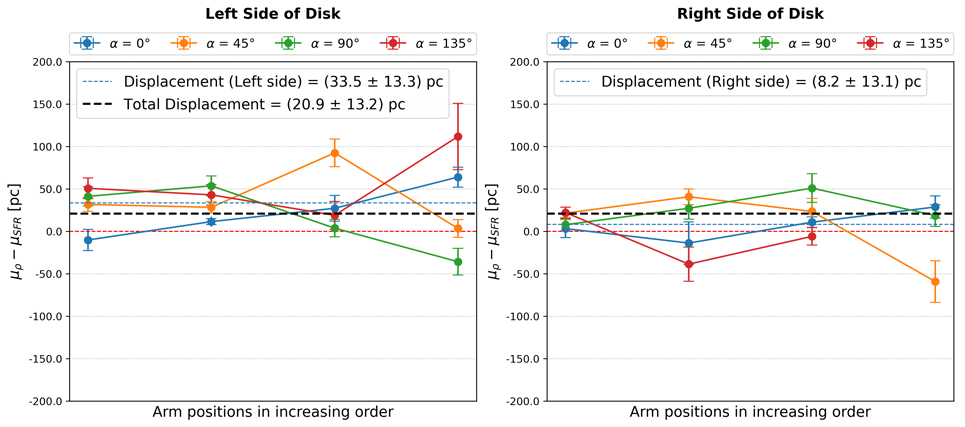}
    \caption[Displacement of the star formation regions in the spiral arms.]{Differences between arm centers measured in density and in SFR. The $x$ axis has no units, and it simply helps the visualization of the data. The $y$ axis shows the difference $\mu_{\rho} - \mu_{SFR}$ for the points in the disk where we measured both $\mu_{\rho}$ and $\mu_{SFR}$, in the same spiral arm. We show the results for all disk orientations (see \autoref{fig:rotations-density}). }
    \label{fig:displacements-rho-sfr}
\end{figure}

To achieve this, we took all the positions in the disk where we measured both arm centers (in the arms measured in density and SFR). From this, me compute the difference $\mu_{\rho} - \mu_{SFR}$. The results are shown in \autoref{fig:displacements-rho-sfr} for both sides of the disk and for every available orientation (see \autoref{fig:rotations-density}). \par

We found a total mean displacement (indicated with the black dashed line) of $\Delta \mu = (20.9 \pm 13.2) $ pc. This result is in agreement with a model where the star formation lags behind the spiral arms due to the dynamical time associated to the star formation process. This result also supports a prediction of the density-wave theory. The regions where the star formation takes place rotate with a fixed spiral pattern. Due to the dynamical time-scale, the new-born stars fall behind (or move ahead) from the gaseous arm (\citealt{pour2016strong}). \par

\subsection{Comparing with observations}
\begin{figure}[t!]
\captionsetup[subfigure]{labelformat=empty}
\begin{subfigure}{0.49\textwidth}
  \centering
  \includegraphics[width=0.99\linewidth]{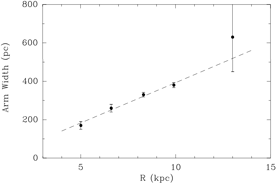}
  \caption{}
  \label{fig:reid2014}
\end{subfigure}%
\begin{subfigure}{.5\textwidth}
  \centering
  \includegraphics[width=0.99\linewidth]{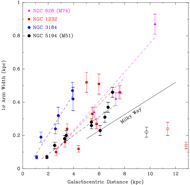}
  \caption{}
  \label{fig:honig2015}
\end{subfigure}
\vspace{-1\baselineskip}
\caption[Observational results of arm-widths by \citealt{reid2014trigonometric} and \citealt{honig2015characteristics}]{Observational results of arm-widths by \citealt{reid2014trigonometric} and \citealt{honig2015characteristics}.}
\label{fig:observational-results-1}
\end{figure}

There are observational records of such behaviour made by \citealt{reid2014trigonometric}, for the spiral arms of the Milky Way. They used trigonometric parallaxes and proper motions from high-mass star forming regions to map the spiral structure of our galaxy and obtain measures of the width of the spiral arms. Their result is presented in \figref{reid2014}. \par 
Other measures of the spiral arm widths have been made by \citealt{honig2015characteristics}, who used the location of giant $\mathrm{H}_{II}$  regions to trace the spiral structure of late-type spiral galaxies NGC 628 (M74), NGC 1232, NGC 3184 and NGC 5194 (M51). They used Hubble Space Telescope, the Jakobus Kapteyn Telescope and VLT (Very Large Telescope) images, with $H_{\alpha}$ and Blue filters. Their method is different, they fit the spiral pattern as a log-periodic spiral, construct the distribution of minimum distances of $\mathrm{H}_{II}$ to the arm, and adopt a Gaussian approximation to that distribution, taking 1-$\sigma$ as the width. Their results are presented in \figref{honig2015}. \par 

\citealt{honig2015characteristics} mention the presence of some outliers in their distribution: points with very low values of width at large radii. This can be seen in the bottom right of the \figref{honig2015}. They attribute this to the narrowing of the spirals arms at their ends. \par

\newpage
\section{Conclusions}

In this work we studied the properties of the gas disk in a simulation of an isolated disk galaxy. The galaxy model contains interaction between stars through gravity. It uses a TreePM algorithm to calculate the gravitational forces between particles (\citealt{springel2003history}) and it includes a formulation of Ideal Magnetohydrodynamics by \citealt{hopkins2016accurate}. We briefly described it in \autoref{Description-of-the-simulations}. \par
\medskip
The galaxy model includes a self-regulated star formation model by \citealt{springel2003cosmological}. This model is an statistical formulation of the process of star formation with a two-phase interstellar medium. We made a detailed description of that model in \autoref{Parameters-of-the-galaxy-model}, where we talk about the mass and energy transfer between the cold and hot phases. The model includes feedback mechanisms from stellar winds and supernovae explosions. \par 
\medskip
The physical parameters of the galaxy in the simulation are based on the main component of a minor merger, AM2322-321, located at the Octans constellation. Its parameters were set based on observations made by \citealt{ferreiro2008sample} and surface photometry performed by \citealt{arboleda2019}. A list with all the parameters for the galaxy model can be found in \autoref{tab:final-parameters} and \autoref{tab:final-parameters-sim}. A whole sequence of the galaxy evolution during the simulation is in \autoref{fig:snapshot_collage}, where the snapshot used in this work is shown. \par 
\medskip
We developed a method for the extraction of the spiral structure that is inside the gas disk. This method consists in measuring a contrast parameter, \autoref{eq:deltax}, that sets a threshold for the over-dense parts of the disk. The entire description of the method is outlined with great detail for the density of the gas in \autoref{extraction}. \autoref{fig:interp-density}, \autoref{fig:density-threshold} and \autoref{fig:density-compared} summarize the procedure for the extraction of the spiral structure in the disk. \par 
\medskip
After obtaining the spiral structure for each property, using each property's threshold (see \autoref{fig:density-compared}, \autoref{fig:sfr-compared}, \autoref{fig:U-compared}, \autoref{fig:ionization-regimes-compared}, \autoref{fig:Hn-compared}), we explain a method to study the local properties of the arms, specifically, a technique to measure the widths of the arm locally. \par
\medskip
We model the transverse structure in each property, using a simple gaussian function, \autoref{eq:gaussian}. The model is discussed and described in \autoref{gaussian-model} We extract the envelopes of those contrast regions in a plot of $x$ vs $\rho$ (or $x$ vs $SFR$, $x$ vs $U$ and the other properties). The results of the fit are presented in \autoref{fig:left-right-arms}, \autoref{fig:left-right-arms-sfr}, \autoref{fig:left-right-arms-U}, \autoref{fig:left-right-arms-Ne} and \autoref{fig:left-right-arms-Hn}. \par 
\medskip
The spiral structure extraction and the arm width measuring is described for the density, SFR, Free Electrons and Neutral Hydrogen in \autoref{Data-extraction}, \autoref{SFR-results}, \autoref{Us-results}, \autoref{Nes-results} and \autoref{H-results}, respectively. \par  
\medskip
To evaluate the arm widths as a functions of $R$ we followed two methods. The first one we called the \enquote{Horizontal Band Method}. This method evaluates the arm widths in a horizontal strip of disk centered at $y=0$. \autoref{fig:band-selection}, \autoref{fig:band-selection-sfr}, \autoref{fig:band-selection-U}, \autoref{fig:band-selection-Ne} and \autoref{fig:band-selection-Hn-log} illustrate this method and summarize it for each property. \par
\medskip
We made four iterations of this procedure for different disk orientations. We rotated the disk an collected data for $\alpha = [0^{\circ}, 45^{\circ}, 90^{\circ}, 135^{\circ}]$, as we illustrate with the density distribution of particles in \autoref{fig:rotations-density}. \par
\medskip
In \autoref{fig:results-grid} and \autoref{fig:results-grid-radial} we show all the collected data in the four disks orientations already mentioned. We did linear fitting to the data, the estimated parameters for the linear fits in \autoref{fig:results-grid-radial} are gathered in \autoref{tab:parameters-band}, with their corresponding errors. \par
\medskip
We describe a different approach to trace the widths for each arm, separately. We called this approach the \enquote{Arm Tracing Method}. The idea is to independently trace the width of the two arms, as shown in \autoref{fig:arm-tracing-disks-grid}. The results of this procedure are shown in \autoref{fig:arm-tracing-results-grid}, along with linear fits for each spiral arm. The parameters from the fits and their errors are listed in \autoref{tab:parameters-arm1-arm2}. We found that the linear fits for the Arm1 and Arm2 are very similar in every property, except for the Free Electrons where Arm2's width decreases faster than Arm1. \par 
\medskip
From applying the two methods we arrived at different conclusion for every property. The widths of the arms measured in density clearly increase with $R$. The SFR arms decrease with $R$. The negative slope of the linear fit for the widths in SFR is similar to the positive slope of the linear fit for the widths of the arms in density ($m_{SFR}=-0.025$ vs $m_{\rho}=0.029$), using the Arm Tracing method. \par
\medskip
For the Internal Energy, we saw that the widths oscillate around a value of $\sigma_{U} = 0.349$, with a high dispersion. For the Free Electrons, we found that the variability of the width is intrinsic to the arms themselves and is not caused by the Horizontal Band Method. It shows an anti-correlation with $R$. For the Neutral Hydrogen our conclusions were similar, since it behaves similarly as the Free Electrons, but with less dispersion. \par
\medskip
We took a closer look at the density and SFR in \autoref{fig:density-vs-sfr} and \autoref{fig:density-vs-sfr-disk}.  We found that despite the arms expand and enlarge their diameter as we move away from the center of the galaxy (in the density distribution), the star formation does not broaden with the arms, on the contrary, star formation regions shrink and only take place in the innermost parts of the spiral arms. From this we can conclude that the abundance of star formation activity depends more on the global density distribution than on how width the spiral arms are. The density threshold to trigger star formation is more easily reached in the inner parts of the gas disk and is only reached in the very centers of the spiral arms at large radii. \par 
\medskip
We found that $\bar{\sigma}_{\rho}[R<5 \ \text{kpc}]$ and $\bar{\sigma}_{\text{SFR}}[R<5 \ \text{kpc}] $ have almost the same value. On the other hand, we found that $\bar{\sigma}_{\rho}[5\ \text{kpc}<R<8 \ \text{kpc}]$ is $\approx 1.8$ times bigger than $\bar{\sigma}_{\text{SFR}}[5\ \text{kpc}<R<8 \ \text{kpc}]$, supporting the conclusion in the previous paragraph. \par 
\medskip
We used the data from the Horizontal Band Method to measure the mean separation between the arms measured in density and in SFR. The results are shown in \autoref{fig:displacements-rho-sfr}. We found that there is a systematic displacement between the arm centers. The mean separation value found was $\Delta \mu = (20.9 \pm 13.2)$ pc. This result is in agreement with a prediction of the density-wave theory where the regions of star formation move with a fixed spiral pattern and the new-born stars appear displaced from the gaseous disk due to the dynamical time associated to the star formation process. \par
\medskip
Finally, we look at some observational results from \citealt{reid2014trigonometric} and \citealt{honig2015characteristics}. The authors of this work found a positive correlation between the arm width and galactic radii in the Milky Way and another four grand-design spiral galaxies. Their results are in agreement with the findings of this work for the the widths of the arms measured in density. \par

\include{Chapters/Chapter5} 

%----------------------------------------------------------------------------------------
%	THESIS CONTENT - APPENDICES
%----------------------------------------------------------------------------------------

\appendix % Cue to tell LaTeX that the following "chapters" are Appendices

% Include the appendices of the thesis as separate files from the Appendices folder
% Uncomment the lines as you write the Appendices

%\include{Appendices/AppendixA}
%\include{Appendices/AppendixB}
%\include{Appendices/AppendixC}

%----------------------------------------------------------------------------------------
%	BIBLIOGRAPHY
%----------------------------------------------------------------------------------------

\printbibliography[heading=bibnumbered]

%----------------------------------------------------------------------------------------

\end{document}